\documentclass[english,twocolumn,10pt]{article}
\usepackage{pdfpages}
\usepackage{import}
\usepackage{amsfonts}
\usepackage{braket}
\usepackage{times}
\usepackage{graphicx}
\usepackage{color}
\graphicspath{{Figures/}}
\usepackage[labelfont=bf]{caption}
\usepackage{siunitx}
\usepackage[caption=false]{subfig}

\usepackage[section]{placeins}
\usepackage{xspace} 
\usepackage{url}

\usepackage[breaklinks]{hyperref}

\usepackage{capt-of}

\usepackage[switch]{lineno}  

\usepackage{titling}  
\usepackage{mathtools}

\newcommand{\lind}{\ensuremath{\mathbb{L}}}

\RequirePackage[T1]{fontenc}
\RequirePackage[utf8]{inputenc}
\RequirePackage{lmodern}

\usepackage[nomain, acronym]{glossaries}

\renewcommand{\figurename}{Fig.}

\DeclareCaptionLabelSeparator{bar}{ | }
\newcommand{\target}[1]{\mathbb{#1}}

\title{Precision tomography of a three-qubit donor quantum processor in silicon}

\usepackage{authblk}

\author[1]{Mateusz T. M\k{a}dzik\thanks{These two authors contributed equally.}\thanks{Currently at QuTech, Delft University of Technology, 2628 CJ Delft, The Netherlands.}}
\author[1]{Serwan Asaad{$^*$}\thanks{Currently at Center for Quantum Devices, Niels Bohr Institute, University of Copenhagen, and Microsoft Quantum Lab Copenhagen, Copenhagen, Denmark.}}
\author[2,3]{Akram Youssry}
\author[1]{Benjamin Joecker}
\author[4]{Kenneth M. Rudinger}
\author[4]{Erik Nielsen}
\author[4]{Kevin C. Young}
\author[4]{Timothy J. Proctor}
\author[5]{Andrew D. Baczewski}
\author[1]{Arne Laucht}
\author[1]{Vivien Schmitt\thanks{Currently at Univ. Grenoble Alpes, Grenoble INP, CEA, IRIG-PHELIQS, F-38000 Grenoble, France.}}
\author[1]{Fay E. Hudson}
\author[6]{Kohei M. Itoh}
\author[7]{Alexander M. Jakob}
\author[7]{Brett C. Johnson}
\author[7]{David N. Jamieson}
\author[1]{Andrew S. Dzurak}
\author[2]{Christopher Ferrie}
\author[4]{Robin Blume-Kohout}
\author[1]{Andrea Morello
	\thanks{To whom correspondence should be addressed; E-mail: a.morello@unsw.edu.au}}

\affil[1]{School of
	Electrical Engineering and Telecommunications, UNSW Sydney, Sydney, NSW 2052, Australia}
\affil[2]{Centre for Quantum Software and Information, University of Technology Sydney, Ultimo, NSW 2007, Australia}
\affil[3]{Department of Electronics and Communication Engineering, Faculty of Engineering, Ain Shams University, Cairo, Egypt
}
\affil[4]{Quantum Performance Laboratory, Sandia National Laboratories, Albuquerque, NM 87185 and Livermore, CA 94550, USA}
\affil[5]{Center for Computing Research, Sandia National Laboratories, Albuquerque, NM 87185, USA}
\affil[6]{School of Fundamental Science and Technology, Keio University, Kohoku-ku, Yokohama, Japan}
\affil[7]{School of Physics, University of Melbourne, Melbourne, VIC 3010, Australia}

\date{}

\begin{document} 
	
	\maketitle 
	\addcontentsline{toc}{section}{Main text}
	
	\textbf{Nuclear spins were among the first physical platforms to be considered for quantum information processing\cite{Kane1998,Vandersypen2005}, because of their exceptional quantum coherence\cite{Saeedi2013} and atomic-scale footprint. However, their full potential for quantum computing has not yet been realized, due to the lack of methods to link nuclear qubits within a scalable device combined with multi-qubit operations with sufficient fidelity to sustain fault-tolerant quantum computation. Here we demonstrate universal quantum logic operations using a pair of ion-implanted $^{31}$P donor nuclei in a silicon nanoelectronic device. A nuclear two-qubit controlled-Z gate is obtained by imparting a geometric phase to a shared electron spin\cite{filidou2012ultrafast}, and used to prepare entangled Bell states with fidelities up to 94.2(2.7)\%. The quantum operations are precisely characterised using gate set tomography (GST)\cite{nielsen2020gate}, yielding one-qubit average gate fidelities up to 99.95(2)\%, two-qubit average gate fidelity of 99.37(11)\% and two-qubit preparation/measurement fidelities of 98.95(4)\%. These three metrics indicate that nuclear spins in silicon are approaching the performance demanded in fault-tolerant quantum processors \cite{fowler2012surface}. We then demonstrate entanglement between the two nuclei and the shared electron by producing a Greenberger-Horne-Zeilinger three-qubit state with 92.5(1.0)\% fidelity. Since electron spin qubits in semiconductors can be further coupled to other electrons\cite{harvey2017coherent,he2019two,madzik2021conditional} or physically shuttled across different locations\cite{hensen2020silicon,yoneda2020coherent}, these results establish a viable route for scalable quantum information processing using donor nuclear and electron spins.
	}
	\ \\
	
	\begin{figure*}[!h]
			 \centering
			 \includegraphics{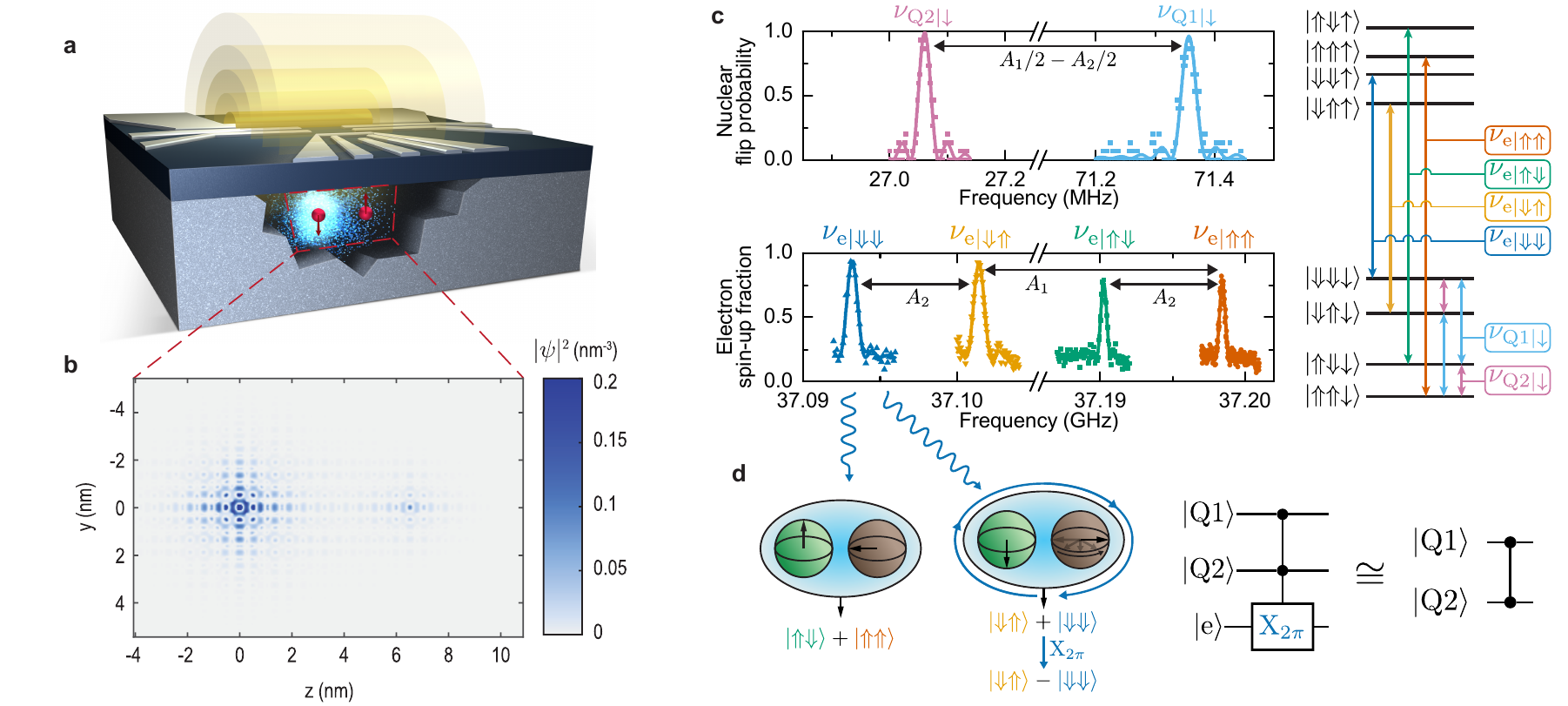}
			\captionof{figure}{
				\textbf{Operation of a one-electron  -- two-nuclei quantum processor.}
				\textbf{a,} Artist's impression of a pair of $^{31}$P nuclei (red), asymmetrically coupled to the same electron (blue). The spins are controlled by oscillating magnetic fields (yellow) generated on-chip. \textbf{b,} Effective-mass calculation of the wavefunction $\psi(y,z)$ of the third electron on the 2P cluster. The observed values of hyperfine coupling are well reproduced by assuming a 6.5~nm spacing between the donors.
				\textbf{c,} Experimental NMR spectrum of the $^{31}$P nuclei (top) and ESR spectrum of the shared electron (bottom) at $B_0=1.33$ T, along with energy level diagram (right) of the eight-dimensional Hilbert space (spacings not to scale).
				The spectra yield the hyperfine couplings $A_1 \approx 95~\mathrm{MHz}$ and $A_2 \approx 9~\mathrm{MHz}$ between the electron and the nuclear qubits Q1, Q2.
				\textbf{d,} Implementation of a geometric two-qubit CZ gate.
				A conditional $\pi$ phase shift is acquired when a $2\pi$ rotation is applied on the electron spin at frequency $
				\nu_{\mathrm{e}|\Downarrow\Downarrow}$, i.e. conditional on the nuclear spins being $\ket{\Downarrow\Downarrow}$.
				This operation corresponds to the CZ gate on the nuclei when restricted to the electron $\ket{\downarrow}$ subspace.}
			\label{fig:figure 1}
		\end{figure*}

	Nuclear spins are the most coherent quantum systems in the solid state \cite{Saeedi2013,zhong2015optically}, owing to their extremely weak coupling to the environment. In the context of quantum information processing, the long coherence is associated with record single-qubit gate fidelities \cite{Muhonen2015}. However, the weak coupling poses a challenge for multi-qubit logic operations. Using spin-carrying defects in diamond \cite{bradley2019ten} and silicon carbide \cite{bourassa2020entanglement}, this problem can be addressed by coupling multiple nuclei to a common electron spin, thus creating quantum registers that can sustain small quantum logic operations and error correction \cite{Waldherr2014}. Exciting progress is being made on linking several such defects via optical photons \cite{bhaskar2020experimental,Pompili2021}.
	
	Still missing, however, is a pathway to exploit the atomic-scale dimension of nuclear spin qubits to engineer scalable quantum processors, where densely-packed qubits are integrated and operated within a semiconductor chip \cite{vandersypen2017interfacing}. This requires entangling the nuclear qubits with electrons that can either be physically moved, or entangled with other nearby electrons. It also requires interspersing the electron-nuclear quantum processing units with spin readout devices \cite{Morello2010}. Here we show experimentally that silicon - the material underpinning the whole of modern digital information technology - is the natural system in which to develop dense nuclear spin based quantum processors \cite{Kane1998}.
	
	\noindent \textbf{One electron -- two nuclei quantum processor}
	
	The experiments are conducted on a system of two $^{31}$P donor atoms, introduced in an isotopically purified $^{28}$Si substrate by ion implantation (see Methods). A three-qubit processor is formed by using an electron (e) with spin $S=1/2$ (basis states $\ket{\uparrow}, \ket{\downarrow}$) and two nuclei (Q1, Q2) with spin $I=1/2$ (basis states $\ket{\Uparrow}, \ket{\Downarrow}$). Metallic structures on the surface of the chip provide electrostatic control of the donors, create a single-electron transistor (SET) charge sensor, and deliver microwave and radiofrequency signals through a broadband antenna (Fig.~\ref{fig:figure 1}a, Extended Data Fig.~\ref{fig ED: device SEM}). With this setup, we can perform single-shot electron spin readout \cite{Morello2010}, and high fidelity ($\approx 99.9\%$) single-shot quantum nondemolition readout of the nuclear spins \cite{Pla2013}, as well as nuclear magnetic resonance (NMR) and electron spin resonance (ESR) \cite{Pla2012} on all spins involved (see Methods).

	The ESR spectra in Fig.~\ref{fig:figure 1}c exhibit four resonances. This means that the ESR frequency depends upon the state of two nuclei, to which the electron is coupled by contact hyperfine interactions $A_1 \approx 95$~MHz and $A_2 \approx 9$~MHz, with a dependence on the gate potentials caused by the Stark shift (Extended Data Fig.~\ref{fig ED: hyperfine shifts}). We adopt labels where, for instance, $\nu_{\mathrm{e}|\Downarrow\Downarrow}$ represents the frequency at which the electron spin undergoes transitions conditional on the two nuclear spin qubits being in the $\ket{\rm Q_1 Q_2} = \ket{\Downarrow\Downarrow}$ state, and so on. The values of $A_1, A_2$ can be independently checked by measuring the frequencies $\nu_{Q1|\downarrow}, \nu_{Q2|\downarrow}$ at which each nucleus responds while the electron is in the $\ket{\downarrow}$ state (Supplementary Information S1). The hyperfine-coupled electron could either be the first or the third electron bound to the donor cluster. Since its spin relaxation time $T_{\rm 1e}$ is three orders of magnitude shorter than expected from a one-electron system (Extended Data Fig.~\ref{fig ED: electron properties}), we interpret the ESR spectrum in Fig.~\ref{fig:figure 1}c as describing the response of the third electron bound to a 2P donor system.

	An effective-mass calculation of the wavefunction of the third electron in a 2P system (see Methods) reproduces the observed values of $A_1$ and $A_2$ by assuming donors spaced $6.5$~nm apart, and subjected to an electric field 2~mV/nm that pulls the electron wavefunction more strongly towards donor 1 (Fig.~\ref{fig:figure 1}b). The $^{31}$P nuclei in this 2P cluster are spaced more widely than those  produced by scanning probe lithography \cite{he2019two,ivie2021impact}, where the sub-nanometre inter-donor spacing causes a strongly anisotropic hyperfine coupling, which randomizes the nuclear spin state each time the electron is removed from the cluster for spin readout \cite{hile2018addressable}. Here, instead, the probability of flipping a nuclear spin by electron ionisation is of order $10^{-6}$ (Extended Data Fig.~\ref{fig ED: nuclear flipping rates}), meaning that our nuclear readout is almost perfectly quantum nondemolition.
	
	\noindent \textbf{Nuclear two-qubit operations}
	\begin{figure}
			\centering
			 \includegraphics{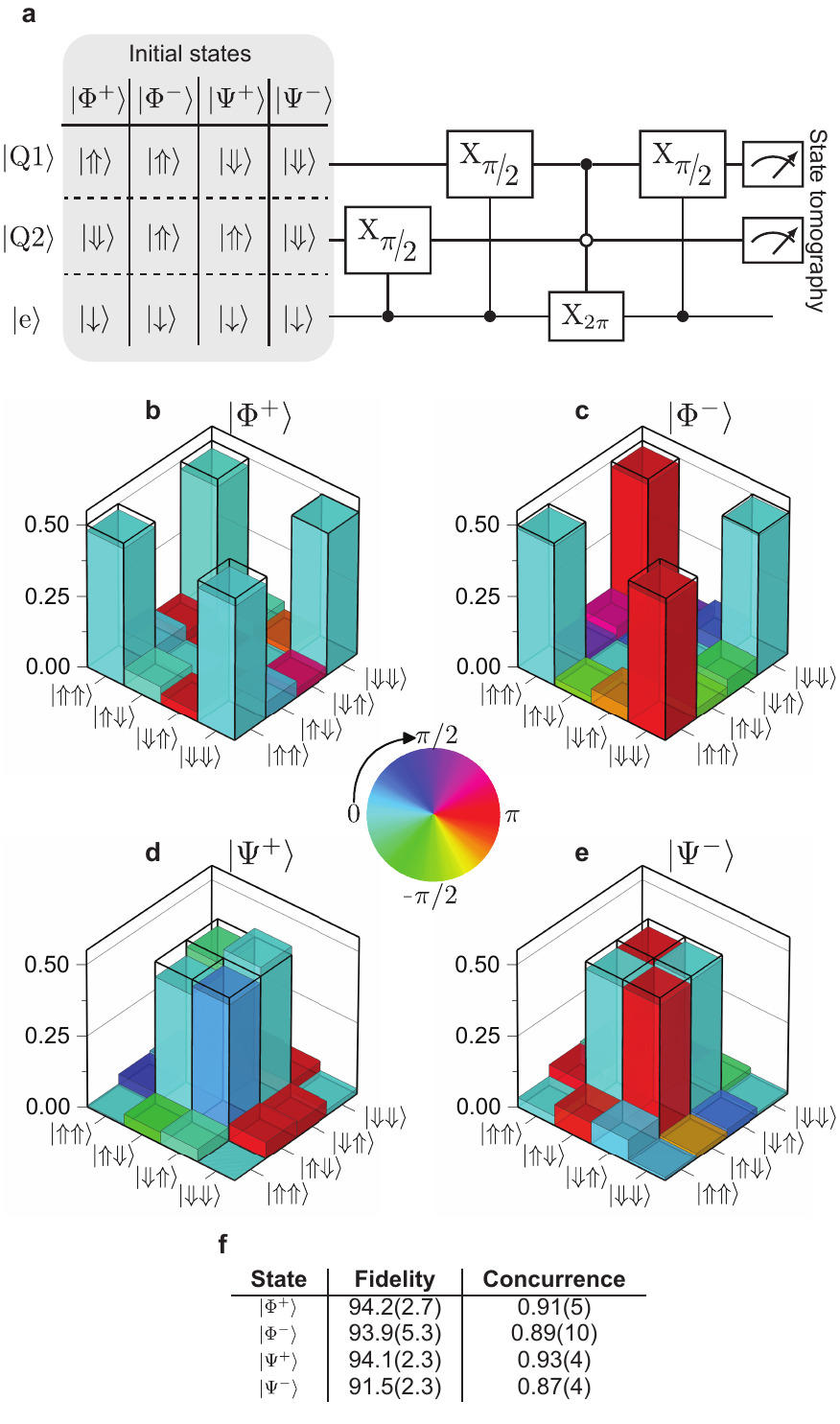}
			\captionof{figure}{\textbf{Tomography of nuclear Bell states.} \textbf{a,} Each of the four Bell states has been generated using the same quantum circuit, only varying the initial spin state. \textbf{b-e,} Quantum state tomography results for (b) $\Phi^+$; (c) $\Phi^-$; (d) $\Psi^+$; (e) $\Psi^-$ Bell state. No corrections have been applied to compensate readout errors. Hollow, black boxes indicate the outcome of an ideal measurement for each Bell state. \textbf{f,} Table of Bell state fidelities and concurrences. The error bars are estimated using Monte Carlo bootstrap re-sampling and represent $1\sigma$ confidence level.
			}
			\label{fig:figure 2}
		\end{figure}
		
	We first consider the two $^{31}$P nuclear spins as the qubits of interest. One-qubit logic operations are trivially achieved by NMR pulses \cite{Pla2013} (Methods and Extended Data Fig.~\ref{fig: nuclear T2}), where $A_1 \neq A_2$ provides the spectral selectivity to address each qubit individually (Fig.~\ref{fig:figure 1}c). Two-qubit operations are less trivial, since the nuclei are not directly coupled to each other (Supplementary Information S1 and S9). They are, however, hyperfine-coupled to the same electron. This allows the implementation of a geometric two-qubit controlled-Z (CZ) gate \cite{filidou2012ultrafast,Waldherr2014}.

	When a quantum two-level system is made to trace a closed trajectory on its Bloch sphere, its quantum state acquires a geometric phase equal to half the solid angle enclosed by the trajectory \cite{anandan1992geometric}. Fig.~\ref{fig:figure 1}d illustrates how an electron $2\pi$-pulse at the frequency $\nu_{\mathrm{e}|\Downarrow\Downarrow}$ (see Fig.~\ref{fig:figure 1}d) constitutes a nuclear CZ 2-qubit gate. Starting from the state $\ket{\Downarrow}\otimes(\ket{\Downarrow}+\ket{\Uparrow})/\sqrt{2} \equiv (\ket{\Downarrow\Downarrow}+\ket{\Downarrow\Uparrow})/\sqrt{2}$, the electron $X_{2\pi}$ pulse at $\nu_{\rm e|\Downarrow\Downarrow}$ introduces a phase factor $e^{i\pi}=-1$ to the $\ket{\Downarrow\Downarrow}$ branch of the superposition, resulting in the state $(-\ket{\Downarrow\Downarrow}+\ket{\Downarrow\Uparrow})/\sqrt{2} \equiv \ket{\Downarrow}\otimes(-\ket{\Downarrow}+\ket{\Uparrow})/\sqrt{2}$, i.e. a rotation of Q2 by 180 degrees around the $z$-axis of its Bloch sphere, which is the output of a CZ operation. Conversely, if the initial state of Q1 were $\ket{\Uparrow}$, the pulse at $\nu_{\rm e|\Downarrow\Downarrow}$ would have no effect on the electron, leaving the nuclear qubits unaffected.
	
	A nuclear controlled-NOT (CNOT) gate is obtained by sandwiching the CZ gate between a nuclear $-\pi/2$ and $\pi/2$ pulse (Extended Data Fig. \ref{fig ED: CNOT zero-CNOT}a). Applying an ESR $X_{2\pi}$ pulse at $\nu_{\mathrm{e}|\Uparrow\Downarrow}$ transforms the sequence into a zero-CNOT gate, i.e. a gate that flips Q2 when Q1 is in the $\ket{0}\equiv\ket{\Uparrow}$ state (Extended Data Fig. \ref{fig ED: CNOT zero-CNOT}b, and Supplementary Information S2).

	We apply this universal gate set (Fig.~\ref{fig:figure 2}a) to produce each of the four maximally-entangled Bell states of the two nuclear spins, $\ket{\Phi^{\pm}}=(\ket{\Downarrow\Downarrow} \pm \ket{\Uparrow\Uparrow})/\sqrt{2}$ and $\ket{\Psi^{\pm}}=(\ket{\Downarrow\Uparrow} \pm \ket{\Uparrow\Downarrow})/\sqrt{2}$. We reconstruct the full density matrices of the Bell states using maximum likelihood quantum state tomography~\cite{James2001} (Supplementary Information S3). The reconstructed states (Fig.~\ref{fig:figure 2}f) have fidelities of up to 94.2(2.7)\%, and concurrences as high as 0.93(4), proving the creation of genuine two-qubit entanglement. Here and elsewhere, error bars indicate $1\sigma$ confidence intervals. Bell fidelities and concurrences are calculated without removing state preparation and measurement (SPAM) errors (Extended Data Table~\ref{fig ED: SPAM errors}).
	
	\noindent \textbf{Gate set tomography}
			\begin{figure*}[!h]
			\centering
			\includegraphics[width=\linewidth]{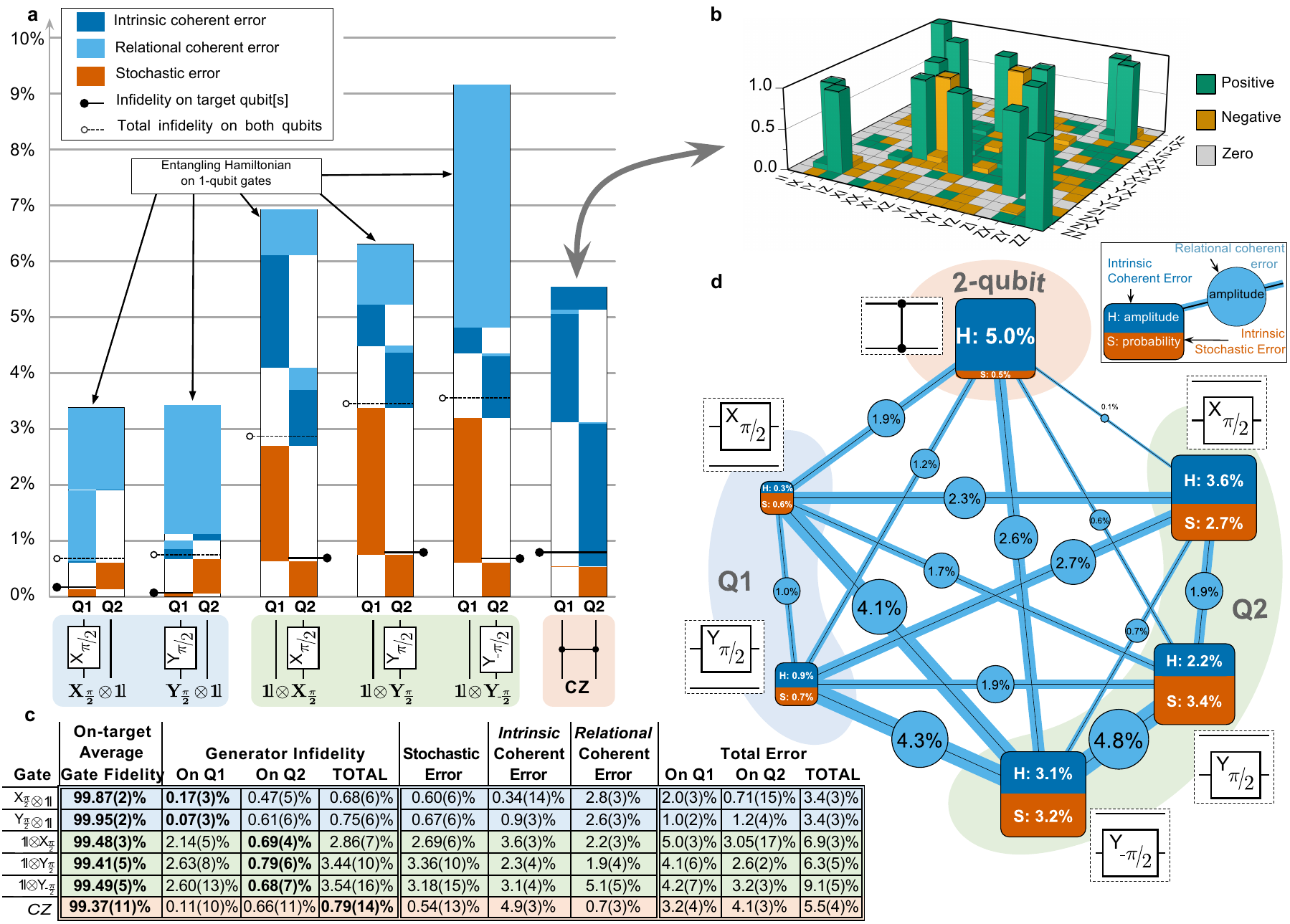}
			\captionof{figure}{\textbf{Precise tomographic characterization of 1- and 2-qubit gate quality}.  Process matrices for all 6 gates (e.g., the CZ gate shown in \textbf{b}) were estimated using gate set tomography (GST) and represented as error generators with associated rates.  \textbf{a}, Each gate’s total error rate (columns) can be partitioned into coherent (blue) and stochastic (orange) components, then further into components acting on Q1 (left), Q2 (right), and on both at once (wide).  Coherent errors are further partitioned into intrinsic (dark) and relational (light), which were assigned to specific gates by fixing a gauge.  Each gate's generator infidelity (see Supplementary Information S9) is shown, on the whole 2-qubit system (hollow pins) and on its target qubit[s] only (black pins). The CZ gate’s total infidelity is only 0.79(14)\%.  Single-qubit gates have on-target infidelities of 0.07(3)-0.79(6)\%, but display significant crosstalk errors on the spectator qubit and unexpected entangling coherent (ZZ) errors.  \textbf{c}, Error metrics for each gate are aggregated by type (stochastic/coherent) and support (Q1/Q2/total). Uncertainties in parentheses represent $1\sigma$ confidence intervals. In addition to generator infidelity, each gate's average gate fidelity on its target qubit[s] is shown, to facilitate comparison with literature.  \textbf{d}, A gauge-invariant representation of relational errors between gates (e.g.~ misalignment of rotation axes) that were assigned to individual gates in \textbf{a,c} by fixing a gauge.  Each gate is labeled with its intrinsic coherent (H) and stochastic (S) errors, while edges between two gates show the total amplitude of relational coherent error (misaligment) between them.  Large gauge-invariant relational errors between single-qubit gates confirm that the entangling coherent errors observed in \textbf{a} are not an artifact of gauge-fixing.}
			\label{fig:figure 3}
		\end{figure*}
		
	We used a customized, efficient gate set tomography (GST) \cite{Dehollain2016b,BlumeKohout2017NC,nielsen2020gate} analysis (see Methods, Extended Data Figs.~\ref{fig ED: GST schematic}, \ref{fig ED: GST results}, \ref{fig ED: Simulated RB} and Supplementary Information S4, S5, S8) to investigate the quality of six logic operations on two nuclear-spin qubits: $X_{\pi/2}$ and $Y_{\pi/2}$ rotations on Q1 and Q2, an additional $Y_{-\pi/2}$ rotation on Q2, and the entangling CZ gate. No two single-qubit operations are ever performed in parallel.
	GST probes these six logic operations and reconstructs a full two-qubit model for their behavior. Earlier experiments on electron spins in silicon used randomized benchmarking (RB) \cite{huang2019fidelity,xue2019benchmarking} to extract a single number for the average fidelity of all logic operations. Characterising specific gates required ``interleaved'' RB, which can suffer systematic errors \cite{KimmelPRX2014,CarignanDugasNJP2019}. Most importantly, RB does not reveal the cause or nature of the errors. Our GST method enables measuring each gate's fidelity to high precision, distinguishing the contributions of stochastic and coherent errors, and separating local errors (on the target qubit) from crosstalk errors (on, or coupling to, the undriven spectator qubit).  
	
	GST estimates a two-qubit process matrix for each logic operation ($G_i:\ i=1\ldots6$) using maximum likelihood estimation. We represent each $G_i$ as the composition of its ideal target unitary process ($\target{G}_i$) with an error process written in terms of a Lindbladian generator ($\lind_i$):  $G_i = e^{\lind_i}\target{G}_i$.  Each gate's error generator (EG) can be written as a linear combination of independent elementary EGs that describe distinct kinds of error \cite{blume2021taxonomy}. Each elementary EG's coefficient in $\lind_i$ is the rate (per gate) at which that error builds up.  Any Markovian error process can be described using just four kinds of elementary EGs: Hamiltonian (H), indexed by a single two-qubit Pauli operator, cause coherent or unitary errors (e.g., $H_{ZZ}$ generates a coherent $ZZ$ rotation); Pauli-stochastic (S), also indexed by a single Pauli, cause probabilistic Pauli errors (e.g. $S_{IX}$ causes probabilistic $X$ errors on Q2); Pauli-correlation (C), and active (A), indexed by two Paulis, describe more exotic errors (see Methods) that were not detected in this experiment.
	We found that each gate's behavior could be described using just 13-14 elementary EGs: 3 local S errors and 3 local H errors acting on each of Q1 and Q2, and 1-2 entangling H errors (discussed in detail below).  Extended Data Figure \ref{fig ED: GST results} shows those errors' rates, along with the process matrices and full EGs used to derive them.  To get a higher-level picture of gate quality, we aggregate the rates of related errors (see Methods) to report total rates of stochastic and coherent errors on each qubit and on the entire 2-qubit system. We present two overall figures of merit in Figure \ref{fig:figure 3}a,c: generator infidelity  and total error.  Generator infidelity is closely related to entanglement infidelity, which accurately predicts average gate performance in realistic large-scale quantum processors and can be compared to fault-tolerance thresholds (see Methods and Supplementary Information S9). Total error is related to diamond norm (see Supplementary Information S9) and estimates worst-case gate performance in any circuit, including structured or periodic circuits. In Fig.~\ref{fig:figure 3}c, we additionally report each gate's average gate fidelity on its target to ease comparison of these results with those from the literature. 
	
	The process matrices estimated by GST are not unique. An equivalent representation of the gate set can be constructed by a \emph{gauge transformation} \cite{Proctor2017PRL,nielsen2020gate} in which all process matrices are conjugated by some invertible matrix, $G_i \rightarrow M G_i M^{-1}$. Some gate errors, such as over/under-rotations or errors on idle spectator qubits, are nearly unaffected by choice of gauge; they are \textit{intrinsic} to that gate.  But other errors, such as a tilted rotation axis, can be shifted from one gate to another by changing gauge.  These \textit{relational} errors cannot be objectively associated with any particular gate.  Recognizing this, we divide coherent errors into intrinsic and relational components (Fig.~\ref{fig:figure 3}a,c). Intrinsic errors perturb a gate's eigenvalues, whereas relational errors perturb its eigenvectors.
	In Fig.~\ref{fig:figure 3}a,c we follow standard GST practice by choosing a gauge that makes the gates as close to their targets as possible. This associates relational errors with individual gates, in a way that depends critically on the choice of gauge.  But the magnitude of a given relational error between a set of gates is gauge-invariant, and Fig.~\ref{fig:figure 3}d illustrates the total relational error between each pair of gates.  In this work, we found evidence only for pairwise relational errors, although more complex multi-gate relational errors are possible.
	
	All 6 gates achieved on-target fidelities $>99\%$, with infidelities as low as $0.07(3)\%$ on Q1 and $0.68(7)\%$ on Q2.  However, we observed significant crosstalk on the spectator qubit during 1-qubit gates, resulting in full logic operations (1-qubit gate and spectator idle operation in parallel) with higher infidelities of $0.68(6)\%-3.5(2)\%$.  Remarkably, the CZ gate's infidelity of $0.79(14)\%$ is almost on par with the single-qubit gates -- a rare scenario in multi-qubit systems (Fig.~\ref{fig:figure 3}a,c). 
	
	SPAM errors were estimated by GST as 1.05(4)\% on average, and as low as 0.25(3)\% for the $\ket{\Uparrow\Uparrow}$ state (Extended Data Table~\ref{fig ED: SPAM errors}). This is a unique feature of nuclear spin qubits, afforded by the quantum nondemolition nature of the measurement process \cite{Pla2013} (Methods and Extended Data Fig.~\ref{fig ED: nuclear flipping rates}).
	
	GST provided unambiguous evidence for a surprising relational error: weight-2 (entangling) $H_{ZZ}$ and/or $H_{\target{G}_i[ZZ]}$ coherent errors on each 1-qubit gate $G_i$,  with amplitudes from $1.8-5.0\%$ (Extended Data Figure \ref{fig ED: GST results}).  These errors are consistent with an intermittent $ZZ$ Hamiltonian during the gate pulses.  
	After ruling out a wide range of possible error channels, we propose that the observed $H_{ZZ}$ error arises from the spurious accumulation of geometric phase by the electron spin, caused by off-resonance leakage of microwave power near the ESR frequencies (Supplementary Information S9). This observation illustrates the diagnostic power of GST, which revealed an error channel we had not anticipated. It also shows GST's ability to unveil correlated and entangling errors, whose detection and prevention is of key importance for the realization of fault-tolerant quantum computers \cite{novais2013surface}.
	
	\noindent \textbf{Three-qubit entanglement}
			\begin{figure*}[!h]
			     \centering
			\includegraphics{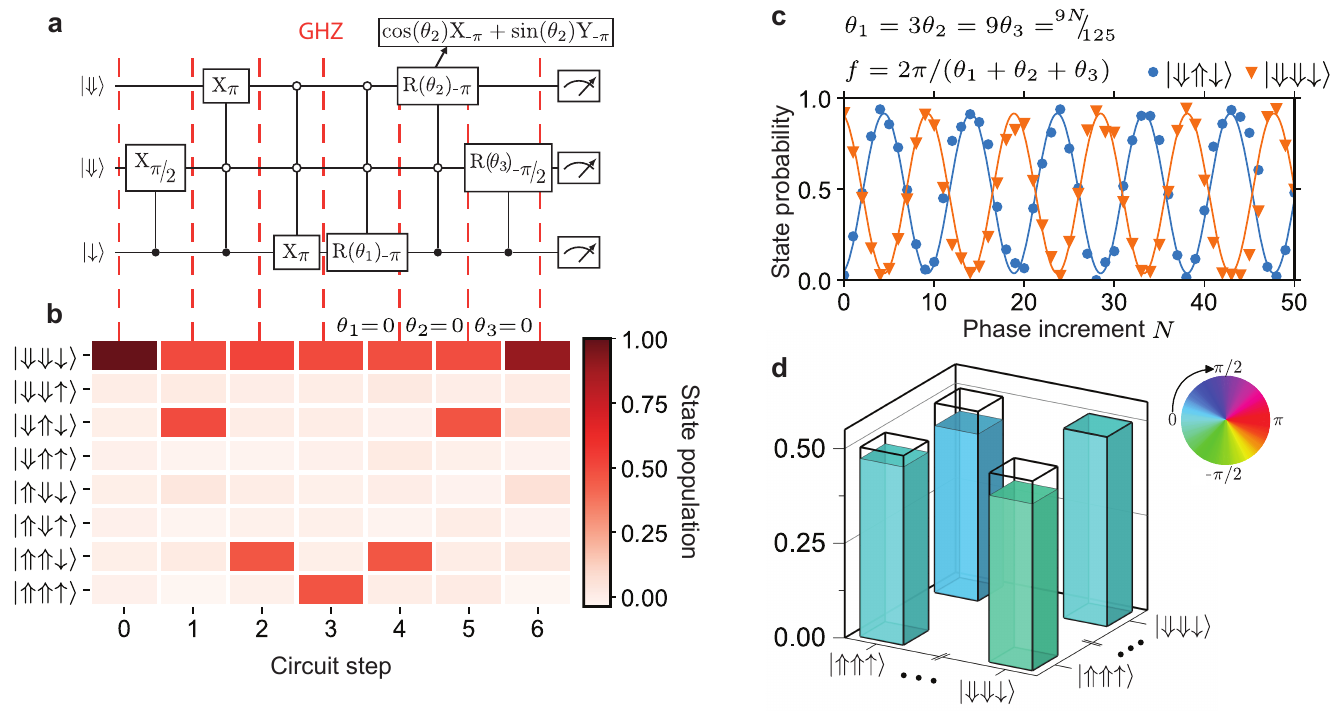}
			\captionof{figure}{\textbf{Creation and tomography of an electron-nuclear three-qubit GHZ state.}
				\textbf{a,} Starting from $\ket{\Downarrow\Downarrow\downarrow}$, the first three gates generate an entangled three-qubit GHZ state. All eight state populations are read out (\textbf{b}) at each circuit step (red dashed lines), and estimated without correcting for SPAM errors (Supplementary Information S7).
				The final three gates $R(\theta_i)_{\phi}$ reverse the operations of the first three if the rotation angles are $\theta_1=\theta_2=\theta_3=0$, returning to the initial state in the absence of errors.
				The two gates that are conditional on $\mathrm{Q_2}$ are composed of multiple pulses (Supplementary Information S6).
				\textbf{c,} The coherence between the GHZ components $\ket{\Downarrow\Downarrow\downarrow}$ and $\ket{\Uparrow\Uparrow\uparrow}$ is probed by incrementing the phases $\theta_i$ of the reversal pulses.
				This induces oscillations at frequency $f=2\pi/(\theta_1+\theta_2+\theta_3)$ whose amplitude and phase correspond to the purity and phase relation between $\ket{\Downarrow\Downarrow\downarrow}$ and $\ket{\Uparrow\Uparrow\uparrow}$.
				\textbf{d,} Density-matrix extrema of the GHZ state.
				The state populations of the GHZ components $\ket{\Downarrow\Downarrow\downarrow}$ and $\ket{\Uparrow\Uparrow\uparrow}$ at circuit step 3 (\textbf{b}) provide the diagonal entries, while the oscillation amplitude and phase (\textbf{c}) provide the off-diagonal entries.
				From these values, the fidelity to the nearest GHZ state is estimated as 92.5(1.0)\%, including SPAM.
			}
			\label{fig:figure 4}
		\end{figure*}
		
	The nuclear logic gates shown above would not scale beyond a single, highly localized cluster of donors. However, adding the hyperfine-coupled electron qubit yields a scalable heterogeneous architecture. Electron qubits decohere faster (see Extended Data Figs. \ref{fig ED: electron properties} and \ref{fig: nuclear T2} for a comparison), but admit faster control. If high-fidelity entanglement between electron and nuclear qubits can be created, electron qubits can enable fast coherent communication between distant nuclei (via electron-electron entanglement, or physical shuttling) or serve as high-fidelity ancilla qubits for quantum error correction. To demonstrate this capability, we produce the maximally entangled three-qubit Greenberger-Horne-Zeilinger (GHZ) state $\ket{\psi_{\rm GHZ}}=(\ket{\Uparrow\Uparrow\uparrow}+\ket{\Downarrow\Downarrow\downarrow})/\sqrt{2}$ using the pulse sequence shown in Fig.~\ref{fig:figure 4}a. Starting from $\ket{\Downarrow\Downarrow\downarrow}$, an NMR $Y_{\pi/2}$ pulse at $\nu_{Q2|\downarrow}$ creates a coherent superposition state of nucleus 2, followed by a nuclear zCNOT gate (as in Fig.~\ref{fig:figure 2}a) to produce a nuclear $\ket{\Phi^+}$ state, and an ESR $X_{\pi}$ pulse at $\nu_{\mathrm{e}|\Downarrow\Downarrow}$ to arrive at $\ket{\psi_{\rm GHZ}}$. Since the ESR frequency directly depends on the state of both nuclei, the latter pulse constitutes a natural 3-qubit Toffoli gate, making the creation of 3-qubit entanglement particularly simple, as in nitrogen-vacancy centres in diamond \cite{neumann2008multipartite}. Executing Toffoli gates on electrons in quantum dots \cite{takeda2021quantum} requires more complex protocols, but can be simplified by a combination of exchange and microwave pulses \cite{gullans2019protocol}.
	
	Measuring the populations of the eight electron-nuclear states (Supplementary Information S7) after each step confirms the expected evolution from $\ket{\Downarrow\Downarrow\downarrow}$ to $\ket{\psi_{\rm GHZ}}$ (Fig.~\ref{fig:figure 4}b). The evolution can be undone by applying the sequence in reverse, yielding a return probability to $\ket{\Downarrow\Downarrow\downarrow}$ of 89.6(9)\%, including SPAM errors. As in the two-qubit case, measuring the populations is a useful sanity check but does not prove multipartite entanglement, which requires knowing the off-diagonal terms of the density matrix $\rho_{\rm GHZ}=\ket{\psi_{\rm GHZ}}\!\bra{\psi_{\rm GHZ}}$. 
	
	Standard tomography methods require measuring the target state in different bases, obtained by rotating the qubits prior to measurement. However, the superposition of $\ket{\Downarrow\Downarrow\downarrow}$ and $\ket{\Uparrow\Uparrow\uparrow}$ dephases at a rate dominated by the electron dephasing time $T_{\rm 2e}^* \approx 100$~$\mu$s (Extended Data Fig.~\ref{fig ED: electron properties}), which is only marginally longer than the nuclear spin operation time $\approx 10-20$~$\mu$s. Therefore, the GHZ state will have significantly dephased by the time it is projected onto each measurement basis.
	
	We circumvent this problem by adopting a tomography method that minimises the time spent in the GHZ state. An extension of a method first introduced for the measurement of electron-nuclear entanglement in spin ensembles \cite{mehring2003entanglement}, it is related to the parity scan commonly used in trapped ions \cite{sackett2000experimental} and superconducting circuits \cite{wei2020verifying}. We repeat the reversal of the GHZ state (Fig.~\ref{fig:figure 4}b) $N=100$ times, each time introducing phase shifts $\theta_{1,2,3}$ to the rotation axes of the three reversal pulses, with $\theta_1 = 3\theta_2 = 9\theta_3 = 9N/125$. The return probability to $\ket{\Downarrow\Downarrow\downarrow}$ oscillates with $N$; the amplitude and phase of the oscillations yield the off-diagonal matrix element $\bra{\Downarrow\Downarrow\downarrow}\rho_{\rm GHZ}\ket{\Uparrow\Uparrow\uparrow}=\rho_{18}$.
	
	Since the ideal $\rho_{\rm GHZ}$ has nonzero elements only on its four corners, the populations $\rho_{11},\rho_{88}$ and the coherence $\rho_{18}$ are sufficient to determine the GHZ state fidelity $\mathcal{F}_{\rm GHZ} = 92.5(1.0)\%$. Also here, SPAM errors remain included in total infidelity. By comparison, an 88\% GHZ state fidelity has been reported in a triple quantum dot after removing SPAM errors, whereas the uncorrected fidelity is 45.8\% \cite{takeda2021quantum}. This highlights the drastic effect of SPAM of multi-qubit entanglement, and the robustness of our system against such errors. The different coherence and operation timescales for electron and nuclei need not be an obstacle for the use of such entangled states in scaled-up architectures, because all further entangling or shuttling operations between electrons will occur on $\simeq 1$~$\mu$s time scales.
	
	\noindent \textbf{Outlook}
	
	The demonstration of 1-qubit, 2-qubit and SPAM errors at or below the 1\% level highlight the potential of nuclear spins in silicon as a credible platform for fault-tolerant quantum computing. An often-quoted example, based on surface code quantum error correction, sets a fault-tolerance threshold of 0.56\% for the entanglement infidelity of 1- and 2-qubit gates and the SPAM errors \cite{fowler2012surface}.
	
	Several avenues are available to harness the high-fidelity operations demonstrated here.  Replacing the $^{31}$P donors with the higher-spin group-V analogues such as $^{123}$Sb ($I=7/2$) or $^{209}$Bi ($I=9/2$) would provide access to a much larger Hilbert space in which to encode quantum information. For example, a cluster of two $^{123}$Sb donors contains the equivalent of six qubits in the nuclear spins, plus an electron qubit. An error-correcting code can be efficiently implemented in high-spin nuclei \cite{gross2021hardwareefficient}, where our method would provide a pathway for universal operations between the logical qubits encoded in each nucleus.
	
	Moving to heavier group-V donors also allows the electrical control of the nuclear spins \cite{asaad2020coherent}. Combined with the electrical drive of the electron-nuclear `flip-flop' transition \cite{Tosi2017}, this implies the ability to control electron and nuclei by purely electrical means. In a two-donor system as shown here, the entangling CZ gate could similarly be obtained by an electrical $2\pi$-pulse on a flip-flop transition.
	
	The electron-nuclear entanglement we have demonstrated can be harnessed to scale up beyond a pair of nuclei coupled to the same electron. Neighbouring donor electrons can be entangled via exchange interaction by performing controlled-rotation resonant gates \cite{madzik2021conditional} or $\sqrt{\mathrm{SWAP}}$ gates \cite{he2019two}. Wider distances could be afforded by physically shuttling the electron across lithographic quantum dots \cite{pica2016surface,buonacorsi2019network}, while preserving the quantum information encoded in it \cite{yoneda2020coherent}. Our methods would apply equally to isoelectronic nuclear spin centres like $^{73}$Ge and $^{29}$Si, where it has been shown that the nuclear qubit coherence is preserved while shuttling the electron across neighbouring dots \cite{hensen2020silicon}. Furthermore, electron spins can mediate the coherent interaction between nuclear spin qubits and microwave photons \cite{tosi2018robust,mielke2020nuclear}. Recent experiments on electron spin qubits in silicon report 1- and 2-qubit gate fidelities above 99\% \cite{xue2021computing,noiri2021fast}. Therefore, the fidelity of electron qubit operations will not constitute a bottleneck for the performance of electron-nuclear quantum processors. These examples illustrate the significance of universal high-fidelity two-qubit operations with nuclear spins in  a platform like silicon, which can simultaneously host nuclear and electron spin qubits, lithographic quantum dots, and dense readout and control devices \cite{vandersypen2017interfacing}.

	\section*{Methods}
	\subsection*{Device fabrication}
	The quantum processor is fabricated using methods compatible with standard silicon MOS processes. We start from a high quality silicon substrate (p-type $\langle$100$\rangle$; 10-20 $\Omega$cm), on top of which a 900~nm thick epilayer of isotopically enriched $^{28}$Si has been grown using low-pressure chemical vapour deposition (LPCVD). The residual $^{29}$Si concentration is 730~ppm. Heavily-doped n$^+$ regions for Ohmic contacts and lightly-doped p regions for leakage prevention are defined by thermal diffusion of phosphorus and boron, respectively. A 200~nm thick SiO$_2$ field oxide is grown in a wet oxidation furnace. In the centre of the device, an opening of 20 $\mu$m $\times$ 40 $\mu$m is etched in the field oxide using HF acid. Immediately after, a 8 nm thick, high quality dry SiO$_2$ gate oxide is grown in this opening. In preparation for ion implantation, a 90~nm $\times$ 100~nm aperture is opened in a PMMA mask using electron-beam-lithography (EBL). The samples are implanted with P$^+$ ions at an acceleration voltage of 10 keV per ion. During implantation the samples were tilted by 8 degrees and the fluence was set at $1.4 \times 10^{12}/\mathrm{cm}^2$. Donor activation and implantation damage repair is achieved through the process of a rapid thermal annealing (5 seconds at 1000 $^{\circ}$C). The gate layout is patterned around the implantation region in three EBL steps, each followed by aluminium thermal deposition (25~nm thickness for layer 1; 50~nm for layer 2; 100~nm for layer 3).  Immediately after each metal deposition, the sample is exposed to a pure, low pressure (100~mTorr) oxygen atmosphere to form an Al$_2$O$_3$ layer, which electrically insulated the overlapping metal gates. At the last step, samples are annealed in a forming gas (400~$^{\circ}$C, 15~min, 95$\%$ N$_2$ / 5$\%$ H$_2$) aimed at passivating the interface traps.
	
	\subsection*{Experimental setup}
	The device was wire-bonded to a gold-plated printed circuit board and placed in a copper enclosure.
	The enclosure was placed in a permanent magnet array \cite{adambukulam2020ultra}, producing a static magnetic field of 1.33~T at the device (see Extended Data Fig.~\ref{fig ED: device SEM} for field orientation).
	The board was mounted on a Bluefors BF-LD400 cryogen-free dilution refrigerator, reaching a base temperature of 14~mK, while the effective electron temperature was $\approx 150$~mK.
	
	DC bias voltages were applied to all gates using Stanford Research Systems (SRS) SIM928 voltage sources.
	A room-temperature resistive combiner was used for the fast donor gates (Extended Data Fig.~\ref{fig ED: device SEM}) to add DC voltages to AC signals produced by the LeCroy Arbstudio 1104, which then passed through an 80~MHz low-pass filter; all other gates passed through a 20~Hz low-pass filter. All filtering takes place at the mixing chamber plate. The wiring includes graphite-coated flexible coaxial cables to reduce triboelectric noise \cite{kalra2016vibration}.
	
	Microwave pulses to induce ESR transitions were applied to an on-chip broadband antenna \cite{dehollain2012nanoscale} using a Rohde \& Schwarz SGS100A vector microwave source combined with an SGU100A upconverter.
	The microwave carrier frequency remained fixed at $37.1004125$~GHz, while the output frequency was varied within a pulse sequence by mixing it with a radiofrequency (RF) signal using double-sideband modulation, i.e. by applying RF pulses to the in-phase port of the SGS100A IQ mixer (the quadrature port was terminated by a 50~$\Omega$ load).
	The carrier frequency was chosen such that whenever one sideband tone was resonant with an ESR pulse, the second sideband was off-resonant with all other ESR frequencies. 
	To suppress microwave signals when not needed, 0~V was applied to the in-phase port of the IQ mixer.
	Under these circumstances, the carrier frequency is expected to be suppressed by 35 dB, according to the source data sheet.
	The RF pulses used for double-sideband modulation were generated by one of the two channels of the Agilent 81180A arbitrary waveform generator; the second channel delivered RF pulses to the microwave antenna to drive NMR transitions.
	The microwave signal for ESR and RF signal for NMR were combined in a Marki Microwave DPX-1721 diplexer.
	
	The SET current passed through a Femto DLPCA-200 transimpedance amplifier ($10^7$ V/A gain, 50~kHz bandwidth), followed by an SRS SIM910 JFET post-amplifier ($10^2$ V/V gain), SRS SIM965 analog filter (50 kHz cutoff low-pass Bessel filter), and acquired via an AlazarTech ATS9440 PCI digitizer card.
	The instruments were triggered by a SpinCore PulseBlasterESR-PRO.
	The measurements instruments were controlled by Python code using the quantum measurement software packages QCoDeS and SilQ.

	\subsection*{System Hamiltonian}
	The static Hamiltonian of our combined electron-nuclei system is
	\begin{equation}
		H_\mathrm{s} = -\gamma_\mathrm{e} B_0 \hat{S}_z - \gamma_\mathrm{n} B_0 (\hat{I}_{1,z} + \hat{I}_{2,z}) + A_1 \vec{S} \cdot \vec{I}_1 + A_2 \vec{S} \cdot \vec{I}_2,
	\end{equation}
	where $\gamma_\mathrm{e}\approx -27.97$ GHz T$^{-1}$ is the electron gyromagnetic ratio~\cite{feher1959electron}, $\gamma_\mathrm{n}\approx 17.23$ MHz T$^{-1}$ is the nuclear gyromagnetic ratio~\cite{steger2011optically}, $\vec{S} = [\hat{S}_x, \hat{S}_y, \hat{S}_z]$ are the electron spin operators, and $\vec{I}_i = [\hat{I}_{i,x}, \hat{I}_{i,y}, \hat{I}_{i,z}]$ are the nuclear spin operators for nucleus $i\in {1,2}$. 
	The static magnetic field $B_0=1.33~\mathrm{T}$ is aligned along $\hat{z}$, and $A_1 \approx 95~\mathrm{MHz}$, ($A_2 \approx 9~\mathrm{MHz}$) is the hyperfine interaction strength between the electron and nucleus 1 (2).
	
	An AC drive applied to the microwave line is used to induce transitions between nuclear spin states and between electron spin states.
	The drive predominantly modulates the transverse magnetic field as
	\begin{equation}
		H_\mathrm{rf}(t) = -\gamma_\mathrm{e} \vec{B}_1 \cdot \vec{S} \sin{ \omega t} - \gamma_\mathrm{n} \vec{B}_1 \cdot (\hat{I}_1 + \hat{I}_2) \sin \omega t,
	\end{equation}
	where $\vec{B}_1$ is the oscillating magnetic field strength, primarily aligned along $\hat{y}$.
	
	\subsection*{Electron spin readout}
	An electron spin readout is realized through the spin to charge conversion \cite{elzerman2004single,Morello2009}. This method utilizes a single electron transistor (SET) as both a charge sensor and an electron reservoir. The electron spin $\ket{\downarrow}$ and $\ket{\uparrow}$ states are separated by the Zeeman energy, which scales linearly with the external magnetic field. Thermal broadening of the SET at 100~mK is much smaller than the Zeeman splitting of two electron spin states. This means that, at the read position, the donor electron spin down state faces only occupied levels in the SET island (tunneling is prohibited) and the spin up state faces only unoccupied states and can freely tunnel out the SET island. This event will shift the energy ladder in the SET island, bringing it out of the Coulomb blockade, thus causing a burst in the current. This burst will last until $\ket{\downarrow}$ electron tunnels to the donor. If the electron has been projected to the $\ket{\downarrow}$ state then no change in the SET current will be recorded, as the electron cannot tunnel to the SET island. At the end of each read phase the electron spin is reinitialized in $\ket{\downarrow}$ for the next single shot cycle. The fidelity of single-shot electron readout and $\ket{\downarrow}$ initialisation by spin-dependent tunnelling is $\approx 80$\% in this device. However, we further increase the initialisation fidelity by letting the electron thermalise to the lattice temperature for a time $\gg T_{\rm 1e}$ (Fig.~\ref{fig ED: electron properties}\textbf{b}) before triggering further operations.
	
	\subsection*{Nuclear spin readout and initialisation}
	The readout of the two nuclear spin qubits is an extension of the well-known method developed for a single donor \cite{Pla2013}, based on the excitation of the electron bound to the nuclei, conditional on a particular nuclear state, followed by electron spin readout \cite{Morello2010}. The same method is used to initialise the nuclei in a known state.
	
	In the present system, consisting of an electron coupled to two $^{31}$P donors with different hyperfine couplings $A_1 \gg A_2$, we find four well-separated electron spin resonance (ESR) frequencies (Fig.~\ref{fig:figure 1}c), conditional on the $\ket{\Downarrow\Downarrow}, \ket{\Downarrow\Uparrow}, \ket{\Uparrow\Downarrow}, \ket{\Uparrow\Uparrow}$ nuclear states. An electron in the $\ket{\downarrow}$ state is initially drawn from a cold charge reservoir onto the donor cluster (independently of nuclear states). We then apply a microwave $\pi$-pulse at a particular ESR frequency, for instance $\nu_{\rm e|\Downarrow\Downarrow}$ corresponding to the $\ket{\Downarrow\Downarrow}$ nuclear spin state, and then measure the electron spin. If it is found in the $\ket{\uparrow}$ state, then the nuclear spins are projected to the $\ket{\Downarrow\Downarrow}$ state. If the electron is $\ket{\downarrow}$ (i.e. the pulse at $\nu_{\rm e|\Downarrow\Downarrow}$ failed to flip it to $\ket{\uparrow}$), the nuclear spins are projected to the subspace orthogonal to  the $\ket{\Downarrow\Downarrow}$ state. This constitutes a nuclear spins single-shot readout, with a fidelity given by the product of the electron single-shot readout fidelity (typically $\approx 80$\%) and the electron $\pi$-pulse fidelity ($\gg 99$\%).
	
	This nuclear readout is a projective, approximately quantum non demolition (QND) process \cite{Pla2013}. The ideal QND measurement relies on the observable $I_z$ to commute with the Hamiltonian $H_{\mathrm{int}}$ describing an interaction between the observable and the measurement apparatus $[I_z,H_{\mathrm{int}}] = 0$ \cite{braginsky1996quantum}. In our case the hyperfine terms $A_1S_zI_{z1}$ and $A_2S_zI_{z2}$ constitute $H_{\mathrm{int}}$. The observation of nuclear spin quantum jumps originating from the electron measurement by spin-dependent tunnelling (ionization shock) hints at a deviation from QND nature of the readout process \cite{Pla2013}. It implies the presence of terms of the form $A_{||}/2(S_+I_- + S_-I_+)$ in the hyperfine coupling, and possibly additional anisotropic terms, which do not commute with $I_z$. In our experiment, the deviation from the ideal QND measurement is extremely small, of order $10^{-6}$, as shown in  Extended Data Figure~\ref{fig ED: nuclear flipping rates}. 
	
	We exploit the near-perfect QND nature of the nuclear spin readout by repeating the cycle [load $\ket{\downarrow}$ -- ESR $\pi$-pulse -- electron readout] between 7 and 40 times, to substantially increase the nuclear single-shot readout fidelity. This is the fundamental reason why our average SPAM errors are $\approx 1$\% (Extended Data Table~\ref{fig ED: SPAM errors}), and we have thus reported Bell and GHZ state fidelities without removing SPAM errors from the estimate. 
	
	\subsection*{ESR and NMR calibration}
	
	\subsubsection*{Gate calibration}
	Both the 1-qubit NMR gates and the 2-qubit ESR gate were iteratively calibrated using a combination of GST and other tuning methods. 
	Rabi flops were first used to obtain roughly calibrated 1-qubit NMR gates. Next, 1-qubit GST was repeatedly
	employed to identify and correct error contributions such as over-/under-rotations and detunings. Other routines such as the repeated application of gates were performed in between GST measurements to
	independently verify the improvements to 1-qubit gate fidelities of GST. 
	The calibrated NMR $\pi/2$ pulse duration of Q1 (Q2) is 12.0~$\mu$s (25.3~$\mu$s). The discrepancy between the two durations is largely due to the hyperfine interaction enhancing the Rabi frequency of Q1 and reducing the Rabi frequency of Q2, combined with line reflections and filtering.
	
	For the geometric 2-qubit gate based upon an electron 2$\pi$ pulse, we found that a trivial calibration using
	Rabi flops already gave a near-optimal result. GST was then used for fine-tuning and for the detection of
	small error contributions such as a minor frequency shift.
	The calibrated ESR $2\pi$ pulse duration of the CZ gate is 1.89~$\mu$s at an output power of 20~dBm.
	
	\subsubsection*{Periodic frequency recalibration}
	To keep the system tuned throughout the measurements, the NMR frequencies $\nu_{Q1|\downarrow}$ and $\nu_{Q2|\downarrow}$ and ESR frequency $\nu_{\mathrm{e}|\downarrow\downarrow}$ were calibrated every ten circuits.
	The ESR frequency was calibrated by measuring the ESR spectrum and selecting the frequency of the ESR peak.
	The NMR frequencies were measured by a variant of the Ramsey sequence, consisting of an $X_{\pi/2}$ and $Y_{\pi/2}$ separated by a wait time $\tau$.
	An off-resonant RF pulse was applied during the wait time to mitigate any frequency shift caused by the absence of an RF drive.
	Since nuclear readout has a near-unity fidelity, this measurement should result in a nuclear flipping probability $P_\mathrm{flip} = 0.5$ if the RF frequency $f_\mathrm{RF}$ matches the average NMR frequency $f_\mathrm{NMR}$ throughout the measurement.
	Therefore, any deviation of $P_\mathrm{flip}$ from 0.5 provides a direct estimate of the frequency mismatch $\Delta f = f_\mathrm{NMR} - f_\mathrm{RF} = \arcsin{\left(2P_\mathrm{flip} - 1)\right)} / (2\pi \tau)$, provided that $|\Delta f/\tau| < 0.25$.
	A higher $\tau$ more accurately estimates $\delta f$, while a lower $\tau$ results in the condition $|\Delta f/\tau| < 0.25$ being valid for a broader range of $\Delta f$.
	The NMR recalibration sequence iteratively increased the wait time $\tau = 40~\mu\mathrm{s} \rightarrow 100~\mu\mathrm{s} \rightarrow 160~\mu\mathrm{s}$ to ensure that the condition $|\Delta f/\tau| < 0.25$ remains satisfied while increasing the accuracy at which the NMR frequency is estimated.
	For each $\tau$, the NMR frequency was estimated by repeating this sequence and updating the RF frequency until $P_\mathrm{flip}$ fell within the range [0.4, 0.6].
	
	\subsection*{Measurement overhead}
	Instrument setups and calibration routines add a significant overhead to the GST measurements.
	An estimate of this overhead can be obtained by comparing the total measurement duration to the duration of a single pulse sequence.
	The 2Q GST measurement shown in Fig.~\ref{fig:figure 3} was acquired over 61 hours, during which 300-503
	shots were acquired for each of the 1593 circuits. This results in an average duration of 340 ms per GST pulse sequence iteration.
	Compared to the average pulse sequence duration of around 121 ms, this corresponds to an
	overhead of ~185\%.
	
	\subsection*{Effective mass theory simulations of the hyperfine interaction}
	To simulate the wave function of the third electron in the 2P system, the effective mass theory (EMT) model of the neutral 2P system in Ref.~\cite{joecker2020full} is extended in a mean-field approach.
	
	For short donor separations, the two inner electrons are tightly bound in a magnetically inactive singlet orbital. The third electron then only interacts with the inner
	ones to the extent that it experiences the Coulomb repulsion of their fixed charge
	distribution
	\begin{equation}
		V(\vec{r})=\frac{e^2}{4\pi\epsilon_{\rm Si}}\int\frac{\rho_{\rm S}(\vec{r}^\prime)}{|\vec{r}^\prime-\vec{r}|}d^3\vec{r}^\prime.
		\label{Eq:MeanFieldPotential}
	\end{equation}
	Here, $e$ is the electron charge, $\epsilon_{\rm Si}$ the dielectric constant in silicon and $\rho_{\rm S}(\vec{r}^\prime)$ is the charge density of the tightly bound electrons found in Ref.~\cite{joecker2020full}. The third electron is then effectively described by the sum of the 2P EMT Hamiltonian in an electric field~\cite{joecker2020full} and the corresponding mean-field potential in Eq.~(\ref{Eq:MeanFieldPotential}).
	
	Here, only 2P configurations along the [100] crystal axis with distances d$\leq$7~nm and realistic fields E$\leq$2~mV/nm are considered. In this regime the inter-donor exchange dominates the on-site exchange and the mean-field approach is justified. 
	
	The chosen basis is a combination of two STO-3G \cite{joecker2020full} orbitals, one variationally optimized at $d$=0.5~nm and the other at $d$=7~nm.
	
	To compute the hyperfine interaction strength, the electron density at the nucleus is rescaled by a bunching factor of 440 \cite{gamble2015multivalley}. The experimentally found hyperfine configuration is found for donors spaced 6.5~nm apart, and subjected to an electric field 2~mV/nm.
	
	\subsection*{Gate set tomography experiments}
	
	We designed a customized GST experiment for a set of 6 logic gates: $X_{\pi/2}$ and $Y_{\pi/2}$ rotations on each qubit, an additional $Y_{-\pi/2}$ rotation on Q2, and the symmetric CZ gate between them.  A basic 2-qubit GST experiment for this gate set comprises a list of quantum circuits defined by: (1) choosing a set of 75 short ``germ'' circuits that, when repeated, collectively amplify every error rate; (2) repeating each germ several times to times to form ``germ power'' circuits whose lengths are approximately $L=1,2,4,\ldots L_{\mathrm{max}}$; and (3) prefacing and appending each germ power with each of 16 ``preparation fiducial'' circuits and each of 11 ``measurement fiducial'' circuits.  We used $L_{\mathrm{max}}=8$, yielding a set of 20606 circuits (this is not a simple multiplication because germ circuits with depth $>1$ do not appear at shorter $L$).  We eliminated 92\% of these circuits using two techniques from \cite{nielsen2020gate}. First, we identified a subset of 18 germs that amplify any dominant errors in each gate (if $L_{\mathrm{max}}$ was very large, subdominant errors would get echoed away by dominant errors).  This yielded a total of 50 germ powers.  Second, for the $L>1$ germ powers, we identified and eliminated pairs of fiducial circuits that provided redundant information.  This trimmed the circuits per germ power from 176 to as few as 16, and the total number of circuits from 8800 to just 1592.  Each of those circuits was repeated 300-500 times to gather statistics.  We used maximum likelihood estimation (MLE) implemented in the \texttt{pyGSTi} software \cite{pyGSTi,Nielsen2020QST} to estimate $16\times 16$ 2-qubit process matrices $\{G_i:\ i=1\ldots 6\}$ for all six operations. 
	
	\subsection*{Constructing and selecting reduced models}
	
	Process matrices are a comprehensive, but not especially transparent, representation of gate errors.  So we used each gate's ideal target (unitary) operation $\target{G}_i$ to construct an error generator \cite{blume2021taxonomy} $\lind_i = \log(G_i \target{G}_i^{-1})$ that presents the same information more usefully.  Representing noisy gates this way enables us to split each gate's total error into parts that act on Q1 only, Q2 only, or both qubits together -- and then further into coherent and stochastic errors -- to reveal those errors' sources and consequences.  It also enables the construction of simple, efficient ``reduced models'' for gate errors, by identifying swaths of elementary error generators whose rates are indistinguishable from zero.
	
	Pinning the coefficients of $k$ elementary error generators to zero yields a reduced model with $k$ fewer parameters, whose likelihood ($\mathcal{L}$) can be found by MLE.  We evaluate the statistical significance of error rates that were pinned by seeing how much $\mathcal{L}$ declines.  If a given error's true rate is zero, then pinning it to zero in the model reduces $2\log\mathcal{L}$, on average, by 1 \cite{Wilks1938AMS}.  So when we pin $k$ rates, we compute the ``evidence ratio'' $r = 2\Delta\log\mathcal{L}/k$, where $\Delta\log\mathcal{L}$ is the difference between the two models' likelihood \cite{nielsen2021efficient}.  If $r\leq 1$, the pinned rates are strictly negligible; if $r\leq 2$, then the smaller model is preferred by Akaike's information criterion (AIC) \cite{akaike1998information}; other criteria (e.g. the Bayesian BIC) impose higher thresholds.  We used a slightly higher threshold and chose the smaller model whenever $r\leq5$. Using this methodology, we constructed a model that describes the data well, in which just 83 (out of $1440$) elementary errors' rates are significantly different from zero.
	
	The rates of all the un-pinned elementary errors form a vector describing the noisy model.  In general, un-physical gauge degrees of freedom \cite{nielsen2020gate} will give rise to a foliation of the model space into gauge manifolds on which the loglikelihood is constant.  In our analysis, we work in the limit of small errors and gauge transformations where the space is approximately linear, and identify the subspace that is gauge invariant.  We are able to construct a basis for the gauge-invariant subspace whose elements correspond to relational or intrinsic errors and have a definite type (H, S, or A), allowing us to decompose the model's total error as shown in Figure \ref{fig:figure 3}.
	
	Extended Data Figure \ref{fig ED: GST results} presents each gate's 13-14 nonzero elementary error rates after projecting the error vector onto the gauge-invariant subspace (column 3), along with the process matrices (column 1) and error generators (column 2) from which they are derived. Here and elsewhere, error bars are $1\sigma$ confidence intervals computed using the Hessian of the loglikelihood function.

	\subsection*{Aggregated error rates and metrics}
	\label{sec:aggregated}
	Our GST analysis aims to identify specific gate errors and understand how these errors affect the overall performance of our system.  It begins with the raw output of GST -- rates of elementary errors on gates.  We aggregate these error rates in different ways, yielding each gate's total error and infidelity, and partitioning those metrics into their components on Q1 or Q2 or both qubits together, in order to summarize different aspects of system performance. We additionally report average gate fidelities to facilitate comparison with the literature. 
	
	Gate errors by definition cause unintended changes in the state of the system.  S error generators produce stochastic errors that transfer \emph{probability} to erroneous states; H generators produce coherent errors that transfer \emph{amplitude} to erroneous states.  We can interpret the \emph{rate} of an error generator, to first order, as the amount of erroneous probability (denoted $\epsilon$ for S generators) or amplitude (denoted $\theta$ for H generators) transferred by a single use of the gate when acting on one half of a maximally entangled state.  
	
	It is useful to group similar errors together and aggregate their rates.  We classify and combine error generators according to:
	\begin{itemize}
		\item Their type (H or S),
		\item Their support (Q1, Q2, or joint),
		\item Whether they are intrinsic to a single gate, or relational between gates (H errors only; relational S errors were negligible).
	\end{itemize}

	The elementary error generators described in the main text have definite type and support.  For example, the $H_{XI}$ generator has type H and support on Q1.  Any error generator on a given gate is intrinsic to that gate if it commutes with the gate, and relational otherwise.
	For example, if single-qubit $X_{\pi/2}$ and $Y_{\pi/2}$ gates produce rotations around axes that are separated by only $89^\circ$ instead of $90^\circ$, then either gate can be considered perfect at the cost of assigning a $1^\circ$ tilt error to the other gate.  This error can be moved between the two gates by a gauge transformation $M$ that rotates both gates by $1^\circ$ around the $Z$-axis. This error is purely relational; it cannot be assigned definitively to one gate or the other, but can be unambiguously observed in circuits containing both gates.
	
	To divide each gate's errors into intrinsic and relational components, we represent the gate's error generator as a vector in a space spanned by the H and S elementary error generators.  Error generators that commute with the target gate form a subspace that is invariant under gauge transformations.  The error generator's projection onto this space is its intrinsic component.  Error generators in the complement of the intrinsic subspace are relational -- they can be changed or eliminated by gauge transformations -- and the projection of the gate's error generator onto this complement is its relational component.
	
	To construct aggregated error metrics, we start by aggregating H and S rates separately.  They add in different ways, because H error rates correspond to amplitudes while S error rates correspond to probabilities.  Rates of S generators add directly ($\epsilon_{\mathrm{agg}} = \sum_i{\epsilon_i}$), while rates of H generators add in quadrature ($\theta_{\mathrm{agg}} = (\sum_i{\theta^2_i})^{1/2}$).  Combining H and S error rates into a single metric is trickier -- there is no unique way to do so because the impact of coherent errors depends on how they interfere over the course of a circuit.  We therefore consider two quantities:  \emph{total error} $\epsilon_{\rm tot} = \epsilon_{\mathrm{agg}}+\theta_{\mathrm{agg}}$ and \emph{generator infidelity} $\hat{\epsilon}=\epsilon_{\mathrm{agg}}+\theta_{\mathrm{agg}}^2$. Total error approximates the maximal rate at which gate errors could add up in any circuit, while infidelity quantifies the same errors' average impact in a random circuit.  
	
	Both of these metrics appear in Fig.~\ref{fig:figure 3}, where in panels a, c, and d we report aggregated error rates that partition the overall error in various ways (see the discussion in S10 of the Supplement).  We report a third metric, the \emph{average gate fidelity} (AGF) on each gate's target qubit[s], in Fig.~\ref{fig:figure 3}c and in the abstract to aid comparison with other published results.  The on-target AGF provides an overall (and gauge-dependent) measure of the average performance of a gate when acting only on the target qubit(s).  For a gate targeting Q1, it is defined as:
	\begin{equation}\label{eq:ontargetagi}
		\bar{\epsilon}^{(Q1)}  = 1 - \frac{1}{2}\int d\psi \,\, \langle\psi| {\mathrm{tr}}_{Q2} \left[e^{\lind}\left( |\psi \rangle \langle \psi| \otimes \mathbb{I} \right)\right] |\psi\rangle
	\end{equation}
	For a two-qubit gate, the on-target AGF is simply the AGF of the two-qubit operation:
	\begin{equation}\label{eq:agi}
		\bar{\epsilon}  = 1 - \int d\psi \,\, \langle\psi| e^{\lind}(|\psi \rangle \langle \psi | ) |\psi\rangle,
	\end{equation}
	In both cases, $d\psi$ is the Haar measure (over 1-qubit states in Eq.~\ref{eq:ontargetagi} and over 2-qubit states in Eq.~\ref{eq:agi}) and $\lind$ is the error generator of the gate.  Although AGF is provided for comparison to the literature, it is not a good predictor of performance in general circuits (see Supplemental Information S9), and when we use the unqualified term ``fidelity'', it always denotes generator fidelity, $\hat{\epsilon}$.  Section S9 of the Supplement includes an extensive discussion of overall gate error metrics and their relationships.

	\section*{Data availability}
	The experimental data that support the findings of this study are available in Figshare with the identifier doi.org/10.6084/m9.figshare.c.5471706.
	
	\section*{Code availability}
	The GST analysis was performed using a developmental version of pyGSTi that requires expert-level knowledge of the software to install and run. A future official release of pyGSTi will support the type of analysis performed here using a simple and well-documented Python script. Until this code is available, interested readers can contact the corresponding author to get help with accessing and running the existing code.
	Multivalley effective mass theory calculations, some of the results of which are illustrated in Fig.~\ref{fig:figure 1}b, were performed using a fork of the code first developed in the production of Ref.~\cite{gamble2015multivalley} that was extended to include multielectron interactions as reported in Ref.~\cite{joecker2020full}.
	Requests for a license for and copy of this code will be directed to points of contact at Sandia National Laboratories and the University of New South Wales, through the corresponding author.
	The analysis code for Bell state tomography is in Figshare with the identifier doi.org/10.6084/m9.figshare.c.5471706.
	
	\section*{Acknowledgements}
	We acknowledge helpful conversations with W. Huang, R. Rahman, S. Seritan, and C. H. Yang and technical support from T. Botzem. The research was supported by the Australian Research Council (Grant no. CE170100012), the US Army Research Office (Contract no. W911NF-17-1-0200), and the Australian Department of Industry, Innovation and Science (Grant No. AUSMURI000002). We acknowledge support from the Australian National Fabrication Facility (ANFF). This material is based upon work supported in part by the iHPC facility at UTS, by the by the U.S. Department of Energy, Office of Science, Office of Advanced Scientific Computing Research’s Quantum Testbed Pathfinder and Early Career Research Programs, and by the U.S. Department of Energy, Office of Science, National Quantum Information Science Research Centers (Quantum Systems Accelerator). Sandia National Laboratories is a multimission laboratory managed and operated by National Technology and Engineering Solutions of Sandia, LLC, a wholly owned subsidiary of Honeywell International, Inc., for the U.S. Department of Energy’s National Nuclear Security Administration under contract DE-NA0003525.  All statements of fact, opinion or conclusions contained herein are those of the authors and should not be construed as representing the official views or policies of IARPA, the ODNI, the U.S. Department of Energy, or the U.S. Government.
	
	\section*{Author contributions}
	M.T.M., V.S. and F.E.H. fabricated the device, with A.M.'s and A.S.D.'s supervision, on an isotopically-enriched $^{28}$Si wafer supplied by K.M.I.. A.M.J., B.C.J. and D.N.J. designed and performed the ion implantation. M.T.M. and S.A. performed the experiments and analysed the data, with A.L. and A.M.'s supervision. B.J. and A.D.B. developed and applied computational tools to calculate the electron wavefunction and the Hamiltonian evolution. A.Y. designed the initial GST sequences, with C.F.'s supervision. K.M.R., E.N., K.C.Y., T.J.P. and R.B.-K. developed the and applied the GST method. A.M., R.B-K., M.T.M. and S.A. wrote the manuscript, with input from all coauthors.
	
	These authors contributed equally: Mateusz T. M\k{a}dzik, Serwan Asaad.
	
	\section*{Author Information}
	
	\textbf{Correspondence and requests for materials} should be addressed to Andrea Morello, a.morello@unsw.edu.au.\\
	\textbf{Competing interest} The authors declare no competing interests.

	\newpage
	
	\renewcommand{\figurename}{Extended Data Fig.}
	\setcounter{figure}{0}    
	\onecolumn
	\section*{Extended data figures and tables}
	
	\begin{center}
		\includegraphics{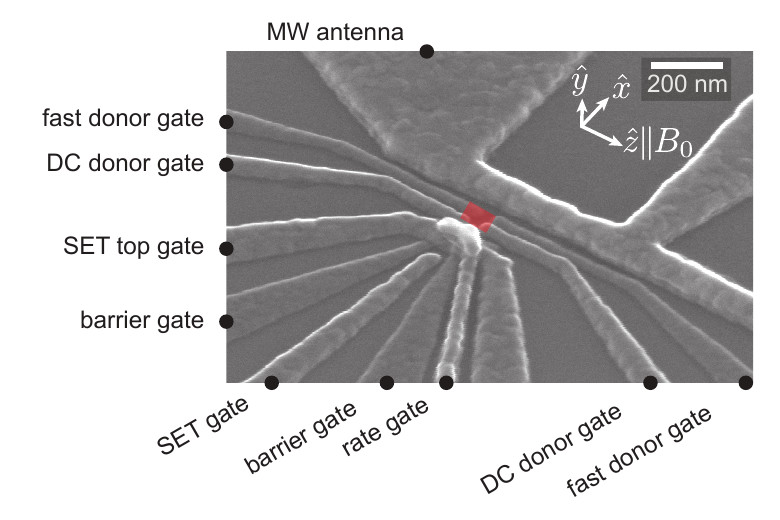}
		\captionof{figure}{\textbf{Device layout.}
			Scanning electron micrograph of a device identical to the one used in this experiment.
			$^{31}$P donor atoms are implanted in the region marked by the orange rectangle, using a fluence of $1.4 \times 10^{12}/\mathrm{cm}^2$ which results in a most probably inter-donor spacing of approximately 8~nm. Four metallic gates are fabricated around the implantation region, and used to modify the electrochemical potential of the donors.
			A nearby SET, formed using the SET top gate and barrier gates, enables charge sensing of a single donor atom, as well as its electron spin through spin-to-charge conversion (Methods).
			The tunnel coupling between the donors and SET is tuned by the rate gate situated between the SET and donor implant region.
			A nearby microwave (MW) antenna is used for ESR and NMR of the donor electron and nuclear spins, respectively.
		}
		\label{fig ED: device SEM}
	\end{center}

	\begin{center}
		\includegraphics{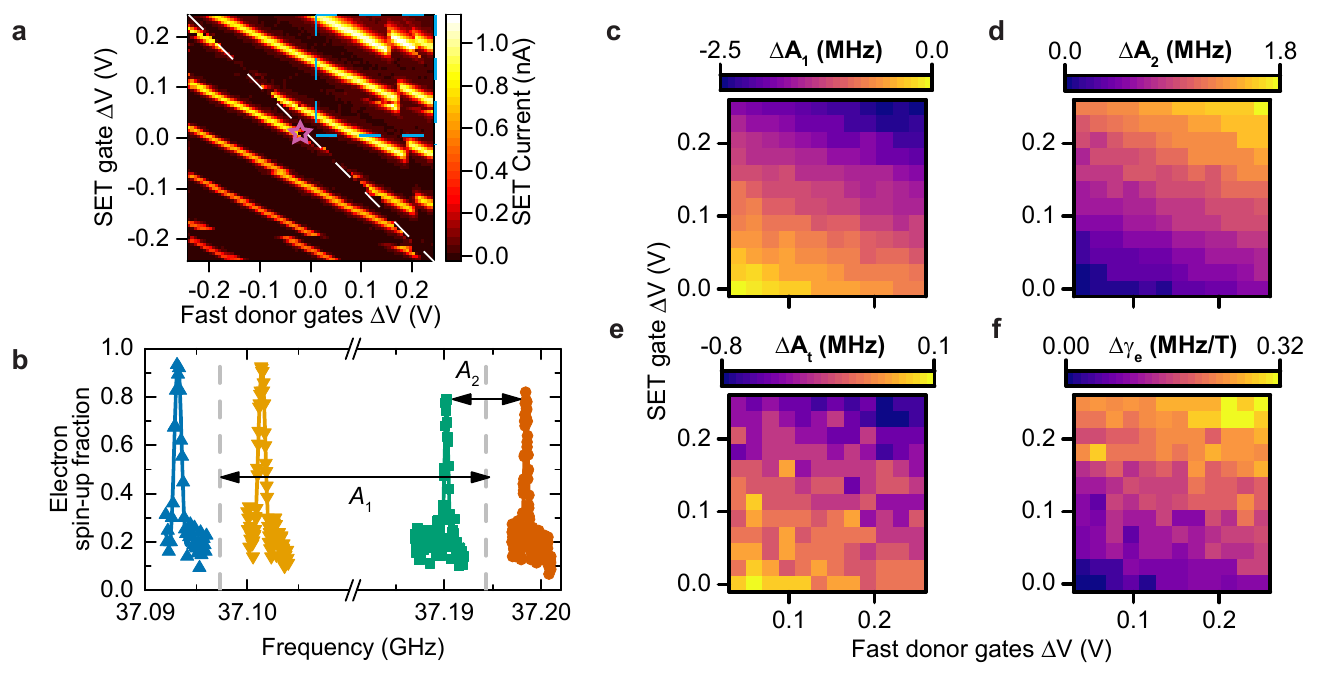}
		\captionof{figure}{
			\textbf{Electrical tunability of the hyperfine interaction and the electron gyromagnetic ratio.}
			\textbf{a,} Map of the SET current as a function of SET gate and fast donor gates (pulsed jointly). The white dashed line indicates the location in gate space where the 2P donor cluster changes its charge state. The third, hyperfine-coupled electron is present on the cluster in the region to the right of the line. Electron spin readout is performed at the location indicated by the pink star. \textbf{b,} ESR spectrum of the electron bound to the 2P cluster, acquired while the system was tuned within the blue dashed rectangle in panel \textbf{a}. The hyperfine couplings $A_1, A_2$ are extracted from ESR frequencies as shown, namely $A_{\rm 1} = (\nu_{\mathrm{e}|\Uparrow\Downarrow} + \nu_{\mathrm{e}|\Uparrow\Uparrow})/2 - (\nu_{\mathrm{e}|\Downarrow\Downarrow} + \nu_{\mathrm{e}|\Downarrow\Uparrow})/2$;  $A_{\rm 2} = \nu_{\mathrm{e}|\Uparrow\Uparrow} - \nu_{\mathrm{e}|\Uparrow\Downarrow}$. \textbf{c-d,} Extracted hyperfine couplings within the marked area. The data shows that $A_1$ decreases and $A_2$ increases upon moving the operation point towards higher gate voltages and away from the donor readout position. \textbf{e,} A small change is also observed in the sum of the two hyperfine interactions $A_{\rm t} = A_{\rm 1} + A_{\rm 2}$. \textbf{f,} Electrical modulation (Stark shift) of the electron gyromagnetic ratio $\gamma_e$, extracted from the shift of the average of the hyperfine-split electron resonances. The ESR frequencies can be tuned with fast donor gates at the rate of  $\Delta\nu_{\mathrm{e}|\Uparrow\Uparrow} = 0.3$~MHz/V;  $\Delta\nu_{\mathrm{e}|\Uparrow\Downarrow} = 5.2$~MHz/V;  $\Delta\nu_{\mathrm{e}|\Downarrow\Uparrow} = 7.6$~MHz/V;  $\Delta\nu_{\mathrm{e}|\Downarrow\Downarrow} = 2.4$~MHz/V.
		}
		\label{fig ED: hyperfine shifts}
	\end{center}
	
	\begin{center}
		\includegraphics[width=\linewidth]{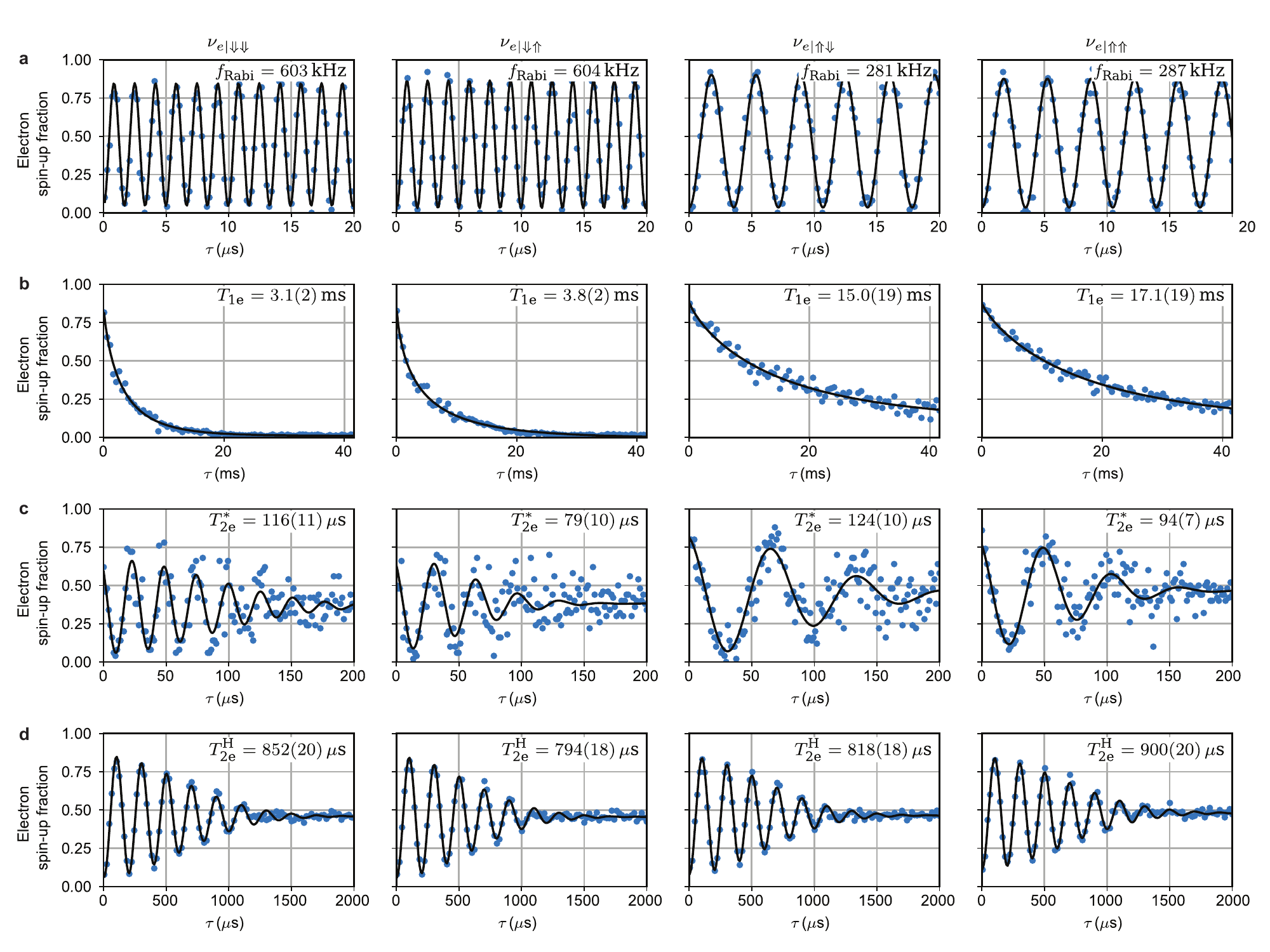}
		\captionof{figure}{
			\textbf{Coherence metrics of the electron spin qubit}.
			The columns correspond to the nuclear configurations $\ket{\Downarrow\Downarrow}$, $\ket{\Downarrow\Uparrow}$, $\ket{\Uparrow\Downarrow}$, $\ket{\Uparrow\Uparrow}$, respectively.
			All measurements start with the electron spin initialized in the $\ket{\downarrow}$ state.
			Error bars are $1\sigma$ confidence intervals.
			\textbf{a,} Electron Rabi oscillations. 
			The measurements were performed by applying a resonant ESR pulse of increasing duration.
			The different Rabi frequencies $f_{\rm Rabi}$ on each resonance are likely due to a frequency-dependent response of the on-chip antenna and the cable connected to it.
			\textbf{b,} Electron spin-lattice relaxation times $T_{\rm 1e}$.
			Measurements were obtained by first adiabatically inverting the electron spin to $\ket{\uparrow}$, followed by a varying wait time $\tau$ before electron readout.
			The observed relaxation times are nearly three orders of magnitude shorter than typically observed in single-electron, single-donor devices~\cite{tenberg2019electron}, and even shorter compared to 1e-2P clusters.
			This strongly suggests that the measured electron is the third one, on top of two more tightly-bound electrons which form a singlet spin state \cite{hsueh2014spin}.
			We also observe a strong dependence of $T_{\rm 1e}$ on nuclear spin configuration.
			\textbf{c,} Electron dephasing times $T_{\rm 2e}^*$.
			The measurements were conducted by performing a Ramsey experiment, i.e. by applying two $\pi/2$ pulses separated by a varying wait time $\tau$, followed by electron readout.
			The Ramsey fringes are fitted to a function of the form $P_\uparrow(\tau) = C_0+C_1\cos(\Delta \omega \cdot\tau + \Delta \phi)\exp[-(\tau/T_{\rm 2e}^*)^2]$, where $\Delta \omega$ is the frequency detuning and $\Delta \phi$ is a phase offset.
			The observed $T_{\rm 2e}^*$ times are comparable to previous values for electrons coupled to a single $^{31}$P nucleus.
			\textbf{d,} Electron Hahn-echo coherence times $T_{\rm 2e}^\mathrm{\rm H}$, obtained by adding a $\pi$ refocusing pulse to the Ramsey sequence. We also varied the phase of the final $\pi/2$ pulse at a rate of one period per $\tau = (5~\mathrm{kHz})^{-1}$., to introduce oscillations in the spin-up fraction which help improve the fitting. 
			The curves are fitted to the same function used to fit the Ramsey fringes, with fixed $\Delta \omega = 5~\mathrm{kHz}$.
			The measured $T_{\rm 2e}^\mathrm{\rm H}$ times are similar to previous observations for electrons coupled to a single $^{31}$P nucleus.
		}
		\label{fig ED: electron properties}
	\end{center}
	
	\begin{center}
		\includegraphics{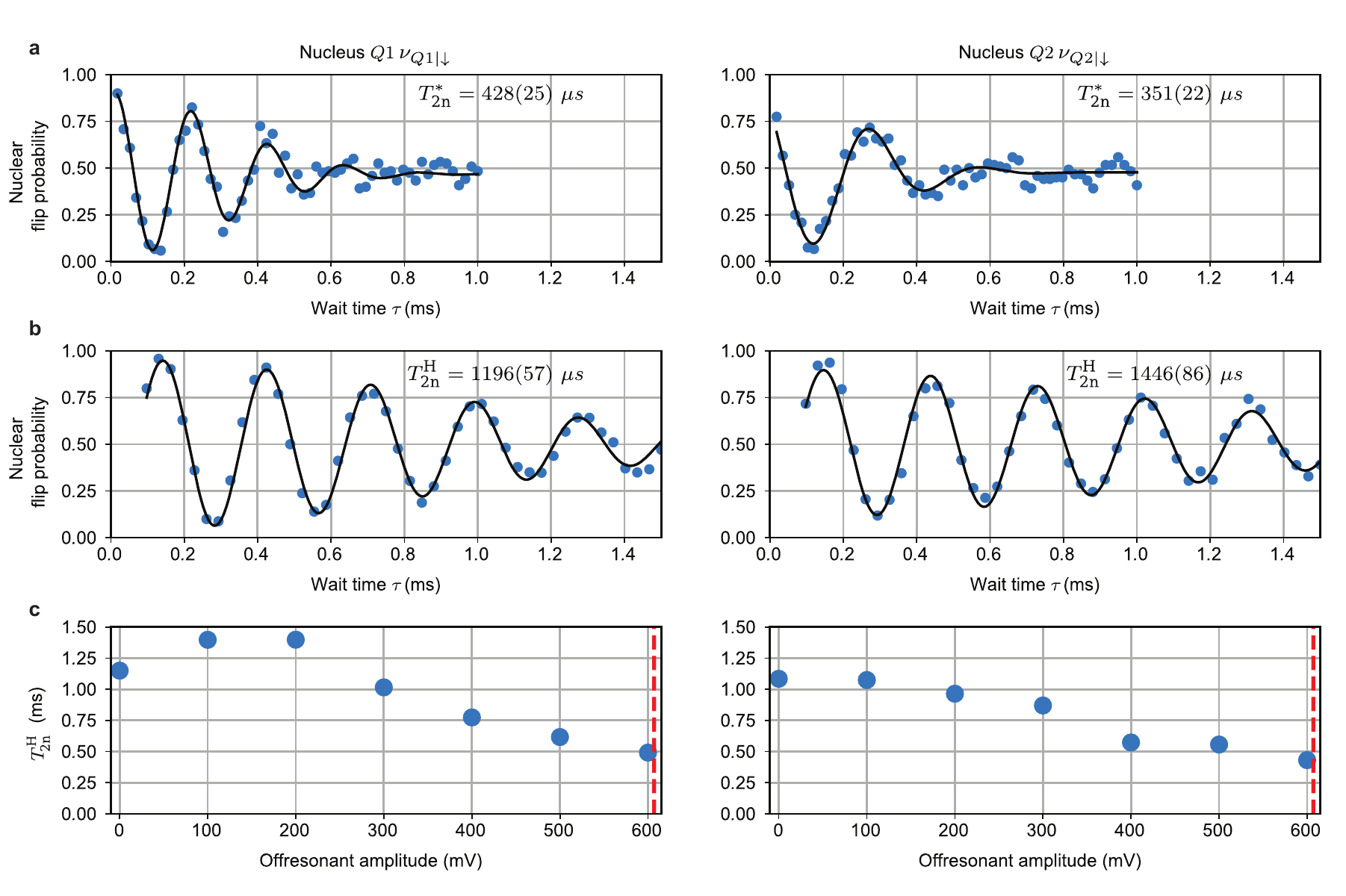}
		\captionof{figure}{\textbf{Nuclear spin coherence times.}
			Panels in column 1 (2) correspond to nucleus $Q1$ ($Q2$).
			Error bars are $1\sigma$ confidence intervals.
			\textbf{a,} Nuclear dephasing times $T_{\rm 2n}^*$, obtained from a Ramsey experiment.
			Results are fitted with a decaying sinusoid with fixed exponent factor 2 (see Extended Data Fig.~\ref{fig ED: electron properties}).
			\textbf{b,} Nuclear Hahn-echo coherence times $T_{\rm 2n}^\mathrm{H}$. To improve fitting, oscillations are induced by incrementing the phase of the final $\pi/2$ pulse with $\tau$ at a rate of one period per $(3.5~\mathrm{kHz})^{-1}$.
			Results are fitted with a decaying sinusoid with fixed exponent factor 2 (see Extended Data Fig.~\ref{fig ED: electron properties}).
			\textbf{c,} Dependence of $T_{\rm 2n}^\mathrm{H}$ on the amplitude of an off-resonance pulse. We perform this experiment to study whether a qubit, nominally left idle (or, in quantum information terms, subjected to an identity gate) is affected by the application of an RF pulse to the other qubit, at a vastly different frequency. Here, during the idle times between NMR pulses, an RF pulse is applied at a fixed frequency $20~\mathrm{MHz}$ -- far off-resonance from both qubits' transitions -- with varying amplitude $V_{\rm RF}$.
			The red dashed line indicates the applied RF amplitude for NMR pulses throughout the experiment.
			We observe a slow decrease of $T_{\rm 2n}^\mathrm{H}$ with increasing $V_{\rm RF}$. This is qualitatively consistent with the observation of large stochastic errors on the idle qubit, as extracted by the GST analysis in Fig.~\ref{fig:figure 3}.
		}
		\label{fig: nuclear T2}
	\end{center}
	
	\begin{center}
		\includegraphics{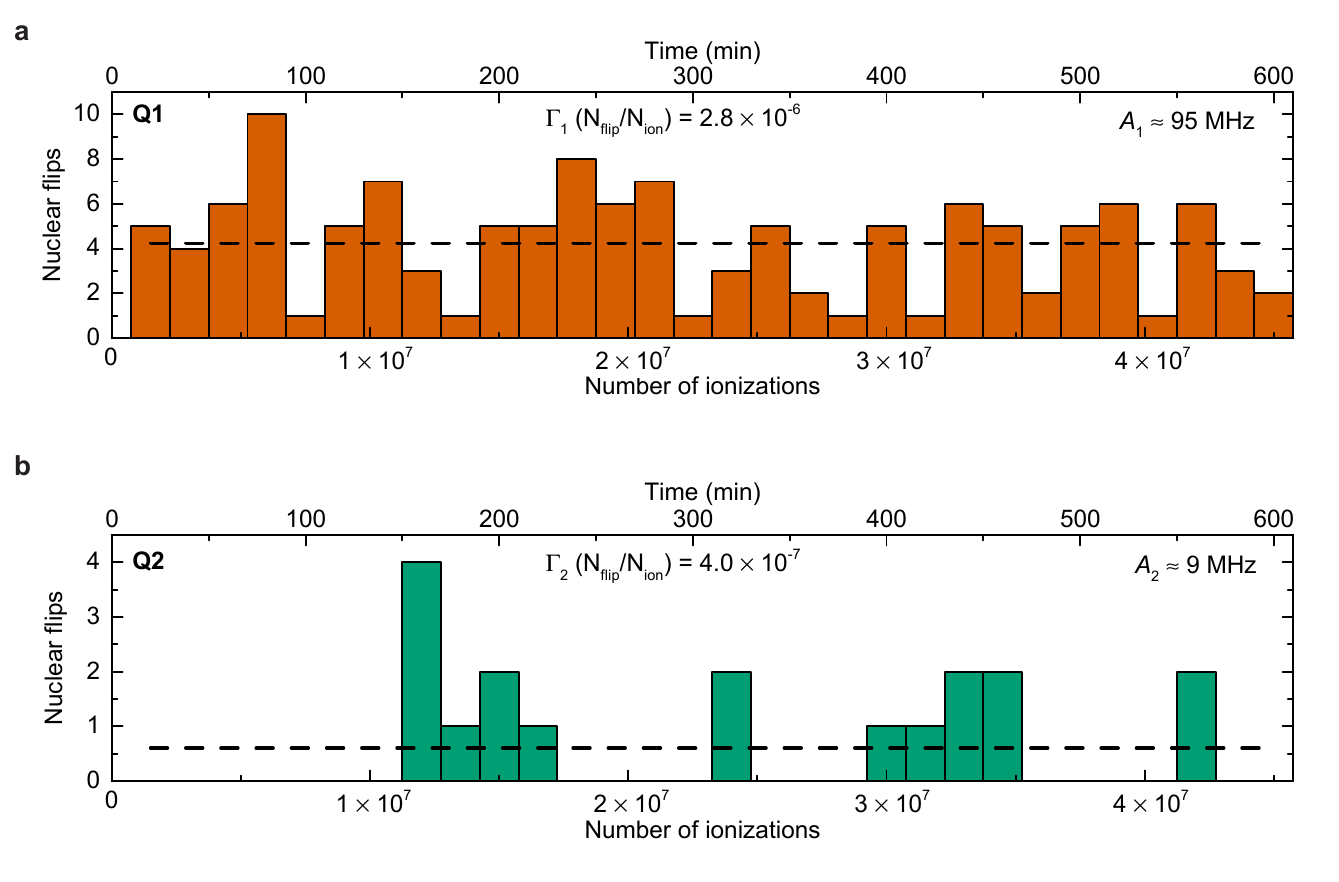}
		\captionof{figure}{
			\textbf{Nuclear spin quantum jumps caused by ionization shock.} 
			The electron and nuclear spin readout relies upon spin-dependent charge tunnelling between the donors and the SET island. If the electron tunnels out of the two-donor system, the hyperfine interactions $A_1,A_2$ suddenly drop to zero. If $A_1$ and $A_2$ include an anisotropic component (e.g. due to the non-spherical shape of the electron wavefunction which results in nonzero dipolar fields at the nuclei), the ionisation is accompanied by a sudden change in the nuclear spin quantisation axes (``ionisation shock''), and can result in a flip of the nuclear spin state. We measure the nuclear spin flips caused by ionisation shock by forcibly loading and unloading an electron from the 2P cluster every 0.8~ms.  \textbf{a,} For qubit 1 with $A_{\rm 1} = 95$~MHz, the flip rate is $\Gamma_1 = 2.8 \times 10^{-6} \frac{\rm N_{flip}}{\rm N_{ion}}$. \textbf{b,} For qubit 2 with $A_{\rm 2} = 9$~MHz, the flip rate is $\Gamma_2 = 4.0 \times 10^{-7} \frac{\rm N_{flip}}{\rm N_{ion}}$. This means that the nuclear spin readout via the electron ancilla is almost exactly quantum non-demolition. From this data, we also extract an average time between random nuclear spin flips of 283 seconds for qubit 1, and 2000 seconds for qubit 2. The extremely low values of $\Gamma$ -- comparable to those observed in single-donor systems -- are the reason why we can reliably operate the two $^{31}$P nuclei as high-fidelity qubits.}
		\label{fig ED: nuclear flipping rates}
	\end{center}
	
	\begin{center}
		\includegraphics{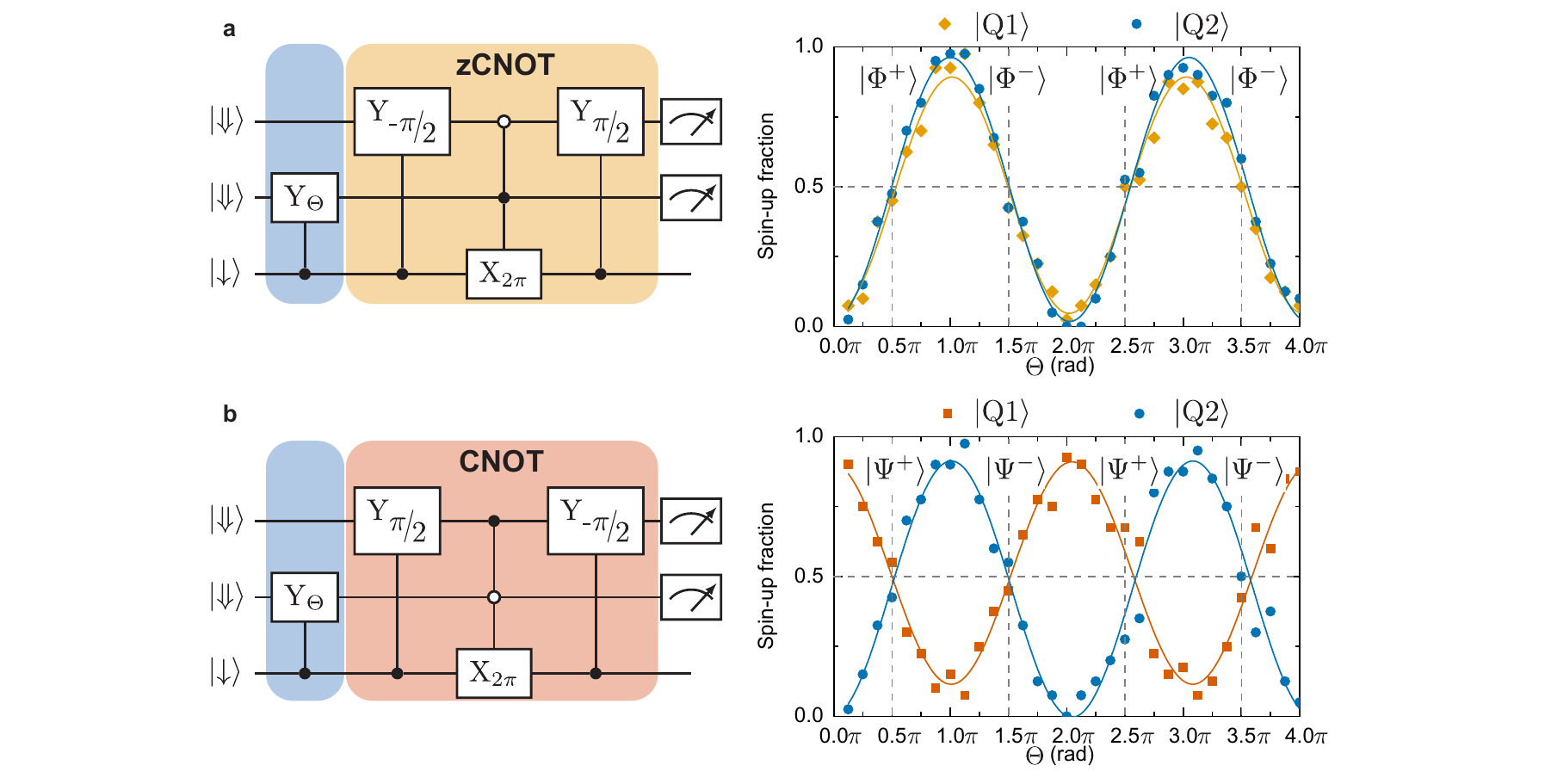}
		\captionof{figure}{\textbf{CNOT and zero-CNOT nuclear two-qubit gates.} We perform Rabi oscillation on the control qubit followed by the application of \textbf{a,} zCNOT or \textbf{b,} CNOT gates. The two qubits are initialized in the $\ket{\Downarrow\Downarrow} \equiv \ket{11}$ state. We observe the Rabi oscillations of both qubits in phase for zCNOT and out of phase for CNOT. At every odd multiple of $\pi/2$ rotation of the control qubit the Bell states are created.
		}
		\label{fig ED: CNOT zero-CNOT}
	\end{center}
	
	\begin{center}
		\includegraphics{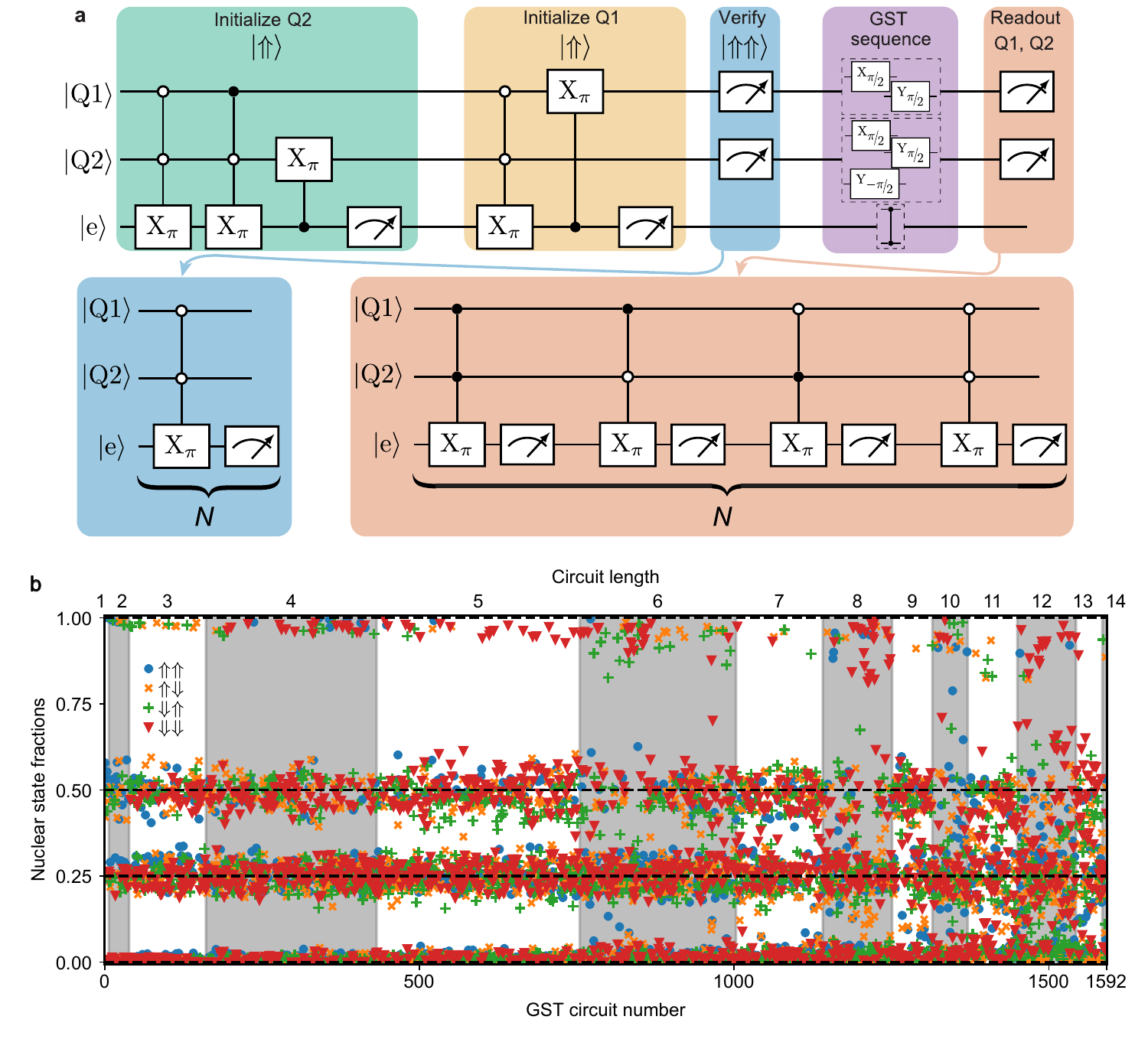}
		\captionof{figure}{\textbf{Two-qubit gate set tomography.}
			\textbf{a,} Measurement circuit for the two-qubit gate set tomography. A modified version of this circuit has been used for Bell state tomography. The green box prepares the qubit 2 in the $\ket{\Uparrow}$ state, then the orange box prepares the qubit 1 in the $\ket{\Uparrow}$ state. The readout step in the blue box (see Methods) determines whether the $\ket{\Uparrow\Uparrow}$ state initialization was successful. Only then the record will be saved. The electron spin is prepared in $\ket{\downarrow}$ during the nuclear spin readout process. Subsequently, the GST sequence is executed. The red box indicates the Q1,Q2 readout step. 
			The total duration of the pulse sequence is 120 ms, of which nuclear spin initialization is 8.6 ms (green and yellow), initial nuclear spin readout is 26.5 ms (blue), 3 ms delay is added for electron initialization (between blue and purple), GST circuit is 10 $\mu$s - 300 $\mu$s (purple), and nuclear readout is 80 ms (orange).
			\textbf{b,} Measurement results for individual two-qubit gate set tomography circuit.
			The first 145 circuits estimate the preparation and measurement fiducials, and the subsequent circuits are ordered by increasing circuit depth.
			At the end of a circuit, there are three situations for the target state populations: 1) the population is entirely in one state, while all others are zero; 2) the population is equally spread over two states, while the other two are zero; 3) the population is equally spread over all four states.
			The measured state populations for the different circuits therefore congregate around the four bands 0, 0.25, 0.5, and 1, as indicated by black dashed lines.
		}
		\label{fig ED: GST schematic}
	\end{center}
	
	\begin{center}
		\includegraphics[width=\linewidth]{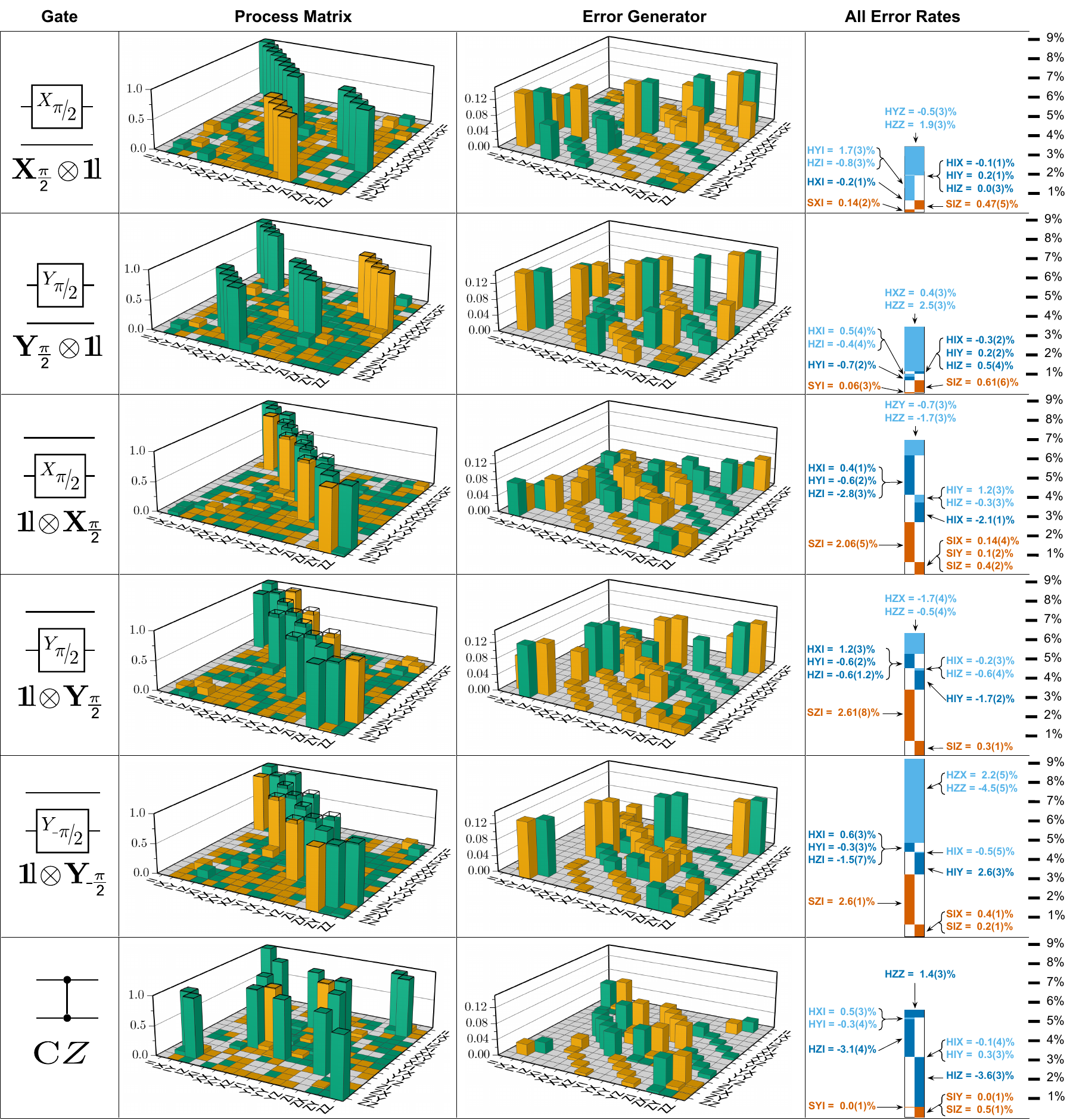}
		\captionof{figure}{\textbf{Estimated gate set, from process matrices to error rates}.  Experimental GST data were analyzed using pyGSTi to obtain self-consistent maximum likelihood estimates of 2-qubit process matrices for all 6 elementary gates.  These are represented (``Process Matrix'' column) in a gauge that minimizes their average total error, as superoperators in the 2-qubit Pauli basis. Green columns indicate positive matrix elements, orange ones are negative.  Wireframe sections indicate differences between estimated and ideal (target) process matrices.  Those process matrices can be transformed to error generators (``Error Generator'' column) that isolate those differences, and are zero if the estimated gate equals its target.  Each gate’s error generator was decomposed into a sparse sum of Hamiltonian and stochastic elementary error generators \cite{blume2021taxonomy}.  Those rates are depicted (``All Error Rates'' column) as contributions to the gate's total error, with $1\sigma$ uncertainties indicated in parentheses. Each non-vanishing elementary error rate (error generators are denoted “H” or “S” followed by a Pauli operator) is listed, and identified with its role in the total error budget (reproduced from Figure 3).  Orange bars indicate stochastic errors, dark blue indicate coherent errors that are intrinsic to the gate, and light blue indicate relational coherent errors that were assigned to this gate.  Total height of the blue region indicates the total coherent error, but because coherent error amplitudes add in quadrature, individual components' heights are proportional to their quadrature.}
		\label{fig ED: GST results}
	\end{center}
	
	\begin{center}
		\includegraphics[width=\linewidth]{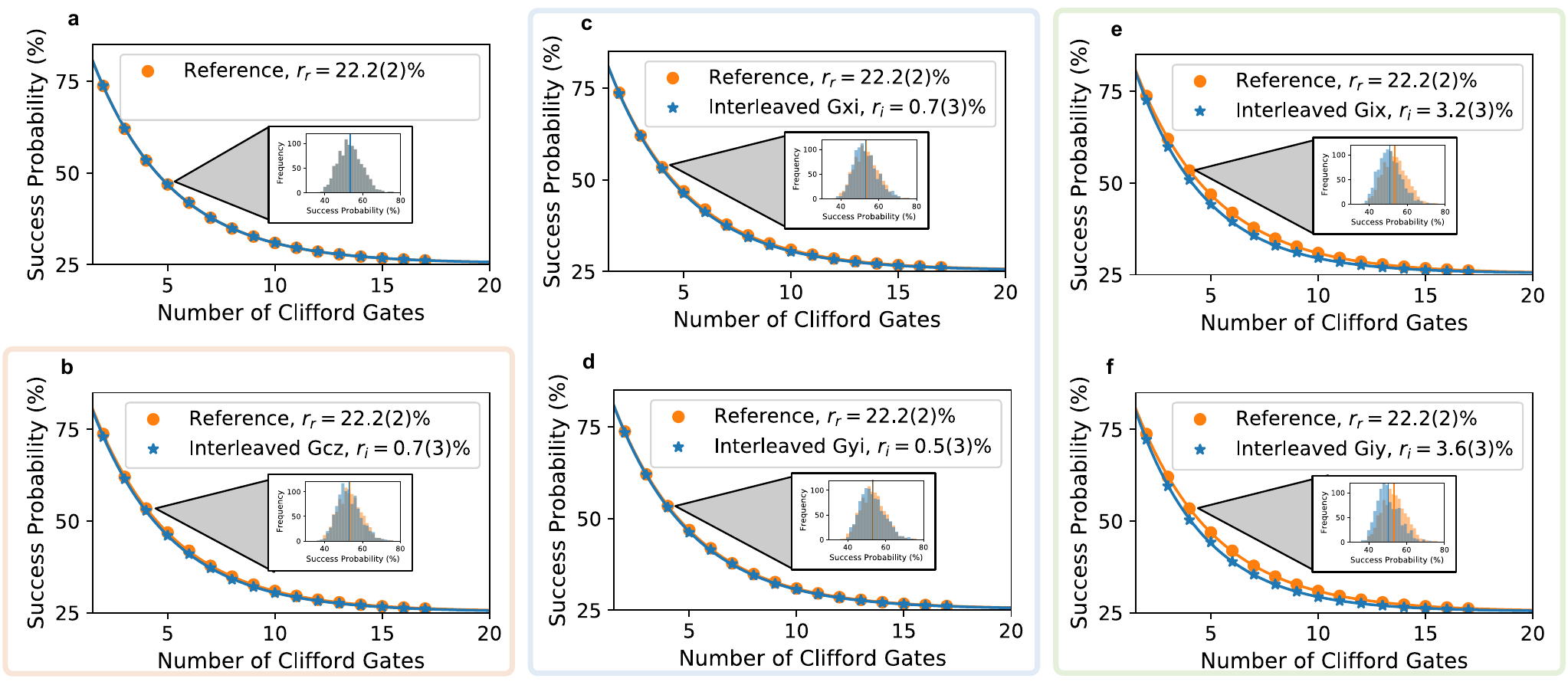}
		\captionof{figure}{\textbf{Simulation of standard and interleaved randomized benchmarking (RB).}  All simulated RB experiments used 2-qubit Clifford subroutines compiled from the 6 native gates, requiring (on average) 14.58 individual gate operations per 2-qubit Clifford.  \textbf{a}, Standard randomized benchmarking, simulated using the GST-estimated gate set, yields a ``reference'' decay rate of $r_r = 22.2(2)\%$, suggesting an average per-gate error rate of $r_r / 14.58 \approx 1.5\%$. $1\sigma$ confidence intervals are indicated in parentheses.   \textbf{b-f}, Simulated interleaved randomized benchmarking for the CZ gate, and 1-qubit $X_{\pi/2}$ and $Y_{\pi/2}$ gates on each qubit, yielded interleaved decay rates $r_r + r_i$.  For each experiment, 1000 random Clifford sequences were generated, at each of 15 circuit depths m, and simulated using the GST process matrices.  Exact probabilities (effectively infinitely many shots of each sequence) were recorded.  Inset histograms show the distribution over 1000 random circuits at m=4.  Observed decays are consistent with each gate’s GST-estimated infidelities -- e.g. $1-F = 0.79\%$ for the C-Z gate (b).  Performing these exact RB experiments in the lab would have required running 90000 circuits to estimate a single parameter ($r_i$) for each gate to the given precision of $\pm0.25\%$.  Using fewer ($<1000$) random circuits at each $m$ would yield lower precision.  GST required only 1500 circuits to estimate \emph{all} error rates to the same precision.}
		\label{fig ED: Simulated RB}
	\end{center}
	
	
	\begin{center}
		\includegraphics{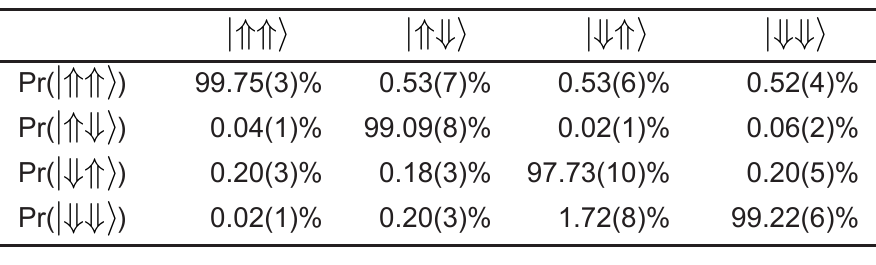}
		\captionof{table}{\textbf{Estimated state preparation and measurement (SPAM) error rates.}  In the GST analysis, the system's initial state was represented by a $4\times 4$ density matrix $\rho$, and the final measurement/readout by a 4-element $4\times 4$ POVM (positive operator-valued measure) $\{E_{\Uparrow\Uparrow},E_{\Uparrow\Downarrow},E_{\Downarrow\Uparrow},E_{\Downarrow\Downarrow}\}$ with $E_j \geq 0$ and $\sum_j{E_j}=I$.  We quantified the overall quality of the SPAM operations by using the GST estimate to compute the table of conditional probabilities shown here.  Each cell shows the estimated probability of a particular readout (e.g.~$\Uparrow\Uparrow$) given (imperfect) initialization in a particular state (e.g.~$\ket{\Downarrow\Downarrow}$).  The $\ket{\Uparrow\Uparrow}$ column can be read out directly from the estimate, since the experiment initalized into $\ket{\Uparrow\Uparrow}$.  Other states must be prepared by applying $X_{\pi/2}$ or $Y_{\pi/2}$ pulses.  These add additional error, which should not be attributed to SPAM operations.  To correct for this, we simulated ideal unitary rotation of the real $\ket{\Uparrow\Uparrow}$ state into each of the other 3 states by (1) taking the GST-estimated $X_{\pi/2}$ gates on each qubit and removing all intrinsic errors from them, and (2) simulating a circuit comprising initialization in $\rho$, an appropriate sequence of those idealized gates, and readout according to $\{E_j\}$.  The resulting analysis shows probabilities of all but one readout error to be below $1\%$, which is unprecedented in semiconductor spin qubit systems, and competitive with the state of the art in other physical platforms.}
		\label{fig ED: SPAM errors}
	\end{center}
	


\section*{Supplementary material}

## S1. ABSENCE OF DIRECT INTERACTION BETWEEN THE NUCLEAR SPIN QUBITS

In this section we provide experimental evidence for the absence of a direct interaction between the nuclear spin qubits Q1 and Q2. First, we remove the outermost electron from the 2P cluster and measure the resonance frequencies of Q1 and Q2. We apply RF pulses at very low power (-21 dBm at the source) to minimise power broadening of the NMR resonances. Supplementary Figure S1a shows that the two qubits have identical resonance frequencies, which are consistent with our estimate of the external magnetic field induced by the permanent magnet board ($\nu_{31\text{P}+} = \gamma_n B_0$, $\gamma_n = 17.23$ MHz/T, $B_0 \approx 1.328$ T). These results are consistent with the remaining two electrons being in a perfect $S = 0$ singlet state, whereby $A_1 = A_2 = 0$. They are also consistent with having removed all electrons from the 2P cluster, but we consider this interpretation less likely, based on the anomalously low value of the electron spin relaxation time $T_{1e}$ (see Extended Data Fig. 2). We can also deduce that the direct dipole-dipole interaction between the two nuclear spins is negligible in comparison to the broadening ($\sim 1$ kHz) of the NMR resonance peaks.

Next, we investigate the NMR spectrum of both Q1 and Q2 with all electrons present on 2P cluster. We measure the frequency response of a target qubit (Q1 - Supplementary Figure S1b; Q2 - Supplementary Figure S1c) with the spectator qubit (Q2 - Supplementary Figure S1b; Q1 - Supplementary Figure S1c) prepared in either $|\Downarrow\rangle$ or $|\Uparrow\rangle$ state. We observe no detectable resonance frequency shift due to a coupling between the nuclei, which in this case might be mediated by the shared electron.

These experiments corroborate the analysis in Section S11 A, wherein it is shown that no plausible value of inter-nuclear interaction can explain the presence of the weight-2 entangling errors unveiled by GST.

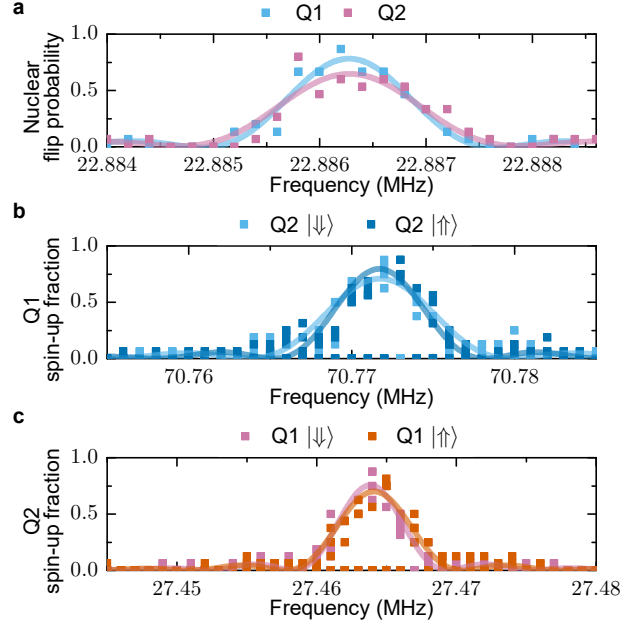

Figure S1. **a,** NMR spectrum of Q1 and Q2 with the third electron removed. The data was acquired at -21 dBm NMR power to minimise spectral broadening. The two remaining electrons are left in the magnetically inactive, $S = 0$ singlet state. The resonance frequencies of the two qubits are equal and consistent with the prediction for the ionized $^{31}$P in the estimated magnetic field produced by the permanent magnet board ($B_0 \approx 1.328$ T). **b,** NMR spectrum of Q1 (**c,** Q2) with all three electrons present on the 2P cluster. Q2 (**c,** Q1) was prepared in either $|\Downarrow\rangle$ or $|\Uparrow\rangle$. In both cases the Q1 (**c,** Q2) resonant frequency remains the same, indicating the absence of significant coupling between the two nuclei.

## S2. CNOT TRUTH TABLES

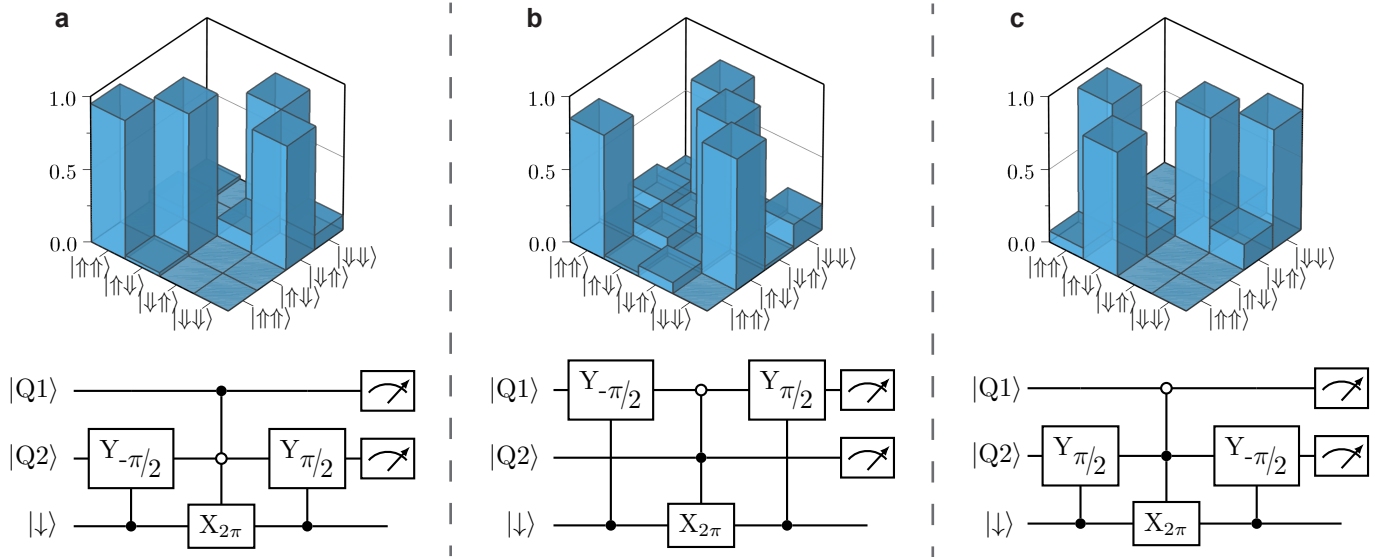

Figure S2. Experimental CNOT truth-tables. The qubits, encoded on the nuclear spins, have been prepared in all four eigenstates, with the electron spin in the $|\downarrow\rangle$ state. We adopt a notation for the computational basis consistent with the standard quantum information conventions, where $|\Downarrow\rangle \equiv |1\rangle$ and $|\Uparrow\rangle \equiv |0\rangle$. **a,** CNOT quantum logic gate, where Q1 serves as a control and Q2 as a target. **b,** CNOT quantum logic gate, where Q1 serves as a target and Q2 as a control. **c,** Zero-CNOT (zCNOT) quantum logic gate, where Q1 serves as a control and Q2 as a target. Here, by selecting a different ESR transition for the electron $2\pi$-pulse and exchanging the single-qubit gates, the target qubit is inverted when the control qubit is in the $|0\rangle$ state.

## S3. BELL STATE TOMOGRAPHY ANCILLARY DATA

The errors in the state preparation fidelity have been calculated using Monte Carlo bootstrap resampling. In this method, we assume a binomial distribution on the measured results and compute the possible measurement outcomes with regards to the experimental sampling (number of single shot readouts for state probability). We sample from the binomial distributions and reconstruct the density matrices 500 times. For each density matrix we calculate state preparation fidelity and concurrence. This allows for estimation of the average values, as well as assessment of the uncertainties in our results.

$$\rho_{\Phi^+} = \begin{pmatrix} 0.4802 + 0.0000i & 0.0600 + 0.0063i & -0.0333 + 0.0119i & 0.4722 - 0.0284i \\ 0.0600 - 0.0063i & 0.0089 + 0.0000i & -0.0054 + 0.0017i & 0.0567 - 0.0097i \\ -0.0333 - 0.0119i & -0.0054 - 0.0017i & 0.0096 + 0.0000i & -0.0219 - 0.0196i \\ 0.4722 + 0.0284i & 0.0567 + 0.0097i & -0.0219 + 0.0196i & 0.5013 + 0.0000i \end{pmatrix} \quad (1)$$

$$\rho_{\Phi^-} = \begin{pmatrix} 0.4880 + 0.0000i & 0.0002 + 0.0248i & -0.0433 + 0.0316i & -0.4804 - 0.0000i \\ 0.0002 - 0.0248i & 0.0079 + 0.0000i & 0.0070 + 0.0060i & -0.0012 + 0.0355i \\ -0.0433 - 0.0316i & 0.0070 - 0.0060i & 0.0125 + 0.0000i & 0.0481 + 0.0408i \\ -0.4804 + 0.0000i & -0.0012 - 0.0355i & 0.0481 - 0.0408i & 0.4916 + 0.0000i \end{pmatrix} \quad (2)$$

$$\rho_{\Psi^+} = \begin{pmatrix} 0.0055 + 0.0000i & 0.0365 + 0.0166i & 0.0464 + 0.0054i & 0.0017 - 0.0000i \\ 0.0365 - 0.0166i & 0.4461 + 0.0000i & 0.4664 - 0.0902i & 0.0121 - 0.0575i \\ 0.0464 - 0.0054i & 0.4664 + 0.0902i & 0.5301 + 0.0000i & 0.0103 - 0.0498i \\ 0.0017 + 0.0000i & 0.0121 + 0.0575i & 0.0103 + 0.0498i & 0.0184 + 0.0000i \end{pmatrix} \quad (3)$$

$$\rho_{\Psi^-} = \begin{pmatrix} 0.0328 + 0.0000i & -0.0724 + 0.0147i & 0.0784 - 0.0149i & 0.0104 - 0.0022i \\ -0.0724 - 0.0147i & 0.4730 + 0.0000i & -0.4728 + 0.0167i & -0.0138 + 0.0131i \\ 0.0784 + 0.0149i & -0.4728 - 0.0167i & 0.4847 + 0.0000i & 0.0236 - 0.0153i \\ 0.0104 + 0.0022i & -0.0138 - 0.0131i & 0.0236 + 0.0153i & 0.0095 + 0.0000i \end{pmatrix} \quad (4)$$

## S4. SINGLE-QUBIT GATE SET TOMOGRAPHY

During the initial calibration stage and in the preparation for the two-qubit gate set tomography, we performed single-qubit GST on both qubits. The 1-qubit GST experiments consist of 448 circuits of length up to 14. We have used these GST error estimates to iteratively correct our control pulses for imperfect calibration, in order to obtain the highest gate fidelities. Supplementary Figure S3a shows an example of data from a single-qubit GST on both Q1 and Q2.

The table in Supplementary Figure S3b compares the estimates of single qubit average gate fidelities obtained during these 1-qubit GST experiments, with those obtained as a subset of the 2-qubit GST experiments described in the main text. We choose to report average gate fidelities, rather than generator infidelities or entanglement infidelities, to facilitate easy comparison with other 1Q result available in the literature. The three datasets were acquired within 3 months from each other. Some variability in the gate fidelities across such a time span is entirely plausible, reflecting slow drifts in the sample caused e.g. by rearrangement of charges in the vicinity of the donors. Nevertheless, these experiments show that the 1-qubit gate fidelities extracted within 2-qubit GST are very close to those obtained by performing 1-qubit GST on each nucleus, and that the sample is remarkably stable over the course of many months.

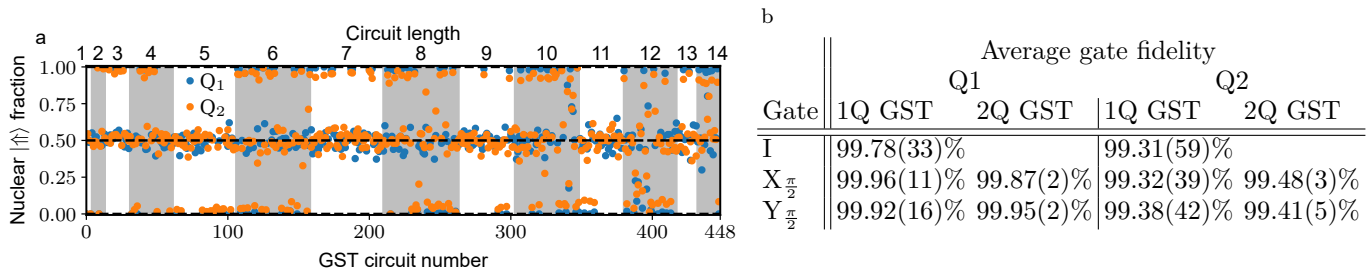

Figure S3. Single-qubit gate set tomography. **a,** Measured nuclear $|\Uparrow\rangle$ fraction of Q1 and Q2 for each of the 448 circuits on Q1 (blue) and Q2 (orange), sorted by circuit length. In the case of perfect gates and perfect measurements, the target $|\Uparrow\rangle$ fractions would be either 0, 0.5, or 1 (dashed lines), depending on the specific sequence. **b,** Average gate fidelities of the identity (I), $X_{\frac{\pi}{2}}$ and $Y_{\frac{\pi}{2}}$ single-qubit gates as obtained from the 1-qubit GST experiments shown in panel **a** (1Q), compared to the same quantities as obtained within the 2-qubit GST experiments described in the main text (2Q). For a fair comparison, we include in the 2Q fidelities only the component acting on the target qubit.

## S5. FREQUENCY RECALIBRATION DURING GST

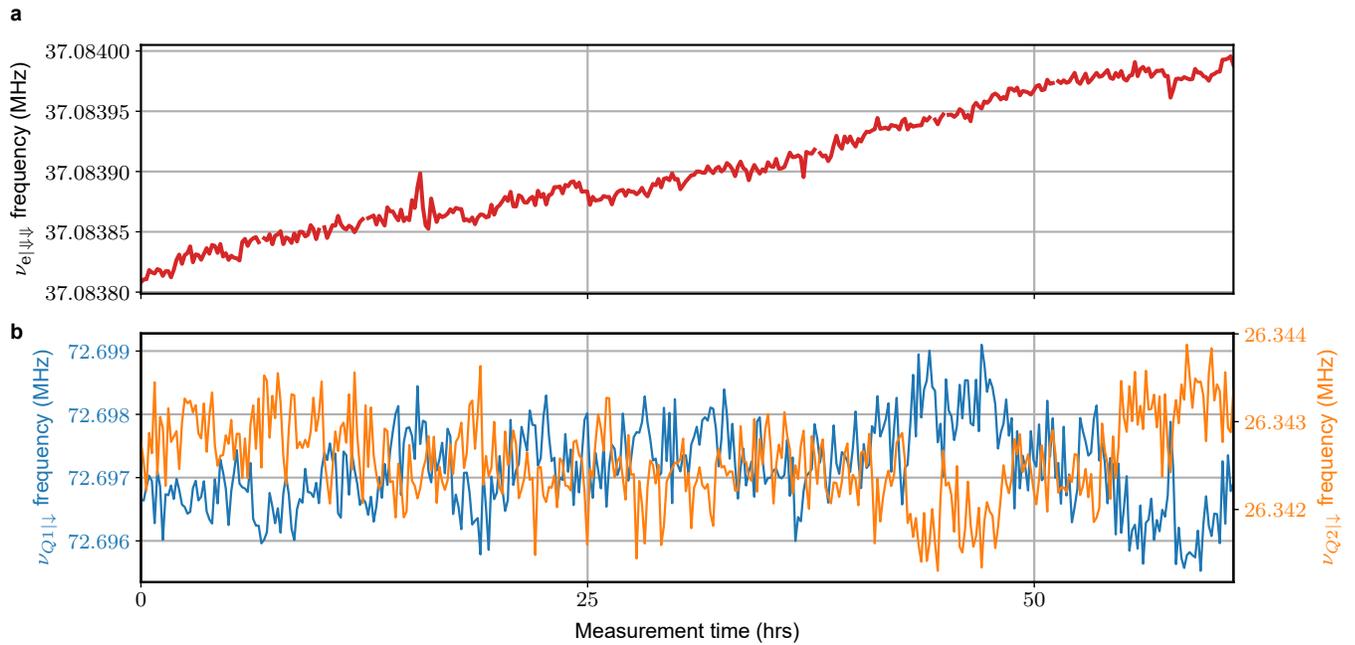

Figure S4. Electron $\nu_{\mathrm{e}|\Downarrow\Downarrow}$ (**a**) and nuclear $\nu_{\mathrm{Q1}|\downarrow}$ and $\nu_{\mathrm{Q2}|\downarrow}$ (**b**) frequency drifts measured while performing two-qubit gate set tomography. The effects of the drifts were periodically cancelled by calibrating the frequencies every tenth GST circuit (see Methods for calibration routines). The electron spin resonance frequency $\nu_{\mathrm{e}|\Downarrow\Downarrow}$ (**a**) experiences a fairly constant upwards drift, consistent with a steady increase of the magnetic field $B_0$. This could be caused by the assembly of the permanent magnet board where the device resides in; in this particular board, some of the NdFeB magnets were purposely oriented in an opposing direction in order to reduce the magnetic field to < 1.4 T (limit set by the maximum frequency of our microwave source). This frustrated magnetic configuration could slowly relax to a lower-energy configuration, with a corresponding higher magnetic field. Conversely, the shifts of the NMR frequencies (**b**) do not appear to be dominated by a magnetic field drift. $\nu_{\mathrm{Q1}|\downarrow}$ (blue) and $\nu_{\mathrm{Q2}|\downarrow}$ (orange) show an evident anti-correlation. This is consistent with hyperfine-induced shifts, since the two hyperfine interactions $A_1$ and $A_2$ were observed to shift in opposite directions in response to applied gate voltages (Extended Data Fig. 2).

## S6. EXPANSION OF GHZ ECHO CIRCUIT

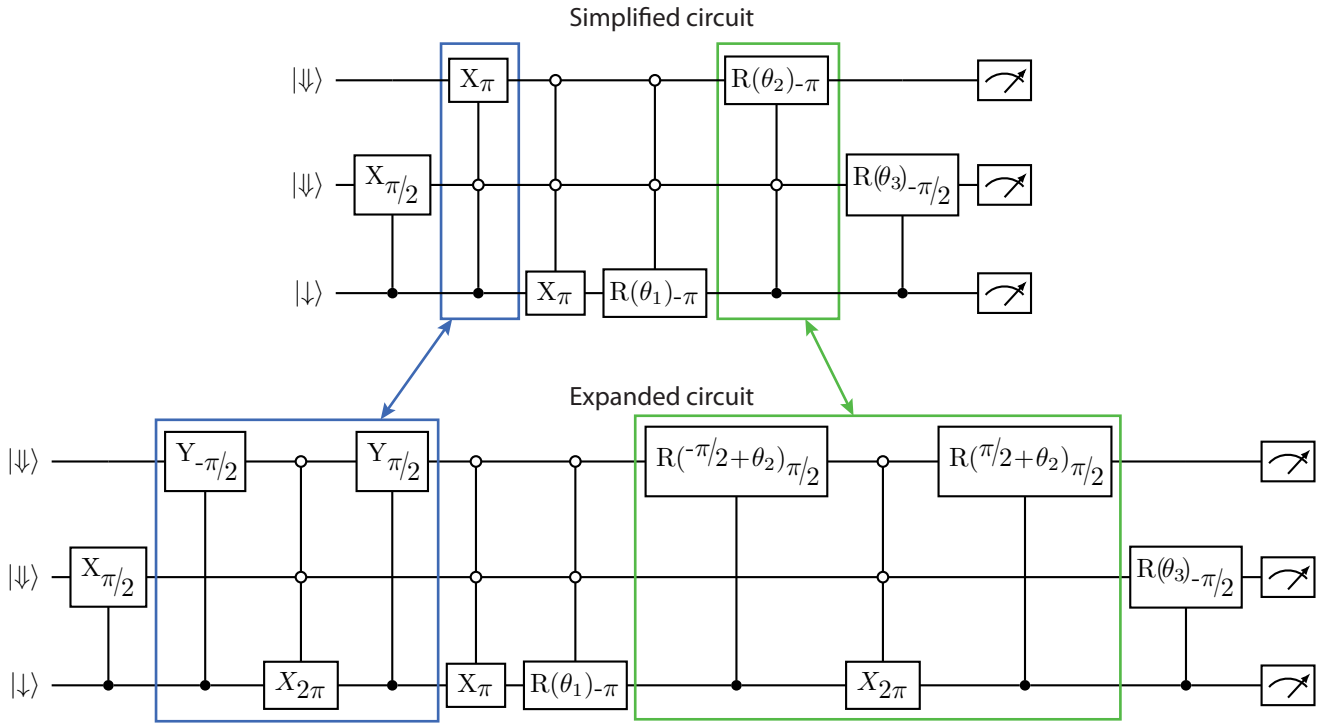

Figure S5. Expansion of the GHZ echo circuit. The circuit used to create a GHZ state and subsequently reverse the operations (top circuit and Fig. 4a in the main text) contains two nuclear operations on $Q_1$ that are conditional on the state of $Q_2$ (blue and green box). These operations are each composed of three pulses that are expanded in the bottom circuit.

**S7. HIGH-FIDELITY MEASUREMENT OF THREE-QUBIT STATES BY MAPPING AND READOUT ON THE NUCLEAR SPINS**

The measurements on the electron-nuclear three-qubit GHZ state (Fig. 4 in the main text) require extracting the populations of all eight basis states of the Hilbert space of the two nuclei and electron. One approach to do so is to first measure the electron spin, and then measure the nuclear spins by mapping their state onto the electron (see Methods). However, in this approach the readout fidelity is limited by the electron readout fidelity $\approx 80\%$, which would therefore limit the observable GHZ fidelity, unless the effect of SPAM errors were removed in post-processing.

In order to prove our ability to produce and measure a high-fidelity GHZ state without removing SPAM errors, we designed a method where specific qubit populations are swapped using conditional NMR or ESR $\pi$ pulses, followed by high-fidelity ($>99\%$) quantum nondemolition nuclear readout (see Methods). Each circuit is repeated three times, applying each of the following additional sets of swapping pulses prior to nuclear readout:

A. No additional pulses,

B. $X_\pi$ at frequency $\nu_{Q2|\downarrow}$,

C. $X_\pi$ at frequency $\nu_{e|\Downarrow\Downarrow}$, then $X_\pi$ at frequency $\nu_{Q1|\downarrow}$.

Each set of swapping pulses $\alpha \in [A, B, C]$ yields four measured nuclear state populations $[P(\Downarrow\Downarrow)_\alpha, P(\Downarrow\Uparrow)_\alpha, P(\Uparrow\Downarrow)_\alpha, P(\Uparrow\Uparrow)_\alpha]$, resulting in a total of twelve measured nuclear state populations. Each of the eight state populations $P(Q_1 Q_2 e)$ can be reconstructed from a minimum of five of the twelve measured state populations in ten different combinations, one example being

$$P(\Uparrow\Uparrow\uparrow) = (P(\Uparrow\Uparrow)_A - P(\Uparrow\Downarrow)_B + P(\Uparrow\Downarrow)_C - P(\Downarrow\Downarrow)_B + P(\Uparrow\Uparrow)_C)/2,$$
$$P(\Uparrow\Uparrow\downarrow) = (P(\Uparrow\Uparrow)_A - P(\Uparrow\Uparrow)_B + P(\Downarrow\Downarrow)_C - P(\Downarrow\Uparrow)_B + P(\Downarrow\Uparrow)_C)/2,$$
$$P(\Uparrow\Downarrow\uparrow) = (P(\Uparrow\Downarrow)_A - P(\Uparrow\Uparrow)_B + P(\Uparrow\Uparrow)_C - P(\Downarrow\Downarrow)_B + P(\Uparrow\Downarrow)_C)/2,$$
$$P(\Uparrow\Downarrow\downarrow) = (P(\Uparrow\Downarrow)_A - P(\Uparrow\Downarrow)_B + P(\Downarrow\Uparrow)_C - P(\Downarrow\Uparrow)_B + P(\Downarrow\Downarrow)_C)/2,$$
$$P(\Downarrow\Uparrow\uparrow) = (P(\Downarrow\Uparrow)_A - P(\Downarrow\Downarrow)_B + P(\Uparrow\Downarrow)_C - P(\Uparrow\Downarrow)_B + P(\Downarrow\Uparrow)_C)/2,$$
$$P(\Downarrow\Uparrow\downarrow) = (P(\Downarrow\Uparrow)_A - P(\Downarrow\Uparrow)_B + P(\Downarrow\Downarrow)_C - P(\Uparrow\Uparrow)_B + P(\Uparrow\Uparrow)_C)/2,$$
$$P(\Downarrow\Downarrow\uparrow) = (P(\Downarrow\Downarrow)_A - P(\Downarrow\Uparrow)_B + P(\Downarrow\Uparrow)_C - P(\Uparrow\Uparrow)_B + P(\Uparrow\Downarrow)_C)/2,$$
$$P(\Downarrow\Downarrow\downarrow) = (P(\Downarrow\Downarrow)_A - P(\Downarrow\Downarrow)_B + P(\Uparrow\Uparrow)_C - P(\Uparrow\Uparrow)_B + P(\Downarrow\Downarrow)_C)/2.$$

Each of the ten combinations produce slightly different state populations dependent on the individual measurement outcomes. The final state populations are then taken as the average of the ten possible state populations.

The validity of this state-readout technique is characterized by preparing each of the eight possible states, and then measuring the populations in all eight states. Figure S6 shows the measured populations, uncorrected for SPAM errors. The results show that for each of the eight initialized states, the measured population of the initialized state is near-unity. With this method we obtained the GHZ fidelity $\mathcal{F}_{\text{GHZ}} = 92.5(1.0)\%$ quoted in the main text, including the residual SPAM errors associated with nuclear readout.

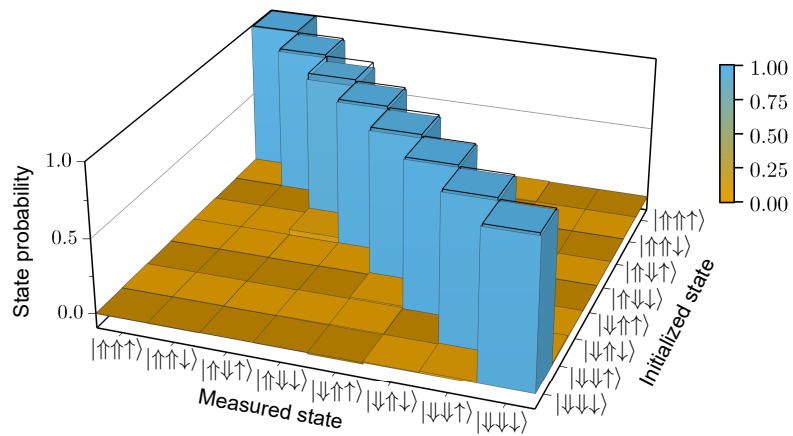

Figure S6. Extraction of eight electron-nuclear states by nuclear readout. Each row corresponds to the spin states of the two nuclei and electron being initialized into one of the eight possible eigenstates. Three sets of swapping pulses are then applied in distinct measurement instances, followed by nuclear readout. The resulting nuclear state populations are then combined to extract the eight state populations of the nuclei and electron. In all eight cases, the measured state population of the initialized state is close to unity, with small deviation arising from SPAM errors.

## S8. COMPARISON OF GST MODEL FITS

The reduced model we analyzed in the main text was selected through a process that compared multiple candidate models (see Fig. S7). In all the candidate models we considered, the errors on gate $G_i$ were described using an error generator $\mathbb{L}_i = \log(G_i \mathbb{G}_i^{-1})$, where $\mathbb{G}_i$ is the ideal target operation. Any 2-qubit $\mathbb{L}_i$ can be written as a linear combination of 240 elementary error generators of four types described in Ref. [1]: Hamiltonian (H), Pauli-stochastic (S), Pauli-correlation (C), and active (A).

- H generators, indexed by a single Pauli operator, cause coherent unitary errors (e.g., $H_{ZZ}$ generates a coherent $ZZ$ rotation).

- S generators, also indexed by a single Pauli, cause probabilistic Pauli errors (e.g. $S_{IX}$ causes probabilistic $X$ errors on Q2).

- C generators, indexed by two Paulis, transform Pauli-stochastic errors into stochastic errors that are not aligned with the Pauli basis.

- A generators, indexed by two Paulis, cause errors requiring feedback from the environment, including cooling (e.g. $T_1$ decay).

The largest model we considered allowed all 240 error generators on each of the 6 gates, and had a total of 1263 parameters (1440 gate + 63 SPAM - 240 gauge). We label this model "CPTP" because it can model almost all gate sets consisting of completely positive trace-preserving operations. We then repeatedly considered smaller (reduced) models, and compared each one to the next larger model using the technique described in the "Constructing and selecting reduced models" Methods section of the main text.

First we pinned all of the A and C generator rates of the CPTP model (1260 parameters) to zero. The resulting "H+S" model had an evidence ratio of $r = 1.1$, indicating that it was clearly preferable to the CPTP model. Next, we pinned all weight-2 S generator rates (correlated stochastic errors, 54 parameters) on all gates. The evidence ratio between this "H+S1" model and the H+S model was $r = 2.8$, and so again we preferred the smaller model. Pinning all of the weight-2 H generator rates to zero in a third "H1+S1" model was unacceptable, as this model had an evidence ratio of $r = 21.1$ when compared with the H+S1 model. By considering several models that added back different sets of weight-2 H rates, we found the final model analyzed in the main text. This model, labeled "H1+S1+ZZ**", includes all the error rates of the H1+S1 model (all weight-1 H and S generators) along with the following weight-2 H rates:

- $H_{ZZ}$ on all 6 gates

- $H_{\mathbb{G}[ZZ]}$ on each single-qubit gate $G$, e.g. $H_{YZ}$ for the $X_{\pi/2} \otimes I$ gate.

With only these specific weight-2 H generators, the H1+S1+ZZ** model has 33 fewer physical parameters (43 fewer rates, of which 10 are gauge degrees of freedom) than the H+S1 model and an evidence ratio of $r = 3.4$, which we accept based on our $r < 5$ criterion. The log-likelihood and evidence ratios for the nested series of candidate models are shown in Fig. S7, along with diagrams showing the decrease in allowed error rates per gate as we consider smaller models.

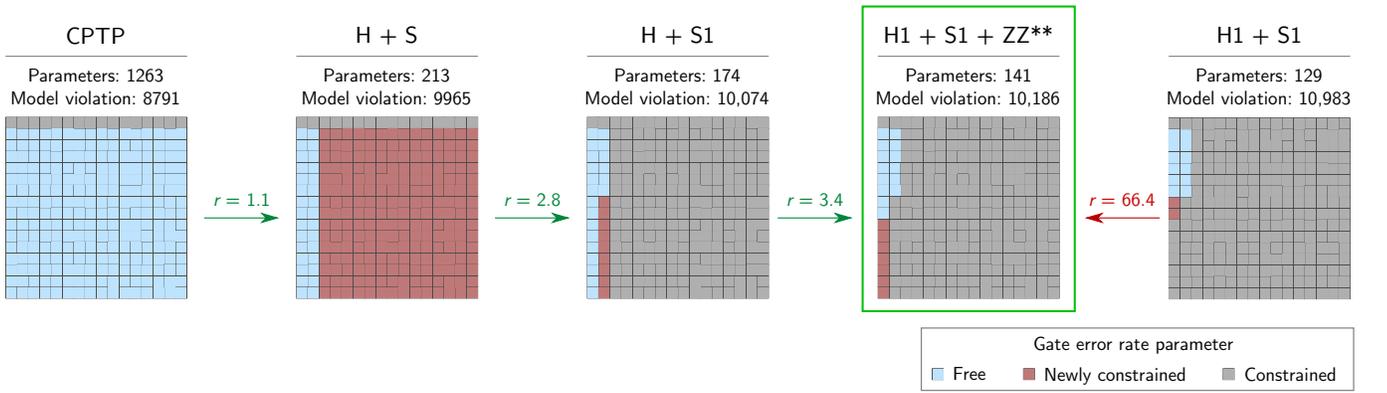

Figure S7. Comparison of candidate gate set models. Models are ordered left-to-right from the largest (most parameters) to the smallest. As explained in the text, the CPTP model allows gates to be arbitrary CPTP maps with all 240 error generator rates per gate. Moving rightward, more and more of these rates are pinned to zero, resulting in smaller models that necessarily provide worse fits to the data. For each model, a gate's allowed error generator rates are depicted as a set of blue boxes against a $15 \times 16$ grid of all the possible error generators for an arbitrary 2-qubit gate. Red boxes indicate the number of error generators that were pinned relative to the next larger model. Gray boxes indicate the number pinned to zero. For each model, the overall number (gates + SPAM - gauge) of model parameters $N$ is given, along with the model violation, $2\Delta \log \mathcal{L}$ (lower = better fit to the data). The evidence ratio $r = \frac{2\Delta \log \mathcal{L}_i - 2\Delta \log \mathcal{L}_j}{N_j - N_i}$ is computed between each pair of models, and we prefer the smaller model when $r \leq 5$. A green rectangle highlights H1+S1+ZZ$^{**}$, the best model among our candidates, which is used for the analysis in the main text.

## S9. REDUCED METRICS OF GATE ERROR

Process matrices are the standard model for errors in quantum gates [2]. They provide a comprehensive description of all possible Markovian errors that can affect a gate's performance, and they can be used to predict the measurement outcome distribution for arbitrary quantum circuits. The price of this flexibility is that process matrices can be very complex. An arbitrary $n$-qubit process matrix is a dense $4^n \times 4^n$ matrix describing a completely positive, trace-preserving map on density matrices. One way to simplify the interpretation of process matrices is through the use of reduced metrics, such as average gate infidelity or the diamond error, that summarize the performance of a quantum gate with a single number. Because the impact of a gate's errors depends strongly on where that gate appears in a quantum circuit, there are many different reduced metrics. In this section, we discuss the most common metrics that appear in the literature, as well as their relation to the novel metrics we introduce in the main text.

Here, as in the main text, we represent a quantum gate by a process matrix $G = e^{\mathbb{L}}\mathbb{G}$, where $\mathbb{G}$ is the process matrix for the perfect unitary implementation of the gate and $\mathbb{L}$ is the error generator. As discussed in the Methods, the error generator is a weighted sum of Hamiltonian (H) and stochastic (S) terms. The rates of the Hamiltonian terms are labeled $\theta$, and those of the stochastic terms are labeled $\epsilon$. Here we define three of the most common reduced metrics of gate quality in terms of the gate's error generator $\mathbb{L}$:

1. The *entanglement infidelity* ($\epsilon_e$):

$$\epsilon_e = 1 - \langle\varphi|(\mathbb{I} \otimes e^{\mathbb{L}})(|\varphi\rangle\langle\varphi|)|\varphi\rangle, \tag{5}$$

where $\varphi$ is any maximally entangled state over a doubled state space (here, a 4-qubit state space) and $\mathbb{I}$ is the identity operator.

2. The *average gate infidelity* ($\bar{\epsilon}$):

$$\bar{\epsilon} = 1 - \int d\psi \; \langle\psi|e^{\mathbb{L}}(|\psi\rangle\langle\psi|)|\psi\rangle, \tag{6}$$

where $d\psi$ is the Haar measure (here, over 2-qubit states).

3. The *diamond error* ($\epsilon_\diamond$):

$$\epsilon_\diamond = \frac{1}{2}\sup_\rho ||(e^{\mathbb{L}} \otimes \mathbb{I})(\rho) - \rho||_1, \tag{7}$$


where $\rho$ is a state over a doubled state space (here, a 4-qubit state space) and $\mathbb{I}$ is the identity operator. The supremum over states can be performed using a semidefinite program [3].

Both the entanglement infidelity ($\epsilon_e$) and the average gate infidelity ($\bar{\epsilon}$) capture a gate's performance in a random context. As discussed below, these metrics are often used to report the results of randomized benchmarking [4]. The diamond error, on the other hand, characterizes worst-case performance. For any quantum circuit, the total variation distance between the observed and ideal probabilities of measurement outcomes is bounded above by the sum of diamond errors over all operations in the circuit [5]. For this reason, thresholds for error correcting codes are often stated in terms of the diamond error [6–8].

In the main text, we introduce two novel reduced metrics that can be computed simply in terms of the error generator rates:

4. The *generator infidelity* ($\hat{\epsilon}$)

$$\hat{\epsilon} = \epsilon_{\text{agg}} + \theta_{\text{agg}}^2, \tag{8}$$

where $\epsilon_{\text{agg}} = \sum_i \epsilon_i$ is the sum of the rates of the stochastic error generators, and $\theta_{\text{agg}} = \sqrt{\sum_i \theta_i^2}$ is the root sum square of the rates of the Hamiltonian error generators. As shown below in Sec. S9 A, the generator infidelity is approximately equal to the entanglement infidelity.

5. The *total error* ($\epsilon_{\text{tot}}$)

$$\epsilon_{\text{tot}} = \epsilon_{\text{agg}} + \theta_{\text{agg}}, \tag{9}$$

where $\epsilon_{\text{agg}}$ and $\theta_{\text{agg}}$ are defined above. The total error captures worst-case performance, and, as shown in Sec. S9 B, is closely related to the diamond error.

In this work, five of the six gates we study are intended to implement single-qubit logic operations. In order to capture crosstalk effects, we model these single-qubit gates with *two-qubit* process matrices. Nonetheless, it can be useful to consider reduced metrics that quantify the performance of these gates when restricted to a single qubit (either the target qubit, or the spectator, which should undergo an identity operation). In the main text we use the following two restricted metrics:

6. The *single-qubit infidelity* ($\hat{\epsilon}^{(Q_j)}$):

$$\hat{\epsilon}^{(j)} = \epsilon_{\text{agg},(j)} + \theta_{\text{agg},(j)}^2, \tag{10}$$

where $\epsilon_{\text{agg},(j)} = \sum_{i,(Q_j)} \epsilon_i$ is the sum of the rates of the stochastic error generators with support solely on qubit $Q_j$, and $\theta_{\text{agg},(j)} = \sqrt{\sum_{i,(Q_j)} \theta_i^2}$ is root sum square of the rates of the Hamiltonian error generators that act solely on the qubit $Q_j$.

7. The *single-qubit average gate infidelity* ($\bar{\epsilon}^{(Q_j)}$):

$$\bar{\epsilon}^{(Q1)} = 1 - \frac{1}{2} \int d\psi \ \langle\psi|\text{tr}_{Q2}\left[e^{\mathbb{L}}\left(|\psi\rangle\langle\psi| \otimes \mathbb{I}\right)\right]|\psi\rangle, \tag{11}$$

$$\bar{\epsilon}^{(Q2)} = 1 - \frac{1}{2} \int d\psi \ \langle\psi|\text{tr}_{Q1}\left[e^{\mathbb{L}}\left(\mathbb{I} \otimes |\psi\rangle\langle\psi|\right)\right]|\psi\rangle, \tag{12}$$

where $d\psi$ is the Haar measure over 1-qubit states, and $\text{tr}_{Q_j}$ is a partial trace over the spectator qubit, $Q_j$. The single-qubit average gate infidelity is equal to the average gate infidelity of the following single-qubit quantum process:

a. Initialize the system in a pure state of target qubit $|\psi\rangle\langle\psi|$ and a completely mixed state of the spectator ($\frac{1}{2}\mathbb{I}$).
b. Apply the error channel ($e^{\mathbb{L}}$) to the two-qubit product state ($|\psi\rangle\langle\psi| \otimes \frac{1}{2}\mathbb{I}$).
c. Trace out the spectator qubit.

The result is a mixed state ($\rho$), whose state infidelity to $\psi$ can be computed simply as $1 - \langle\psi|\rho|\psi\rangle$. Averaging this quantity over Haar-random initial states yields the single-qubit average gate infidelity.

Each of the infidelities defined above has a corresponding *fidelity* equal to one minus the infidelity. In the literature, it is common to see the terms "infidelity" and "process infidelity" used to refer to either the entanglement infidelity or the average gate infidelity, often without specification. These two quantities are related to one another by a dimension-dependent proportionality factor [9], $\epsilon_e = \frac{2^n+1}{2^n}\bar{\epsilon}$, for an operation on $n$ qubits. The average gate fidelity $\bar{\epsilon}$ is not a "stable" metric [10] – it does not compose nicely upon combination of multiple qubits – but it is nonetheless commonly used to report results from randomized benchmarking [4]. Randomized benchmarking has been broadly adopted to experimentally characterise the performance of quantum gates [11–15], so we have reported the on-target average gate fidelities to ease comparison of our results to previous work. Whenever "infidelity" appears unqualified in the main text, it refers to $\hat{\epsilon}$, as defined in Eq. 8.

In Supplementary Information S9 A, we show that the infidelity $\hat{\epsilon}$ is equal to the entanglement infidelity $\epsilon_e$ at lowest order in $\epsilon_{\text{agg}}$ and $\theta_{\text{agg}}$. Unlike the average gate infidelity, the entanglement infidelity is a stable metric [10] – if two gates are performed on separate qubits in parallel, the entanglement fidelity of the composite layer is simply the product of the entanglement fidelities of the two individual gates. The stability of the entanglement infidelity partially motivated its adoption in the context of "cycle benchmarking," which has been used, e.g., for estimating gate errors in a 10-qubit ion-trap quantum processor [16].

In Supplementary Information S9 B, we show that the total error $\epsilon_{\text{tot}}$ provides an upper bound on the diamond error for single-qubit process matrices. This bound must be relaxed slightly for multi-qubit processes.

The infidelity $\hat{\epsilon}$ and the total error $\epsilon_{\text{tot}}$ also share the appealing property of being equal to each other when $\theta_{\text{agg}} = 0$, i.e., in the absence of coherent Hamiltonian errors ("perfect gate calibration"). This property recalls a similar relationship between the diamond error and entanglement infidelity, which coincide for Pauli stochastic error channels [4].

In Table S1, we list the infidelity, the entanglement infidelity, the total error, and the diamond error for the six gates considered in the main text. For all gates, the infidelity and the entanglement infidelity are nearly equal, and the total error and the diamond distance agree to within 22% relative error.

| Gate | Infidelity $\hat{\epsilon}$ | Entanglement infidelity $\epsilon_e$ | Total error $\epsilon_{\text{tot}}$ | Diamond error $\epsilon_\diamond$ |
|---|---|---|---|---|
| $X_{\frac{\pi}{2}} \otimes \mathbb{I}$ | 0.68(6)% | 0.66(6)% | 3.4(3)% | 3.2(2)% |
| $Y_{\frac{\pi}{2}} \otimes \mathbb{I}$ | 0.75(6)% | 0.73(7)% | 3.4(3)% | 3.6(4)% |
| $\mathbb{I} \otimes X_{\frac{\pi}{2}}$ | 2.86(7)% | 2.78(7)% | 6.9(3)% | 6.3(3)% |
| $\mathbb{I} \otimes Y_{\frac{\pi}{2}}$ | 3.44(10)% | 3.34(11)% | 6.3(5)% | 5.9(1.6)% |
| $\mathbb{I} \otimes Y_{-\frac{\pi}{2}}$ | 3.54(16)% | 3.39(18)% | 9.1(5)% | 7.7(5)% |
| CZ | 0.79(14)% | 0.80(16)% | 5.5(4)% | 7.0(4)% |

Table S1. Comparison between the various metrics used to summarize quantum gate performance. The infidelity ($\hat{\epsilon}$) is very close to the entanglement infidelity ($\epsilon_e$) – all the relative differences, $|\hat{\epsilon} - \epsilon_e|/\epsilon_e$, are less than 5%. The $1\sigma$ confidence intervals show that the two quantities take values consistent with each other, indicating that the higher order terms discussed in the text are insignificant. The the total error ($\epsilon_{\text{tot}}$) and the diamond error ($\epsilon_\diamond$ show larger disagreement, but all relative differences, $|\epsilon_{\text{tot}} - \epsilon_\diamond|/\epsilon_\diamond$, are less than 22%.

### A. Relationship between generator infidelity and entanglement infidelity

As part of our analysis within the main text, infidelities are computed for each gate in the H1+S1+ZZ** model. As stated in the "Aggregated error rates and metrics" Methods section, the infidelity we report, $\hat{\epsilon}$, is computed by summing the rates of the S generator rates and the squares of the rates of the H generator rates:

$$\hat{\epsilon} = \sum_i \epsilon_i + \sum_j \theta_j^2 = \epsilon_{\text{agg}} + \theta_{\text{agg}}^2, \tag{13}$$

where $\epsilon_i$ and $\theta_j$ run over all the S and H generator rates respectively. The infidelity ($\hat{\epsilon}$) is closely related to the entanglement infidelity ($\epsilon_e$), defined in Eq. 5. This relationship can be seen by direct calculation. The entanglement infidelity [9] of a quantum process $\rho \to \mathcal{E}(\rho)$ is equal to:

$$\epsilon_e = 1 - \frac{1}{d^3} \sum_Q \text{tr}\left(Q\mathcal{E}(Q)\right) \tag{14}$$

where $Q$ is a Pauli operator and $d$ is the Hilbert space dimension. We express the quantum process using the error generator framework introduced in Ref. [1] as $\mathcal{E}(Q) = \exp(\mathbb{L})(Q)$ and we assume the Lindblad generator $\mathbb{L}$ contains only Hamiltonian (H) and stochastic (S) generators. Each of these generators is indexed by a single Pauli operator:

$$H_P(Q) = -i[P, Q] \tag{15}$$
$$S_P(Q) = PQP - Q \tag{16}$$

Expanding the error process to second order in the generator, we have:

$$\epsilon_e = 1 - \frac{1}{d^3} \sum_Q \text{tr}\left(Q \exp(L)(Q)\right) \tag{17}$$

$$\simeq 1 - \frac{1}{d^3} \sum_Q \text{tr}\left(Q \left[1 + \sum_P (\theta_P H_P + \epsilon_P S_P) + \frac{1}{2} \sum_{PP'} (\theta_P H_P + \epsilon_P S_P)(\theta'_P H'_P + \epsilon'_P S'_P)\right](Q)\right) \tag{18}$$

The terms in this sum evaluate to:

$$\frac{1}{d^3} \sum_Q \text{tr}(Q^2) = 1 \tag{19}$$

$$\frac{1}{d^3} \sum_{P,Q} \theta_P \text{tr}(QH_P(Q)) = \frac{-i}{d^3} \sum_{P,Q} \theta_P \text{tr}(Q[P,Q]) = 0 \tag{20}$$

$$\frac{1}{d^3} \sum_{P,Q} \epsilon_P \text{tr}(QS_P(Q)) = \frac{1}{d^3} \sum_{P,Q} \epsilon_P \text{tr}(QPQP - Q^2) = -\sum_P \epsilon_P \tag{21}$$

$$\frac{1}{2d^3} \sum_{Q,P,P'} \theta_P \theta_{P'} \text{tr}(QH_P(H_{P'}(Q))) = -\frac{1}{2d^3} \sum_{Q,P,P'} \theta_P \theta_{P'} \text{tr}(Q[P,[P',Q]]) = -\sum_P \theta_P^2 \tag{22}$$

$$\frac{1}{2d^3} \sum_{Q,P,P'} \epsilon_P \theta_{P'} \text{tr}(QS_P(H_{P'}(Q))) = \frac{1}{2d^3} \sum_{Q,P,P'} \epsilon_P \theta_{P'} \text{tr}(QP[P',Q]P - Q[P',Q]) = 0 \tag{23}$$

$$\frac{1}{2d^3} \sum_{Q,P,P'} \theta_P \epsilon_{P'} \text{tr}(QH_P(S_{P'}(Q))) = \frac{-i}{2d^3} \sum_{Q,P,P'} \theta_P \epsilon_{P'} \text{tr}(Q[P, P'QP' - Q]) = 0 \tag{24}$$

$$\frac{1}{2d^3} \sum_{Q,P,P'} \epsilon_P \epsilon_{P'} \text{tr}(QS_P(S_{P'}(Q))) = \frac{1}{2d^3} \sum_{Q,P,P'} \epsilon_P \epsilon_{P'} \text{tr}(QP(P'QP' - Q)P - Q(P'QP' - Q)) = O(\epsilon^2) \tag{25}$$

Including the terms that are lowest-order in $\epsilon$ and $\theta$ gives:

$$\epsilon_e \simeq \sum_P \left(\epsilon_P + \theta_P^2\right) = \hat{\epsilon} \tag{26}$$

In evaluating the traces above, we have used the following facts for $d$-dimensional Pauli matrices $Q, P, P'$:

- $\text{tr}(Q) = 0$ unless $Q$ is the identity matrix, in which case the trace is equal to $d$.

- $\text{tr}(Q^2) = d$

- $\text{tr}(PP') = d$ if $P = P'$. Otherwise this trace vanishes.

- $\text{tr}(QPQP - Q^2) = -2d$ if $\{P, Q\} = 0$. Otherwise, this trace vanishes.

- $\text{tr}(QP'QP - P'P) = -2d$ if $P = P'$ and $\{P, Q\} = 0$. Otherwise, this trace vanishes.

- There are $d^2$ Pauli matrices, and any non-identity Pauli matrix commutes with exactly half of all Pauli matrices, and anti-commutes with the other half.

A comparison of $\hat{\epsilon}$ and $\epsilon_e$ for the gates discussed in the main text is shown in Table. S1.

**B. Relationship between the total error and the diamond error**

In this section, we show that the total error ($\epsilon_{\text{tot}}$) upper bounds the diamond error ($\epsilon_\diamond$) for single-qubit gates in the H+S model. For gates on multiple qubits, the diamond error can, in the worst case, exceed the total error by a factor that grows exponentially in the number of qubits.

The error generator for a gate in the H+S model is a sum of stochastic and Hamiltonian terms. S and H generators are each indexed by a single $n$-qubit Pauli matrix ($P$ or $Q$):

$$\mathbb{L} = \sum_P \epsilon_P S_P + \sum_Q \theta_Q H_Q. \tag{27}$$

By the Trotter-Suzuki formula,

$$\exp(\mathbb{L}) = \lim_{t \to \infty} \left[ \exp\left(\sum_Q \frac{\theta_Q}{t} H_Q\right) \prod_P \exp\left(\frac{\epsilon_P}{t} S_P\right) \right]^t \tag{28}$$

We can now compute the diamond error of the error map $\exp(\mathbb{L})$. By subadditivity of the diamond norm,

$$\epsilon_\diamond(\exp(\mathbb{L})) \leq \lim_{t \to \infty} t \left[ \epsilon_\diamond(\exp(\sum_Q \theta_Q H_Q/t)) + \sum_P \epsilon_\diamond(\exp(\epsilon_P S_P/t)) \right] \tag{29}$$

We can bound the contributions of the stochastic terms:

$$\epsilon_\diamond(\exp(\epsilon_P S_P/t)) = \frac{1}{2} \sup_\rho || \exp(\epsilon_P S_P/t)(\rho) - \rho ||_1 \tag{30}$$

$$= \frac{1}{2} \sup_\rho || \exp(-\epsilon_P/t)\rho + (1 - \exp(-\epsilon_P/t) P\rho P - \rho ||_1 \tag{31}$$

$$= (1 - \exp(-\epsilon_P/t)) \frac{1}{2} \sup_\rho || P\rho P - \rho ||_1 \tag{32}$$

$$= (1 - \exp(-\epsilon_P/t)) \tag{33}$$

$$\leq \epsilon_P/t, \tag{34}$$

where $P$ is the Pauli operator associated with the $S_P$ error generator. The supremum in Eq. 32 is equal to 2, and is achieved by $\rho = \frac{1}{2}(\mathbb{I} + Q)$ for any Pauli $Q$ such that $\{P, Q\} = 0$.

The Hamiltonian term, $\exp(\sum_Q \theta_Q H_Q/t)$ describes unitary evolution about an effective Hamiltonian $H = \sum_Q \theta_Q Q/t$, where $Q$ is the Pauli operator corresponding to the $H_Q$ error generator. In order to evaluate this term's contribution to Eq. 29, we rely on the well-known formula for the diamond error of a unitary channel:

$$\epsilon_\diamond(U) = \sqrt{1 - \min_{|\psi\rangle} \left( \langle \psi | U | \psi \rangle^2 \right)}, \tag{35}$$

Because of the Trotter expansion, the rotation induced by the effective Hamiltonian is very small. Expanding the unitary operator to second order in the effective Hamiltonian, we see:

$$\epsilon_\diamond(U) = \max_{|\psi\rangle} \left( \langle \psi | H^2 | \psi \rangle - \langle \psi | H | \psi \rangle^2 \right) \tag{36}$$

$$= \max_{|\psi\rangle} \text{Var}_{|\psi\rangle}(H) \tag{37}$$

$$= \frac{1}{2}(E_{\max} - E_{\min}) \tag{38}$$

The state that maximizes the variance of the Hamiltonian is an equal superposition of the eigenstates corresponding to the highest ($E_{\max}$) and lowest ($E_{\min}$) eigenvalues of the Hamiltonian. For a single qubit, eigenvalues of the effective Hamiltonian have magnitude $E = \sqrt{\sum_Q \theta_Q^2}/t = \theta_{\text{agg}}/t$. This results from the fact that all single-qubit Pauli matrices anti-commute with one another. For multiple qubits, it is possible for the Hamiltonian to be a sum of commuting terms. In this case, the extremal eigenvalue corresponds to the simultaneous +1 eigenstate of the Pauli operators

comprising the Hamiltonian, with magnitude $E = \sum_Q \theta_Q/t$. There are $2^n - 1$ traceless, mutually commuting Pauli operators on an $n$-qubit system, so the largest eigenvalue can be as high as $\sqrt{2^n - 1}\theta_{\text{agg}}/t$. The magnitude of the other extreme eigenvalue generally doesn't grow quite as quickly because it corresponds to a frustrated state. For local error channels (no crosstalk), the discrepancy will instead be proportional to $\sqrt{n}$. For few qubits, this correction is relatively small, and so $\theta_{\text{agg}}$ can nonetheless serve as a useful heuristic for aggregating the contributions from the Hamiltonian terms.

For a single qubit, we can combine the above results to see:

$$\epsilon_\diamond(\exp(\mathbb{L})) \leq \lim_{t \to \infty} t \left[ \sum_Q \epsilon_\diamond(\exp(\theta H_Q/t)) + \sum_P \epsilon_\diamond(\exp(\epsilon_P S_P/t)) \right] \tag{39}$$

$$\leq \lim_{t \to \infty} t \left[ \theta_{\text{agg}}/t + \sum_P \epsilon_P/t \right] \tag{40}$$

$$= \epsilon_{\text{agg}} + \theta_{\text{agg}} \tag{41}$$

$$= \epsilon_{\text{tot}}(\exp(\mathbb{L})) \tag{42}$$

For two or more qubits, the total error no longer bounds the diamond error. Instead, the significantly weaker bound follows from the above discussion:

$$\epsilon_\diamond(\exp(\mathbb{L})) \leq \epsilon_{\text{agg}} + \sum_Q \theta_Q \tag{43}$$

This bound is achievable (at first order). For example, an $n$-qubit channel where each qubit is subject to small, single-qubit, coherent $Z$-rotations will saturate this bound. A comparison of $\epsilon_{\text{tot}}$ and $\epsilon_\diamond$ for the gates discussed in the main text is shown in Table. S1.

## S10. DETAILED ANALYSIS OF THE $X_{\pi/2}$ GATE ON Q2

The analysis presented in Fig. 3 utilized several distinct metrics of gate performance. In this section, we detail how these metrics emerge by taking a detailed look at one exemplary gate, the $X_{\pi/2}$ gate on qubit Q2. Recall that the errors on this gate (and all the others) are given by a Lindbladian generator ($\mathbb{L}$), so that if $G$ and $\mathbb{G}$ are the noisy estimated and ideal $\mathbb{I} \otimes X_{\frac{\pi}{2}}$ gate, respectively, then $G = e^{\mathbb{L}}\mathbb{G}$. As in the main text, $\mathbb{L}$ is a weighted sum of particular (see S8) Hamiltonian (H) and Pauli-stochastic (S) elementary error generators:

$$\mathbb{L} = \sum_P \epsilon_P S_P + \sum_Q \theta_Q H_Q, \qquad (44)$$

where $P$ and $Q$ are two-qubit Pauli operators and the sums range over allowed elementary generators. Rates $\epsilon_P$ and $\theta_Q$ are given by our GST estimate. This estimate possesses gauge degrees of freedom which we resolve by either 1) choosing a gauge that minimizes the sum-of-squared error rate estimates, or 2) computing linear combinations of the estimated error rates that are, in the limit of small errors, gauge invariant.

The first option is more straightforward, and leads to the results shown in panels (a) and (b) of Fig. 3. Choosing a sensible gauge gives meaning to $\epsilon_P$ and $\theta_Q$, which are then categorized by their support and whether they are intrinsic (those that commute with the gate) or relational (those that don't, see Methods). The categorization of errors on the $\mathbb{I} \otimes X_{\frac{\pi}{2}}$ gate are shown in Fig. S8 and listed in Table S9, the former of which is an excerpt from Extended Data Fig. 8 and identical to a column of Fig. 3b except for the presence of labeled rates. For example, we see that $H_{XI}$, with rate $\theta_{XI} = 0.4\%$, is categorized as having support on Q1 (the bar is in the left "lane") and being an intrinsic error ($XI$ commutes with the $\mathbb{I} \otimes X_{\frac{\pi}{2}}$ gate).

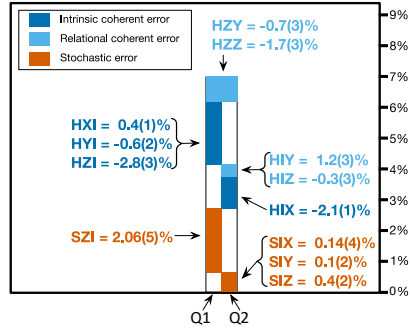

Figure S8. Distribution of error for the $X_{\pi/2}$ gate on Q2. Figure excerpted from Extended Data Fig. 8.

| Elementary error generator | rate | type | support |
|:---:|:---:|:---:|:---:|
| $H_{ZY}$ | -0.7% | relational | joint |
| $H_{ZZ}$ | -1.7% | relational | joint |
| $H_{XI}$ | 0.4% | intrinsic | Q1 |
| $H_{YI}$ | -0.6% | intrinsic | Q1 |
| $H_{ZI}$ | -2.8% | intrinsic | Q1 |
| $H_{IY}$ | 1.2% | relational | Q2 |
| $H_{IZ}$ | -0.3% | relational | Q2 |
| $H_{IX}$ | -2.1% | intrinsic | Q2 |
| $S_{ZI}$ | 2.06% | intrinsic | Q1 |
| $S_{IX}$ | 0.14% | intrinsic | Q2 |
| $S_{IY}$ | 0.1% | intrinsic | Q2 |
| $S_{IZ}$ | 0.4% | intrinsic | Q2 |

Figure S9. Elementary error rates of the $X_{\pi/2}$ gate on Q2

By aggregating these elementary rates we obtain the values in the cells of Fig. 3c. Stochastic errors add directly, and so the total stochastic error is $0.0206 + 0.0014 + 0.001 + 0.004 = 2.7\%$. Coherent errors add in quadrature, so aggregating the intrinsic H rates results in $\sqrt{0.004^2 + (-0.006)^2 + (-0.028)^2 + (-0.021)^2} = 3.6\%$. Similarly the relational H rates combine to give $\sqrt{(-0.007)^2 + (-0.017)^2 + 0.012^2 + (-0.003)^2} = 2.2\%$. The total coherent error (not shown in Fig. 3c), obtained by adding in quadrature all eight of the Hamiltonian error rates, is 4.2%.

Following the formulas for total error and infidelity given in S9 (and in Methods), we compute a total error of $0.027 + 0.042 = 6.9\%$ and an infidelity of $0.027 + 0.042^2 = 2.9\%$. Columns of Fig. 3c headed by (Q1) or (Q2) restrict the

aggregated error rates to those with the given support. For example, the total error for support (Q1) is found by summing the stochastic contribution, 2.06%, with the coherent contribution of $\sqrt{0.004^2 + (-0.006)^2 + (-0.028)^2} = 2.9\%$.

The second way we resolve the issue of gauge freedoms is by computing gauge-invariant error rates. These are linear combinations of the elementary error rates of one or more gates that are, to first order, unaffected by small gauge transformations (i.e., the gauge transformation matrix $M$ is close to the identity). These gauge-invariant error rates are shown Fig. 3d. Gauge-invariant rates that are local to a single gate are placed on the nodes of the graph. These are identical to the intrinsic rates from before, and so we see that the $\mathbb{I} \otimes X_{\frac{\pi}{2}}$ node displays the same intrinsic stochastic rate (2.7%) and intrinsic coherent rate (3.6%) that were computed above.

Gauge-invariant rates that are linear combinations of elementary generators from two or more gates are called gauge-invariant relational error rates. In our analysis, all the gauge-invariant relational error rates were pairwise – they involved error generators from just 2 gates – and they are displayed as edges in Fig. 3d. These are a different type of error rate: they cannot be attributed to any one gate and have "units" different from those of the gauge-fixed on-gate relational errors of Fig. 3a,c (e.g., an error in the angle between two gates' rotation axes). The largest relational error involving $\mathbb{I} \otimes X_{\frac{\pi}{2}}$ is a 2.7% coherent error relative to the $Y_{\frac{\pi}{2}} \otimes \mathbb{I}$ gate. This indicates that by different choices of the gauge, error can be moved between these gates.

## S11. RATIONALIZING TWO-QUBIT ERRORS ON ONE-QUBIT GATES

In Section S11 A we discuss the evidence for the surprising weight-2 entangling errors that GST infers for our one-qubit gates, concluding that it is not caused by non-Markovian effects or outlier data points. This motivates a search for the physical origins of these errors. In Section S11 B we will conclude that the direct entangling interaction between $^{31}$P nuclear spins is far too weak. Then, in Section S11 C we will account for the presence of an indirect electron-mediated entangling interaction between the nuclear spins through concatenated Schrieffer-Wolff transformations [17] that leave us with dressed nuclear spin qubit states. Finally, in Section S11 D we will consider the impact of driving fields in the doubly Schrieffer-Wolff transformed frame that defines our nuclear spin qubit states. We will find that plausible magnetic and electric fluctuations in the effective Hamiltonian, occurring due to the NMR drive, do not rationalize the GST-inferred coherent entangling errors. However, in Section S11 E, we invoke the leakage of microwaves near the ESR transition frequencies to rationalize the effective entangling interaction. The effect of this microwave leakage is qualitatively different from the types of errors studied in the context of fluctuations in the NMR drive, as they result in the accumulation of a geometric phase on the electron spin qubit that is conditioned on the nuclear spin state. The sizes of the observed errors are consistent with plausible levels of leakage in our experimental setup.

### A. Validation of entangling errors

Analysis of the GST data using model selection (see "Construction of reduced model" in the Methods section of the main text) indicated strong evidence for weight-2 entangling H errors on the single-qubit gates. Our final 83-parameter model contained 2 entangling H errors on each of the 5 single-qubit gates. Eliminating them from the model decreased $2\log\mathcal{L}$ by 803. This evidence ratio of $r = 80$ constitutes compelling evidence that entangling errors on single-qubit gates are necessary to explain the data (we consider $r > 5$ significant). Eliminating $H_{ZZ}$ and $H_{\mathbb{G}[ZZ]}$ on any single gate yielded evidence ratios between 46 and 257, which constitutes compelling evidence for entangling errors on every gate.

We investigated whether this evidence could be caused by non-Markovian effects (e.g. slow drift or persistent environments), or whether it might be the result of outlier data points. The quality of the GST fit indicated a moderate degree of non-Markovianity – i.e., variations in the data that are inconsistent with any Markovian gate set – as follows. Relative to a "maximal model" constructed to fit each one of the 1592 circuit's observed frequencies exactly, the full GST model displayed a total $2\Delta\log\mathcal{L}$ of 8793. Perfectly Markovian data would yield $2\Delta\log\mathcal{L} \approx 3 \times 1592 - 1263 = 3513$ (the number of free parameters in the data minus the number in the model), so these data are overdispersed by a factor of $\sqrt{8793/3513} \approx 1.6$. In principle this much overdispersion could create spurious effects at the scale we observed, but only if the deviations are concentrated on a small number of circuits. We did not observe such concentration. Instead, the observed deviations on almost all of the 1592 circuits were consistent with an overdispersed $\chi^2_3$ distribution, indicating that the physical effects causing the deviations affected all circuits fairly uniformly. However, we observed between 6 and 15 clear outliers. Eliminating the largest 6 or 15 outliers and reanalyzing the data did not change conclusions.

Finally, we looked for correlation between (1) the circuits that provided evidence for non-Markovianity, and (2) the circuits that provided evidence for entangling errors. For each circuit, we computed:

1. Its contribution to $2\Delta\log\mathcal{L}$ between (a) the maximal model and (b) the full GST model;

2. Its contribution to $2\Delta\log\mathcal{L}$ between (a) the GST model including entangling errors and (b) the GST model without entangling errors.

We observed no evidence at all of correlation.

In summary, we found no reason to suspect that the evidence of entangling errors on single-qubit gates is an artifact of outliers, data analysis, or systematic non-Markovian effects. This does not rule out such explanations, but it motivates searching for physical (Hamiltonian) causes in what follows.

### B. The direct entangling interaction between nuclear spins is too weak

The only direct entangling interaction between the two nuclear spin qubits is the mutual interaction of their magnetic dipole moments through the magnetic fields that they generate. Due to the dipolar nature of these fields the strength of this interaction decays cubically with the distance between the nuclei, realizing a coupling of $\approx 1\,\text{Hz}$ at $5\,\text{nm}$. For a one-qubit gate time of $\approx 10\,\mu\text{s}$, this translates into an always-on ZZ error of $\approx 10\,\mu\text{rad}$ in size. Instead, the coherent entangling errors are predicted to be $\approx 10\,\text{mrad}$, which would require an internuclear separation of $0.5\,\text{nm}$. This is approximately the lattice constant of silicon, meaning that the nuclear spins would have to be within only a couple

of lattice sites of each other to rationalize this direct interaction as the source of the coherent entangling errors on our one-qubit gates. Such proximity is inconsistent with the degree of controllability of the relative contact hyperfine interactions of the two nuclei (Extended Data Fig. 2), the weak anisotropic hyperfine interactions evident in the non-demolition nature of our readout (Extended Data Fig. 5), and the NMR spectra reported in Section S1. We can thus conclude that the direct interaction between the two nuclear spins is not the origin of these errors.

### C. Accounting for electron-mediated internuclear entanglement

This still leaves the possibility of an indirect effective interaction between the two nuclear spins that is mediated by the electron that they share. To study the impact of such an effective interaction, we apply a sequence of two transformations to the static Hamiltonian for the two nuclear spins and their shared electronic spin. The first transformation will mix certain states with opposite nuclear spin parity and opposite electronic spin. This will give us an effective Hamiltonian in which there is a XX+YY coupling between the nuclear spins. We can then project onto the electron spin-down subspace of this Hamiltonian and apply a second transformation to reduce the remaining off-diagonal coupling even further. Identifying this doubly dressed basis will be critical to analyzing the impact of the electron-mediated effective entangling interaction in the presence of the driving fields that implement one-qubit gates in Section S11 D.

We first separate the static lab frame three-qubit Hamiltonian ($\hat{H}$) into the diagonal Zeeman term ($\hat{H}_0$) and a perturbation due to the contact hyperfine interaction ($\hat{V}$),

$$\hat{H} = \hat{H}_0 + \hat{V}, \tag{45a}$$

$$\hat{H}_0 = -\gamma_n B_0 \left(\hat{I}_{1,z} + \hat{I}_{2,z}\right) - \gamma_e B_0 \hat{S}_z, \text{ and} \tag{45b}$$

$$\hat{V} = \left(A_1 \hat{I}_{1,x} + A_2 \hat{I}_{2,x}\right) \hat{S}_x + \left(A_2 \hat{I}_{1,y} + A_2 \hat{I}_{2,y}\right) \hat{S}_y + \left(A_1 \hat{I}_{1,z} + A_2 \hat{I}_{2,z}\right) \hat{S}_z. \tag{45c}$$

The first transformation that we will apply is a Schrieffer-Wolff transformation to third order in $\hat{V}$ accompanied by a projection onto the electron spin-down subspace. The unitary component of this transformation is generated by a Hermitian operator of the form

$$\begin{aligned}\hat{G}_a =& \theta_1 \left[1 + \frac{\theta_2^2}{4}\right] \left(\hat{I}_{1,y}\hat{S}_x - \hat{I}_{1,x}\hat{S}_y\right) + \theta_2 \left[1 + \frac{\theta_1^2}{4}\right] \left(\hat{I}_{2,y}\hat{S}_x - \hat{I}_{2,x}\hat{S}_y\right) \dots \\ &\dots + \theta_1 \theta_2 \left(\hat{I}_{1,y}\hat{I}_{2,z}\hat{S}_x + \hat{I}_{1,z}\hat{I}_{2,y}\hat{S}_x - \hat{I}_{1,x}\hat{I}_{2,z}\hat{S}_y - \hat{I}_{1,z}\hat{I}_{2,x}\hat{S}_y\right), \end{aligned} \tag{46}$$

where

$$\theta_1 = \frac{A_1}{(\gamma_e - \gamma_n) B_0} = -2.55 \times 10^{-3} \text{ rad and} \tag{47a}$$

$$\theta_2 = \frac{A_2}{(\gamma_e - \gamma_n) B_0} = -2.42 \times 10^{-4} \text{ rad}, \tag{47b}$$

in which the numerical values for the coefficients are $A_1$=95 MHz, $A_2$=9 MHz, $\gamma_n$=17.23 MHz T$^{-1}$, $\gamma_e$=-27.97 GHz T$^{-1}$, and $B_0$=1.33 T.

This unitary transforms the bare eigenbasis of $\hat{H}_0$ into a dressed basis that incorporates the effects of $\hat{V}$ such that we can project out the electron spin-up subspace, leaving only the two nuclear spin qubits. From the form of $\hat{G}_a$, it is evident that this change of basis will mix states in which the electron and exactly one of the nuclear spins have both flipped. This dressed basis will be closer to the true qubit states than the eigenstates of $\hat{H}_0$, though the effective Hamiltonian will still not be strictly diagonal even in the absence of the driving fields that perform one- and two-qubit gates. Before writing down the effective Hamiltonian in the dressed basis, it is instructive to consider the form of the dressed ground state in terms of the bare eigenbasis. Up to a constant for normalization, the leading contributions are

$$|\Uparrow\Uparrow\downarrow'\rangle \sim |\Uparrow\Uparrow\downarrow\rangle + \theta_1 |\Downarrow\Uparrow\uparrow\rangle + \theta_2 |\Uparrow\Downarrow\uparrow\rangle + \mathcal{O}(\theta_1\theta_2). \tag{48}$$

This indicates that the dominant effect of the off-diagonal component of the contact hyperfine interaction is to weakly mix the bare ground state with *both* of the states with opposite nuclear parity *and* opposite electronic spin. This also suggests that the always-on nature of the contact hyperfine interaction means that our qubit states aren't *strictly* the up and down states of the individual nuclear spins, but they are to a very good approximation thanks to the smallness

of $\theta_1$ and $\theta_2$ evident in Eq. 47. It is this weak mixing that is at the heart of effective entangling interactions between the nuclear spins. While the effective Hamiltonian expanded to third order will only include second order terms from $\hat{G}_a$, we note that we included third order terms in Eq. 46 to illustrate that the next order of corrections will not introduce any new mixing among the bare eigenstates. Thus it will be safe to project the third order Hamiltonian onto the electron spin-down subspace and to concatenate a second transformation.

Projecting onto the electron spin-down subspace, the effective Hamiltonian in the dressed basis defined by $\hat{G}_a$ is

$$\hat{H}'_\downarrow = \hat{H}_{0,\downarrow} + \hat{V}_{eff,\downarrow} + \mathcal{O}(\hat{V}^4), \tag{49a}$$

$$\hat{H}_{0,\downarrow} = \left(-\gamma_n B_0 - \frac{A_1}{2}\left[1 - \frac{\theta_1}{2} - \frac{\theta_2^2}{4}\right]\right)\hat{I}_{1,z} + \left(-\gamma_n B_0 - \frac{A_2}{2}\left[1 - \frac{\theta_2}{2} - \frac{\theta_1^2}{4}\right]\right)\hat{I}_{2,z}, \text{ and} \tag{49b}$$

$$\hat{V}_{eff,\downarrow} = \left(\frac{A_1\theta_2 + A_2\theta_1}{4}\left[1 - \frac{\theta_1 + \theta_2}{4}\right]\right)\left(\hat{I}_{1,x}\hat{I}_{2,x} + \hat{I}_{1,y}\hat{I}_{2,y}\right) + \left(\frac{A_1\theta_2 + A_2\theta_1}{4}\frac{\theta_1 + \theta_2}{2}\right)\hat{I}_{1,z}\hat{I}_{2,z}, \tag{49c}$$

where we have shifted away terms proportional to the identity by redefining the zero of energy. We see that the change of basis and projection has renormalized the diagonal component of the contact hyperfine interaction experienced by either nuclear spin. We also note that there are now two always-on entangling interactions between the dressed nuclear spin qubit states:

1. A relatively strong XX+YY interaction, with strength $\approx 3\,\text{kHz}$.

2. A relatively weak ZZ interaction, with strength $\approx 4\,\text{Hz}$.

We note that the quoted strengths include factors of 2 from spin matrices.

On the timescale of an $\approx 10\,\mu\text{s}$ one-qubit gate, these always-on interactions translate into a rotation of $\approx 30\,\text{mrad}$. While this rotation is the same order of magnitude as the $\approx 10\,\text{mrad}$ coherent errors inferred by GST, most of this interaction is mediated by a term of the form $(\hat{I}_{1,x}\hat{I}_{2,x} + \hat{I}_{1,y}\hat{I}_{2,y})$ and the coefficient of the $\hat{I}_{1,z}\hat{I}_{2,z}$ term is three orders of magnitude too small to be consistent with the ZZ errors inferred by GST. The stronger XX+YY interaction seems to be at odds with the GST model that gives the best rationalization of our data. But these simplistic considerations do not account for the fact that the inferred two-qubit errors occur in the presence of driving RF fields that implement one-qubit gates.

To facilitate the analysis of the impact of driving fields on this effective interaction we need to reduce the remaining off-diagonal coupling due to the XX+YY interaction. The first Schrieffer-Wolff transformation mixed states with opposite nuclear parity *and* opposite electronic spin, whereas a second unitary that reduces this coupling will mix states with fixed electronic spin and the same nuclear spin parity. While one might naively expect the size of this mixing to be much smaller than the first transformation, the generator $\hat{G}_b$ will have comparable weights because the ratio of the off-diagonal coupling to the diagonal matrix elements separating the coupled states are roughly the same order of magnitude. In other words, while the effective XX+YY coupling is three orders of magnitude weaker than the contact hyperfine interaction, it is between states that are separated by an energy scale that is three orders of magnitude weaker than the electronic Zeeman splitting, namely the difference in the contact hyperfine interaction between the two nuclei.

The desired form of $\hat{G}_{b\downarrow}$ is

$$i\hat{G}_{b\downarrow} = i\theta_3\left(\hat{I}_{1,y}\hat{I}_{2,x} - \hat{I}_{1,x}\hat{I}_{2,y}\right) \tag{50}$$

with the value of the mixing angle being

$$\theta_3 = \frac{(A_1\theta_2 + A_2\theta_1)\left(1 - \frac{\theta_1+\theta_2}{4}\right)}{4\left(E'_{\Downarrow\Uparrow} - E'_{\Uparrow\Downarrow}\right)} \approx \frac{A_1\theta_2 + A_2\theta_1}{2(A_1 - A_2)} = 2.67 \times 10^{-4}\text{ rad}, \tag{51}$$

where $E'_{\Downarrow\Uparrow}$ and $E'_{\Uparrow\Downarrow}$ are the diagonal components of the effective Hamiltonian for the subscripted basis states. This second transformation eliminates the residual off-diagonal coupling in the effective Hamiltonian at leading order in the XX+YY interaction strength. The concatenation of the two transformations thus provides a description of the nuclear spin qubit states that incorporates dressing of the bare qubit basis states by a coherent electron-mediated internuclear interaction.

### D. Analysis of the impact of the electron-mediated interaction on one-qubit gates

Finally, we consider the impact of the driving fields in terms of the doubly dressed Schrieffer-Wolff basis that defines our nuclear spin qubit states. This change of basis naturally includes the electron-mediated entangling interaction

between nuclear spins that might explain the presence of coherent entangling errors during our one-qubit gates. There are five distinct oscillatory contributions to the effective Hamiltonian expressed in this basis that might rationalize these errors.

1. Resonant magnetic drive of the nucleus on which one-qubit gates are targeted.

2. Off-resonant magnetic drive of the nucleus on which one-qubit gates are not targeted.

3. Off-resonant magnetic drive of the electronic spin that mediates the entangling interaction between the nuclei.

4. Electrical drive of both nuclear spins, through the modulation of their contact hyperfine couplings.

5. Electrical drive of the electron spin, through the modulation of its gyromagnetic ratio.

The first three contributions are due to the $B_1$ field that implements one-qubit NMR gates and the last two contributions are due to an errant electric component to that driving field, which is henceforth indicated by a parametric dependence on the electric potential ($V$). Each of these will contributions will have the form

$$\hat{H}_{drive}(t) = \Delta \hat{H}_{gen} \cos(\omega t - \phi), \tag{52}$$

where $\Delta$ is the drive amplitude in units of frequency, $\hat{H}_{gen}$ is a unitless operator through which the drive generates a rotation, $\omega$ is the frequency of the drive chosen to be resonant with driving exactly one of the nuclear spin flip transitions, and $\phi$ is the phase of the drive that determines the precise rotation being implemented. At linear order in the Schrieffer-Wolff generators, these terms will introduce coherent errors of the form

$$\hat{H}_{err}(t) = \Delta \left( \left[ i\hat{G}_{a\downarrow}, \hat{H}_{gen} \right] + \left[ i\hat{G}_{b\downarrow}, \hat{H}_{gen} \right] \right) \cos(\omega t - \phi). \tag{53}$$

Here we have extended the subscript notation for $\hat{G}_{b\downarrow}$ to $\hat{G}_{a\downarrow}$ to represent the combined action of $\hat{G}_a$ and a projection onto the electron spin-down subspace. The effect of this projection is to discard terms that are proportional to $\hat{S}_x$ or $\hat{S}_y$, which lead to stochastic errors when tracing out the electron. It will be evident that terms that are higher order in the generators will be too small to plausibly rationalize the errors of interest, as we will be unable to find sufficiently strong errors at leading order. We provide the form for each of these errors in Table S2.

| $\Delta$ | $\hat{H}_{gen}$ | $\left[ i\hat{G}_{a\downarrow}, \hat{H}_{gen} \right]$ | $\left[ i\hat{G}_{b\downarrow}, \hat{H}_{gen} \right]$ |
|---|---|---|---|
| $-\gamma_n B_1 \approx -25\,\text{kHz}$ | $\hat{I}_{1,x/y}$ | 0, $\left[ i\hat{G}_a, \hat{H}_{gen} \right]$ strictly acts on $\hat{S}_{x/y}$ | $\theta_3 \hat{I}_{1,z} \hat{I}_{2,x/y}$ |
| $-\gamma_n B_1 \approx -25\,\text{kHz}$ | $\hat{I}_{2,x/y}$ | 0, $\left[ i\hat{G}_a, \hat{H}_{gen} \right]$ strictly acts on $\hat{S}_{x/y}$ | $\theta_3 \hat{I}_{1,x/y} \hat{I}_{2,z}$ |
| $-\gamma_e B_1 \approx 40\,\text{MHz}$ | $\hat{S}_{x/y}$ | $-\frac{\theta_1}{2} \hat{I}_{1,x/y} - \frac{\theta_2}{2} \hat{I}_{2,x/y} - \frac{\theta_1 \theta_2}{2} \left( \hat{I}_{1,x/y} \hat{I}_{2,z} + \hat{I}_{1,z} \hat{I}_{2,x/y} \right),$ | 0, due to electron spin-down projection |
| $\delta A_1(V) \approx -100\,\text{kHz}$ | $\hat{\vec{I}}_1 \cdot \hat{\vec{S}}$ | $\frac{\theta_1}{2} \hat{I}_{1,z} - \frac{\theta_2}{2} \left[ 1 + \frac{\theta_1}{2} \right] \left( \hat{I}_{1,x} \hat{I}_{2,x} + \hat{I}_{1,y} \hat{I}_{2,y} \right) - \frac{\theta_1 \theta_2}{4} \left( \hat{I}_{2,z} - 2\hat{I}_{1,z} \hat{I}_{2,z} \right)$ | $-\theta_3 \left( \hat{I}_{1,x} \hat{I}_{2,x} + \hat{I}_{1,y} \hat{I}_{2,y} \right)$ |
| $\delta A_2(V) \approx 70\,\text{kHz}$ | $\hat{\vec{I}}_2 \cdot \hat{\vec{S}}$ | $\frac{\theta_2}{2} \hat{I}_{2,z} - \frac{\theta_1}{2} \left[ 1 + \frac{\theta_2}{2} \right] \left( \hat{I}_{1,x} \hat{I}_{2,x} + \hat{I}_{1,y} \hat{I}_{2,y} \right) - \frac{\theta_1 \theta_2}{4} \left( \hat{I}_{1,z} - 2\hat{I}_{1,z} \hat{I}_{2,z} \right)$ | $-\theta_3 \left( \hat{I}_{1,x} \hat{I}_{2,x} + \hat{I}_{1,y} \hat{I}_{2,y} \right)$ |
| $\delta\gamma_e(V) B_1 \approx 13\,\text{Hz}$ | $\hat{S}_{x/y}$ | $-\frac{\theta_1}{2} \hat{I}_{1,x/y} - \frac{\theta_2}{2} \hat{I}_{2,x/y} - \frac{\theta_1 \theta_2}{2} \left( \hat{I}_{1,x/y} \hat{I}_{2,z} + \hat{I}_{1,z} \hat{I}_{2,x/y} \right)$ | 0, due to electron spin-down projection |
| $\delta\gamma_e(V) B_0 \approx 12\,\text{kHz}$ | $\hat{S}_z$ | $\left[ i\hat{G}_a, \hat{H}_{gen} \right]$ strictly acts on $\hat{S}_{x/y}$ | 0, due to electron spin-down projection |

Table S2. A summary of the forms of the error terms in Eq. 53 for each of the five mechanisms described at the beginning of Section S11 D. Estimates for the electric modulation of the hyperfine couplings and electron gyromagnetic ratio are taken from experiment, $\delta A_1(V) \approx -10\,\text{MHz}\,\text{V}^{-1} \times 10\,\text{mV} \approx -100\,\text{kHz}$, $\delta A_2(V) \approx 7\,\text{MHz}\,\text{V}^{-1} \times 10\,\text{mV} \approx 70\,\text{kHz}$, and $\delta\gamma_e(V) \approx -0.9\,\text{MHz}\,\text{V}^{-1}\,\text{T}^{-1} \times 10\,\text{mV} \approx 9\,\text{kHz}\,\text{T}^{-1}$. We note that contributions that are third order in the generators are not indicated because they strictly renormalize errors that already occur at lower orders.

We next examine each entry of Table S2 to assess its impact on the execution of one-qubit gates. It is worth noting that the terms that resolve to zero because they act on $\hat{S}_{x/y}$ will still contribute to stochastic errors, but the focus of this assessment is on rationalizing the relatively strong coherent entangling errors. The only terms that will contribute appreciably to one-qubit gates are those that involves generators that flip exactly one of the nuclear spins. This is essential because of the oscillatory time dependence on all terms except for the term in the sixth row, which involves simultaneous electric and magnetic modulation resulting in an always-on term and a term that oscillates at twice the

driving frequency. For all other terms, those that flip exactly one nuclear spin will be constant in the rotating frame associated with driving a one-qubit gate on that particular nucleus.

We now summarize the strength and form of the coherent errors in Table S2. We will consider errors occurring during XI and IX gates for convenience and generalization to YI and IY gates is straightforward.

1. Applying an XI (IX) gate leads to an error of the form ZX (XZ) as is evident in the first two rows of Table S2. However, because the resonance frequencies of the two nuclei are different, this error will remain oscillatory in the rotating frame of the gate being applied. As the amplitude of this error is ≈2 Hz, this effect does not rationalize the errors inferred in the GST model.

2. However, the off-resonant magnetic drive on the *other* qubit when applying an XI (IX) gate will generate a XZ (ZX) error that is constant in the rotating frame of the gate being applied. However, over the course of an ≈10 µs one-qubit gate, this will only lead to an XZ (ZX) rotation of ≈17 µrad, which is still three orders of magnitude weaker than the plausible XZ (ZX) errors in the GST analysis.

3. Off-resonant magnetic drive of the electron that mediates the entangling interaction between nuclei gives rise to a number of errors evident in the third row of Table S2. The leading order contribution when driving an XI (IX) gate itself an XI (IX) "error". Really, this is only an error in as far as it shifts the Rabi frequency and comparable physics has been predicted and observed in diamond NV centers [18, 19]. In spite of the smallness of $\theta_1$ ($\theta_2$), this shift is still 10s of kHz because of the largeness of $\gamma_e$ and it is evident in our experiments. There is also a XZ (ZX) error proportional to $\theta_1\theta_2$. Over the course of an ≈10 µs one-qubit gate, this will lead to an XZ (ZX) rotation of ≈31 µrad, which is three orders of magnitude weaker than the plausible XZ (ZX) errors in the GST analysis.

4. Electric drive of the contact hyperfine couplings also gives rise to a number of errors, evident in the fourth and fifth rows of Table S2. However, each of them is rapidly oscillatory in the frame of any one-qubit X or Y gate. The amplitude of the aggregate XX+YY error is ≈19 Hz, so this effect does not rationalize the errors inferred in the GST model.

5. Electric drive of the electron gyromagnetic ratio gives rise to errors evident in rows six and seven of Table S2. Modulation of the coefficient on the oscillatory $B_1$ drive will give rise to a constant term and a term that oscillates at twice the drive frequency in a non-rotating frame. However the strength of the former will be exceptionally small, with XZ and ZX terms that have totally negligible amplitudes of $\approx 1$ µHz. Modulation of the coefficient on the static $B_0$ coupling will strictly lead to stochastic errors.

In summary, we have shown that a microscopic model that accounts for all of the known physics in the NMR drive of our three-qubit system fails to rationalize the relatively large coherent entangling errors on one-qubit gates that were inferred by GST. The error mechanisms that come the closest to rationalizing these errors are due to an indirect electron-mediated entangling interaction between the nuclei, and the coupling of the magnetic drive to either the shared electron or the nucleus that isn't being resonantly driven. However, the strength of the former (latter) mechanism is two (three) orders of magnitude weaker than the comparable errors that were inferred by GST.

### E. Leakage of the ESR carrier signal leads to coherent entangling errors

In order to rationalize the observed ZZ errors, we instead need to turn to the leakage of microwaves near the ESR transition frequencies. The ESR pulses are driven by the modulation of a drive at a carrier frequency chosen to be off resonance relative to all four of the electron spin flip (ESR) transitions associated with the distinct nuclear spin basis states. When the ESR pulses are being driven, it is one of the sidebands generated by this modulation that is on resonance with one of these transitions. However, when these pulses aren't being driven, the carrier frequency persists and some fraction of it will leak through the IQ-mixer.

The off-resonance drive provided by the leaked microwaves at the carrier frequency persist even during the NMR gates, which are produced by a separate signal generator. This leads to oscillations in the electron spin state, with a small amplitude proportional to the ratio of the intensity of the oscillating magnetic field produced by the carrier leakage and the detuning of the carrier frequency relative to any individual ESR transition. The frequency of these oscillations is determined by this same detuning, which is on the order of 10s of MHz allowing for the accumulation of many of these small oscillations over the course of an NMR gate. This detuning and the leakage magnetic field will differ for each of the individual nuclear spin basis states upon which the electron transitions are conditioned. Each basis state will then experience the accumulation of a distinct and spurious geometric phase due to these small but fast oscillations.

That these spurious phases depend on the support of the nuclear spin state on each of the relevant basis states during an NMR gate gives rise to the ZZ nature of the error that these leakage fields induce, rationalizing the coherent ZZ errors inferred by GST. Note also that the double-sideband modulation that is used to drive any individual ESR transition frequency introduces another source of off-resonance driving that introduces a comparable spurious accumulation of a geometric phase during the ESR pulse, i.e. the CZ gate, as well.

We next provide quantitative estimates of the size of errors due to leakage of the ESR carrier through the IQ-mixer. It is useful to begin by considering the impact of off-resonantly driving one of the ESR transitions for a fixed nuclear spin configuration. Without loss of generality, we consider this drive within the subspace spanned by $\{|\Uparrow\Uparrow\downarrow''\rangle, |\Uparrow\Uparrow\uparrow''\rangle\}$, where the $''$ notation indicates that these basis states are in the doubly Schrieffer-Wolff transformed frame. We provide the caveat that we are no longer considering projection onto the electron spin-down subspace prior to the second transformation and will thus require augmentation of the generating unitary to accommodate removal of the effective XX+YY interaction on the electron spin-up space, as well.

It is then straightforward to adapt the solution to the off-resonant Rabi problem to the evolution of the dressed electron spin qubit states within this particular subspace. The effective rotating-frame Hamiltonian within this subspace is given as

$$H''_{\Uparrow\Uparrow} = \Delta_{\Uparrow\Uparrow} \hat{S}''_z + \Omega \hat{S}''_x, \tag{54}$$

where $\Delta_{\Uparrow\Uparrow}$ is the detuning of the carrier frequency relative to the ESR spin-flip transition frequency when the nuclei are in the $\Uparrow\Uparrow$ state (positive when the carrier frequency is less than the transition frequency), $\Omega$ is the effective Rabi frequency due to the coupling of the leakage field to the electron, and the notation $''$ indicates that the electron spin matrices are defined relative to the natural rotating frame in the doubly transformed basis. $\Omega$ is proportional to the effective $B_1$ field produced by the leaking carrier signal, which we will estimate based on the outcomes of our GST analysis. It is worth noting that because we are in the doubly Schrieffer-Wolff transformed frame $\Omega$ will be dressed by a nuclear-spin-dependent coupling of the leaked field to the nuclei beyond leading order. But this is negligible for the purposes of our estimates and we will assume that $\Omega$ is identical for all nuclear spin configurations.

The leakage field will drive small oscillations on the Bloch sphere associated with the dressed electron spin qubit basis states with frequency $\Omega_{\text{eff},\Uparrow\Uparrow} = \sqrt{\Omega^2 + \Delta_{\Uparrow\Uparrow}^2}$. The amplitude of these oscillations will be such that the maximum probability of the electron spin being observed as $|\uparrow\rangle$ is $\Omega^2/\Omega^2_{\text{eff},\Uparrow\Uparrow}$. The angle of the axis of this rotation is defined by the usual effective magnetic field, $\vec{B}_{\text{eff}} = \Omega\hat{x} + \Delta_{\Uparrow\Uparrow}\hat{z}$ (see the arrows indicated in Fig. S10). Over the period of one of these oscillations, $1/\Omega_{\text{eff},\Uparrow\Uparrow}$, the electron spin qubit will accumulate a geometric Aharonov-Anandan phase [20] as it moves in a small circle on the Bloch sphere starting and ending at $|\Uparrow\Uparrow\downarrow''\rangle$. The associated geometric phase is given as

$$\beta_{\Uparrow\Uparrow} = \pi\left(1 - \frac{\Delta_{\Uparrow\Uparrow}}{\Omega_{\text{eff},\Uparrow\Uparrow}}\right). \tag{55}$$

We will find it convenient to introduce a tilde to indicate the total geometric phase accumulated over a time $t$,

$$\widetilde{\beta_{\Uparrow\Uparrow}} = \beta_{\Uparrow\Uparrow}\Omega_{\text{eff},\Uparrow\Uparrow} t = \pi \Omega t \left(\frac{\Omega}{2\Delta_{\Uparrow\Uparrow}} + \mathcal{O}\left(\left[\frac{\Omega}{\Delta_{\Uparrow\Uparrow}}\right]^3\right)\right). \tag{56}$$

This picture can be extended to the other nuclear spin basis states noting that each will accumulate distinct geometric phases, $\beta_{\Downarrow\Downarrow}$, $\beta_{\Downarrow\Uparrow}$, $\beta_{\Uparrow\Downarrow}$, and $\beta_{\Uparrow\Uparrow}$. In the rotating frame of an NMR pulse implementing an $X_{\pi/2} \otimes I$ gate, the effective Hamiltonian accounting for the geometric phase due to leakage is

$$H_{X_{\pi/2}\otimes I} = \frac{\pi}{4}\left(e^{i(\widetilde{\beta_{\Downarrow\Downarrow}} - \widetilde{\beta_{\Uparrow\Downarrow}})}|\Downarrow\Downarrow''\rangle\langle\Uparrow\Downarrow''| + e^{i(\widetilde{\beta_{\Uparrow\Uparrow}} - \widetilde{\beta_{\Downarrow\Uparrow}})}|\Uparrow\Uparrow''\rangle\langle\Downarrow\Uparrow''| + h.c.\right), \tag{57}$$

which can be decomposed into tensor products of Pauli matrices as

$$H_{X_{\pi/2}\otimes I} = \frac{\pi}{8}\Big(\text{Re}\{e^{i(\widetilde{\beta_{\Downarrow\Downarrow}} - \widetilde{\beta_{\Uparrow\Downarrow}})} + e^{i(\widetilde{\beta_{\Uparrow\Uparrow}} - \widetilde{\beta_{\Downarrow\Uparrow}})}\}X \otimes I - \text{Im}\{e^{i(\widetilde{\beta_{\Downarrow\Downarrow}} - \widetilde{\beta_{\Uparrow\Downarrow}})} - e^{i(\widetilde{\beta_{\Uparrow\Uparrow}} - \widetilde{\beta_{\Downarrow\Uparrow}})}\}Y \otimes I \ldots$$
$$\ldots + \text{Re}\{e^{i(\widetilde{\beta_{\Downarrow\Downarrow}} - \widetilde{\beta_{\Uparrow\Downarrow}})} - e^{i(\widetilde{\beta_{\Uparrow\Uparrow}} - \widetilde{\beta_{\Downarrow\Uparrow}})}\}X \otimes Z - \text{Im}\{e^{i(\widetilde{\beta_{\Downarrow\Downarrow}} - \widetilde{\beta_{\Uparrow\Downarrow}})} + e^{i(\widetilde{\beta_{\Uparrow\Uparrow}} - \widetilde{\beta_{\Downarrow\Uparrow}})}\}Y \otimes Z\Big), \tag{58a}$$

being careful to note that we have factored out a common coefficient of $1/2$. This expression can be further simplified at first order in the geometric phases,

$$H_{X_{\pi/2}\otimes I} = \frac{\pi}{4}\left(X \otimes I - \frac{1}{2}\left(\widetilde{\beta_{\Downarrow\Downarrow}} - \widetilde{\beta_{\Uparrow\Downarrow}} - \widetilde{\beta_{\Uparrow\Uparrow}} + \widetilde{\beta_{\Downarrow\Uparrow}}\right)Y \otimes I - \frac{1}{2}\left(\widetilde{\beta_{\Downarrow\Downarrow}} - \widetilde{\beta_{\Uparrow\Downarrow}} + \widetilde{\beta_{\Uparrow\Uparrow}} - \widetilde{\beta_{\Downarrow\Uparrow}}\right)Y \otimes Z\right). \tag{59}$$

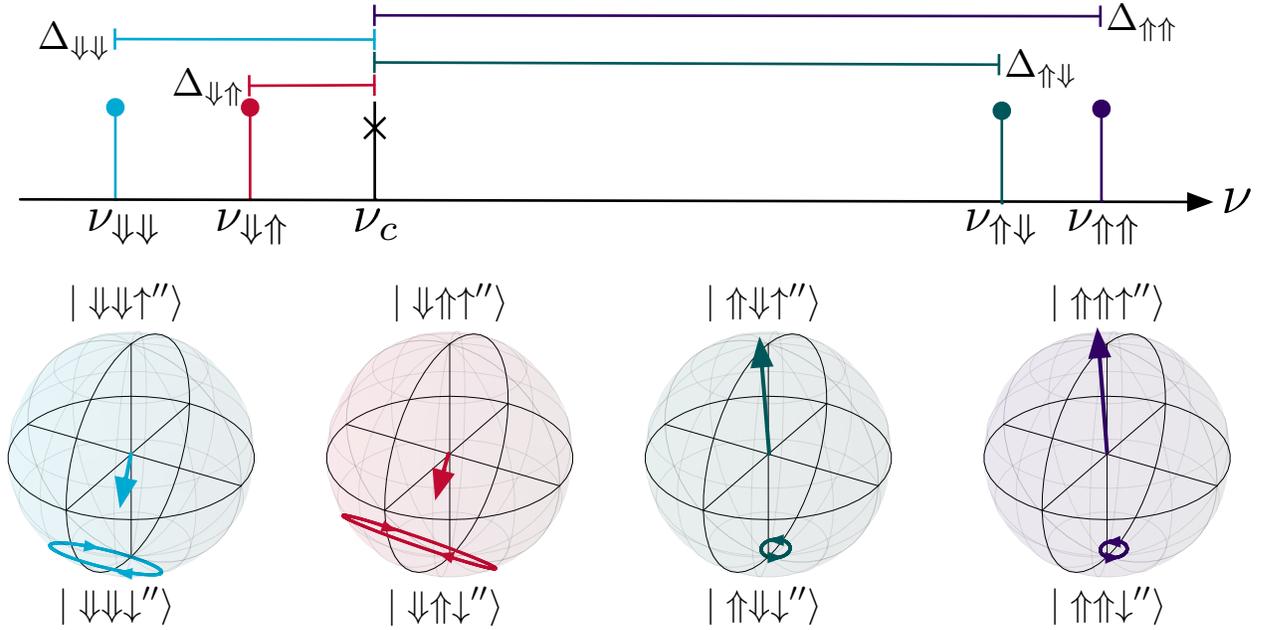

Figure S10. (Bottom panel) Bloch spheres for the electron spin qubit in each of the four nuclear spin basis states. The coloured vectors represent the effective driving field in the rotating frame, $\vec{B}_{eff}$, composed of the vector sum of the detuning and the driving field $\Omega$, caused by the leakage of the carrier signal of the microwave source, at frequency $\nu_c$, and the detunings $\Delta_{**}$ (where $**$ indicates any of the four nuclear orientations) from the individual ESR frequencies ESR spin-flip transitions $\nu_{**}$ (top panel). Starting from an electron spin in the $|\downarrow\rangle$ state (south pole of the Bloch sphere), the microwave leakage will lead to off-resonant driving of the electron spin through a trajectory indicated by the circles on each sphere, about an axis of rotation determined by $\vec{B}_{eff}$. These oscillations are quite small ($\approx 1$ mrad): for illustrative purposes, the value of $\Omega$ is drawn 100 times larger than the true experimental value. Even though the oscillations are 100 times smaller than illustrated, the associated geometric phases accumulate over the hundreds of cycles of the single-qubit NMR gates. While the effective Rabi frequency due to the leakage field is approximately the same for each nuclear spin basis state, the associated detunings vary over a range of 100 MHz (top). The basis state with the smallest detuning (in magnitude) will accumulate the largest geometric phase, in this case $|\Downarrow\Uparrow\uparrow''\rangle$ The accumulation of different geometric phases across each nuclear spin basis states leads to coherent entangling errors on the nominally single-qubit gates. This figure was generated with the help of plotting routines in QuTiP [21].

This form of the rotating frame Hamiltonian makes it evident as to how leakage of the ESR carrier will lead to coherent errors on the nominally single-qubit gate. A unitary model for the gate itself is realized by the exponentiation of Eq. 59, leading to coherent single-qubit YI and ZI errors at first order in the geometric phase factors, as well as coherent YZ and ZZ errors. The ZI and ZZ errors arise due to the commutator between the XI term (rate $\approx 0.8$) and the YI and YZ terms, appearing at first order in the geometric phases in the Zassenhaus formula. For a $\pi/2$ rotation the size of the ZI and ZZ errors will then be comparable to the YI and YZ errors, respectively, but they will be proportionally smaller for a rotation by a smaller angle (e.g., a rate $\ll 0.8$). It is straight forward to work out these same details for the other single-qubit gates, yielding two-qubit coherent errors of the form predicted by GST.

The carrier frequency used in our experiment was $\nu_c = 37\,100.4125$ MHz, and typical detunings from one set of measurements are $\Delta_{\Downarrow\Downarrow}=-18.8225$ MHz, $\Delta_{\Downarrow\Uparrow}=-11.6625$ MHz, $\Delta_{\Uparrow\Downarrow}=80.4375$ MHz, and $\Delta_{\Uparrow\Uparrow}=87.5475$ MHz. For these values, we can rationalize $\approx 1\%$ two-qubit coherent errors on a 20 µs NMR gate with an effective leakage $B_1$ field of 5.8 µT. This translates to a 22 dB attenuation of the carrier relative to the 725 kHz ESR Rabi frequency driven with double-sideband modulation. This is a high but not implausible level of leakage, considering that we did not calibrate the IQ-mixer prior to the experiments.

For readers most accustomed to the standard picture of spin resonance, it is worth highlighting that this leakage has negligible effect on the electron spin state itself, as one can easily verify from a simple Rabi formula: the far-off-resonance drive on the electron resets to zero periodically and often. Conversely, the geometric phase imparted by electron on the nuclei does not reset – it *accumulates* with time. Since the imparted geometric phase differs for each of the four

ESR frequencies, due to the different values of $\Delta_{**}$, it does not amount to an irrelevant global phase.

	
\end{document}